\documentclass[aps,twocolumn,superscriptaddress]{revtex4} %È¥µôdvips¿ÉÒÔÖ±½Ópdf ±àÒë¿´µ½ÑÕÉ«
\usepackage{feynmp}
\usepackage{amssymb}
\usepackage{amsmath}
\usepackage{epsfig}
\usepackage{graphicx}
\usepackage{subfigure}
\usepackage{mathrsfs}
\usepackage{hyperref}
\usepackage{cleveref}
\usepackage{cases}
\usepackage{bbm}
\usepackage{xcolor}
\usepackage{tabularx}
\usepackage{array}
\usepackage{multirow}
\usepackage{CJK}
\usepackage{lmodern}
\usepackage{textcomp}
\usepackage{makecell}
\usepackage{threeparttable}
\usepackage{ulem}

\newcommand{\red}[1]{\textcolor{red}{#1}}
\newcommand{\blue}[1]{\textcolor{blue}{#1}}

\definecolor{light-gray}{gray}{0.65}

\allowdisplaybreaks[3]

\UseRawInputEncoding

\begin{document}

%\begin{CJK*}{GBK}{song}

%%%%%%%%%%%%%%%%%%%%%%%%%%%%%%%%%%%%%%%%%%%%%%%%%%%%%%%%%%%%%%%%%%%%%%%%%%%%%%%%%%%%%%
%%%%%%%%%%%%%%%%%%%%%%%%%%%%%%%%%%%%%%%%%%%%%%%%%%%%%%%%%%%%%%%%%%%%%%%%%%%%%%%%%%%%%%

\title{Critical fates induced by the interaction competition in three-dimensional tilted Dirac semimetals}

\date{\today}

\author{Jing Wang}
\altaffiliation{Corresponding author: jing$\textunderscore$wang@tju.edu.cn}
\affiliation{Department of Physics, Tianjin University, Tianjin 300072, P.R. China}
\affiliation{Tianjin Key Laboratory of Low Dimensional Materials Physics and
Preparing Technology, Tianjin University, Tianjin 300072, P.R. China}

\author{Jie-Qiong Li}
\affiliation{Department of Physics, Tianjin University, Tianjin 300072, P.R. China}

\author{Wen-Hao Bian}
\affiliation{Department of Physics, Tianjin University, Tianjin 300072, P.R. China}
\affiliation{School of Physics, Nanjing University, Nanjing, Jiangsu 210093, P.R. China}

\author{Qiao-Chu Zhang}
\affiliation{Department of Physics, Tianjin University, Tianjin 300072, P.R. China}

\author{Xiao-Yue Ren}
\affiliation{Department of Physics, Tianjin University, Tianjin 300072, P.R. China}

%Ä¿Ç°À´¿´£¬Ö»Ðè°Ñrevtex4-1¸Ä³Érevtex4¾Í¿ÉÒÔ½â¾ö×÷ÕßµÄindexÎÊÌâ

\begin{abstract}

The interplay among Coulomb interaction, electron-phonon coupling, and phonon-phonon coupling
has a significant impact on the low-energy behavior of three-dimensional type-I tilted Dirac semimetals.
To investigate this phenomenon, we construct an effective theory, calculate one-loop
corrections arising from all these interactions, and
establish the coupled energy-dependent flows of all associated interaction parameters
by adopting the renormalization-group approach. Deciphering such coupled evolutions allows us to
determine a series of low-energy critical properties for these materials.
At first, we present the low-energy tendencies of all interaction parameters.
The tilting parameter exhibits distinct tendencies that
depend heavily upon the initial anisotropy of fermion velocities.  In comparison,
the latter is mainly dominated by its initial value but is less sensitive to the former.
Variations in these two quantities drive certain interaction parameters toward the strong anisotropy in the low-energy regime,
indicating the screened interaction in specific directions, and others toward an approximate isotropy.
Additionally, we observe that the tendencies of interaction parameters can be qualitatively clustered into three distinct
types of fixed points, accompanied by the potential instabilities that induce an interaction-driven phase transition to a certain
superconducting state. Furthermore, approaching these fixed points leads to the critical behavior
of physical quantities, such as the density of states,
compressibility, and specific heat, which exhibit quite
different from their noninteracting counterparts and even deviate slightly
from Fermi-liquid behavior. Our investigation sheds light on the intricate
relationship between different types of interactions in these semimetals and provides
useful insights into their fundamental properties.

\end{abstract}

\maketitle

%%%%%%%%%%%%%%%%%%%%%%%%%%%%%%%%%%%%%%%%%%%%%%%%%%%%%%%%%%%%%%%%%%%%%%%%%%%%%%%%%%%%%
%%%%%%%%%%%%%%%%%%%%%%%%%%%%%%%%%%%%%%%%%%%%%%%%%%%%%%%%%%%%%%%%%%%%%%%%%%%%%%%%%%%%%

\section{Introduction}

In recent years, the study of Dirac semimetals (DSM), featuring the intermediate properties between
metals and insulators, has become one of the most active fields in contemporary condensed-matter physics~\cite{Novoselov2005Nature,Castro2009RMP,Moore2010Nature,Hasan2010RMP,Qi2011RMP,
Vafek2014ARCMP,Wehling2014AP,Wang2012PRB,Young2012PRL,Steinberg2014PRL,Liu2014NM,Liu2014Science,
Xiong2015Science,Roy2009PRB,Roy2016,Roy-2014-2016,Savary2014PRB,Moon2014PRX,Montambaux}. Typically,
DSMs possess the Dirac cones and reduced Fermi surfaces composed of discrete Dirac points,
exhibiting gapless low-energy excitations irrespective of the microscopic details. These materials display
linear energy dispersions along two or three directions~\cite{Novoselov2005Nature,
Castro2009RMP,Moore2010Nature,Roy2009PRB,Wang2012PRB,Young2012PRL,Young2012PRL,Liu2014NM,Liu2014Science,
Xiong2015Science,Korshunov2014PRB,Hung2016PRB,Nandkishore2013PRB,Potirniche2014PRB,Nandkishore2017PRB,Roy2016,Roy-2014-2016}.
In particular, their unique properties are guaranteed and protected by kinds of symmetries, including
time-reversal, space-reversal symmetry, etc.~\cite{Castro2009RMP,Vafek2014ARCMP}. However, these Dirac cones
can be stretched and thus tilted by breaking
%fundamental Lorentz symmetry or so-called $t$-Lorentz symmetry~\cite{Jafari2019PRB-t}
a certain fundamental symmetry (such as $t$-Lorentz symmetry)~\cite{Trescher2015PRB,Soluyanov2015Nature,Jafari2019PRB-t} or via an
additional force in a certain direction~\cite{Mao2011ACS}. Consequently, the energy dispersions become anisotropic, resulting in unequal fermion velocities along distinct directions. Henceforth, these materials are known as
tilted Dirac semimetals (tDSM)~\cite{Trescher2015PRB,Soluyanov2015Nature,Lee2018PRB,Lee2019PRB,Peres2010RMP,Jafari2019PRB-t,Peres2010RMP,
Soluyanov2015Nature,Noh2017PRL,Fei2017PRB,Mao2011ACS}. Besides two-dimensional (2D) tilted Dirac cones were reported in an organic compound
$\alpha-(\mathrm{BEDT-TTF})_{2}\mathrm{I}_{3}$ and certain mechanically deformed
graphene~\cite{Katayama2006JPSJ,Kobayashi2007JPSJ,Goerbig2008PRB},
three-dimensional (3D) tilted %Weyl cones were also realized later in $\mathrm{WTe}_{2}$~\cite{Soluyanov2015Nature}
cones have also been realized later in $\mathrm{WTe}_{2}$~\cite{Soluyanov2015Nature}
and the Fulde-Ferrell ground state of a spin-orbit coupled fermionic
superfluid~\cite{Xu2015PRL} or a cold-atom optical lattice~\cite{Xu2016PRA}.
Conventionally, the tDSM can be categorized into two
distinct types based on the tilted angles. Type-I tDSM retains
analogous Dirac cones as long as the tilted angle is insufficient to destroy
the point-like Fermi surface~\cite{Peres2010RMP,Jafari}. In contrast, for type-II tDSM,
such as $\mathrm{PdTe_2}$~\cite{Noh2017PRL,Fei2017PRB}
and $\mathrm{PtTe_2}$~\cite{Yan2017NC}, the Dirac point would be replaced by two straight lines,
indicating the open Fermi surface once the tilted angle reaches a sufficient magnitude~\cite{Soluyanov2015Nature,Lee2018PRB,Jafari2019PRB-t}.

These tilted materials have recently garnered significant attention owing to their unique
low-energy excitations and tilted Dirac cones~\cite{Shekhar2015NP,Parameswaran2014PRX,
Potter2014NC,Baum2015PRX,Arnold2016NC,Zhang2016NC,Lee2018PRB,Lee2019PRB,Fritz2017PRB,Fritz2019arXiv,
Jafari2018PRB,Alidoust2019arXiv,Yang2018PRB,Trescher2015PRB,Proskurin2015PRB,
Brien2016PRL,Zyuzin2016JETPL,Ferreiros2017PRB,Qiong2019NPJB}. Particularly, the effects
of Coulomb interaction and impurities on the low-energy properties of tDSM
were investigated by many groups~\cite{Lee2018PRB,Lee2019PRB,Fritz2017PRB,Fritz2019arXiv}.
However, previous studies on 3D tDSM have insufficiently considered several physical ingredients, such as
phonons and electron-phonon interactions, and their interplay with Coulomb interaction.
These additional degrees of freedom may play a critical role in determining the low-energy behavior of 3D tDSM.
Neglecting them could result in the partial or incomplete capture of important physical information that is
closely associated with such interactions. Therefore, to enhance our understanding of 3D tilted materials, it is essential
to carefully examine how the interplay between Coulomb interaction and electron-phonon
coupling affects the low-energy behavior of 3D tDSMs.

Without loss of generality, we within this work concentrate on the type-I 3D tDSM.
Compared with their 2D counterparts, these materials are more complicated but interesting. On one hand,
as the density of the state of fermionic quasiparticles vanishes as approaching the Dirac point
of type-I tDSM~\cite{Peres2010RMP,Jafari,Fritz2019arXiv},
it is necessary to take into account the effects of long-range
Coulomb interaction between low-energy fermionic excitations, which is marginal at the tree
level in the RG language. On the other hand, the lattice vibrations in 3D materials are more
intricate and lead to the emergence of (acoustic or optical) phonons.
These phonons exhibit different internal properties for ionic and covalent crystals
(For the sake of simplicity, this work is restricted to the latter, wherein the
coupling between phonon and Coulomb auxiliary field vanishes)~\cite{Ruhman2019PRX}.
Phonons not only interact with each other via phonon-phonon
interactions but also inevitably entangle with low-energy fermions, potentially
competing indirectly with Coulomb interaction. In this context, we expect
phonons and their related consequences to play an essential role in governing
the low-energy physics of 3D tDSM. To investigate the unusual behavior
of 3D tDSM in the low-energy regime, we are therefore forced to take into account all these items
on the same footing.

To this end, we employ the powerful renormalization-group (RG)
approach~\cite{Wilson1975RMP,Polchinski9210046,Shankar1994RMP} to treat all physical ingredients
mentioned above on the same footing. Specifically, we construct the effective theory for type-I tDSM and derive the coupled
energy-dependent evolutions of all interaction parameters by carrying out
the energy-shell RG analysis. Subsequently, after performing a numerical analysis of such evolutions,
we systematically investigate the low-energy fates of these interactions and elucidate their implications on the 3D tDSM.

At first, we examine the behavior of various interaction parameters in the low-energy regime, considering
the intimate competition among them. To be specific, we find that the tilting parameter is
insensitive to its initial value but is significantly influenced by the starting anisotropy of
fermion velocities. Regarding the fermion velocities, their ratio can either increase, decrease, or remain
nearly unchanged in the low-energy regime, depending upon their initial anisotropy and tilting parameter.
Turning attention to the dielectric constant, which characterizes the strength of
Coulomb interaction, we notice its tendency to exhibit a strong anisotropy in the low-energy regime, indicating the
screened Coulomb interaction in a certain direction. In comparison, both phonon velocities
and phonon-phonon interaction can either flow towards the approximate isotropy or exhibit the basic results
of the dielectric constant by flowing towards an extreme anisotropy in the low-energy regime.
It is of also noteworthy to point out that the coupling strength of electron-auxiliary bosonic interaction
and the electron-phonon interactions bear similar behavior to that of the dielectric constant and
phonon velocities, respectively.

In addition, we identify three kinds of fixed points by categorizing the energy-dependent
tendencies of all interaction parameters and then investigate the leading instabilities around such
fixed points. Furthermore, the critical tendencies of physical quantities,
including the density of states and compressibility as well as specific heat, are carefully studied as the system
approaches these three distinct types of fixed points. They present different but interesting behavior
compared with their noninteracting counterparts.

The rest of this work is organized as follows. In Sec.~\ref{Sec_model}
we present the microscopic model and outline the low-energy effective theory.
The Sec.~\ref{Sec_RGEqs} is followed to bring out the RG transformations
and then derive the coupled energy-dependent RG equations of all interaction parameters in
our effective theory after taking into account all one-loop corrections.
Afterwards, we within Sec.~\ref{Sec_SM-tend-fates} delve into the
tendencies and fates of these interaction parameters in the low-energy regime by examining evolved the RG flows.
Moving forward to Sec.~\ref{Sec_FP-instab} and Sec.~\ref{Sec_critical_implications},
we present the dominant instabilities and behavior of physical implications as the system approaches
the fixed points, respectively. Finally, a brief summary is provided in Sec.~\ref{Sec_summary}.

\section{Microscopic model and effective theory}\label{Sec_model}

We focus our attention on the 3D tDSM, characterized by an additional term that tilts the energy bands.
Armed with the microscopic structure of 3D tDSM, the noninteracting
Hamiltonian density in the low-energy regime can be formally
expressed as follows~\cite{Goerbig2008PRB,Lee2018PRB,Fritz2019arXiv}:
\begin{eqnarray}
\mathcal{H}_{0}=\zeta v_{z}k_{z}\sigma_{0}+\chi[v_{z}k_{z}\sigma_{z}
+v(k_{x}\sigma_{x}+k_{y}\sigma_{y})],\label{Eq H_0}
\end{eqnarray}
where $v$ and $v_z$ denote the fermion velocities along the $oxy$ plane
and $z$ direction perpendicular to such plane, respectively. In addition, $\chi=\pm1$
stands for the chirality symmetry of Dirac point, and $\sigma_i$ with
$i=1,2,3$ corresponds to the Pauli matrices that act on the lattice space, while $\sigma_0$ represents the identity
matrix.

Hereby, the dimensionless parameter $\zeta$ in Eq.~(\ref{Eq H_0}) serves as a tilting parameter that
enters into the energy dispersions as follows
\begin{eqnarray}
E_{\pm}(\mathbf{k})=\zeta v_zk_{z}\pm
\sqrt{(v_zk_{z})^2+(vk_x)^2+(vk_y)^2}\label{Eq dispersion}.
\end{eqnarray}
As a consequence, the presence of this very parameter can modify and reshape the
overall structure of the Fermi surface by tilting the Dirac cones~\cite{Goerbig2008PRB,Lee2018PRB,Fritz2019arXiv}.
In practice, this necessitates the categorization of two distinct types of tDSM~\cite{Lee2018PRB}.
The first type, termed type-I tDSM, occurs when $|\zeta|<1$. As to type-I tDSM, the point-like Fermi surface remains robust, and the renormalized Dirac cones are preserved against tilted contributions. Instead, the other type, dubbed type-II tDSMs,
emerges when the tilted term plays a more significant role in shaping the Fermi surface at $|\zeta|>1$.
In this case, the point-like structure of the Fermi surface is sabotaged and replaced by two crossed nodal lines~\cite{Lee2018PRB,Lee2019PRB}.

\begin{figure}
\centering
\includegraphics[width=3.2in]{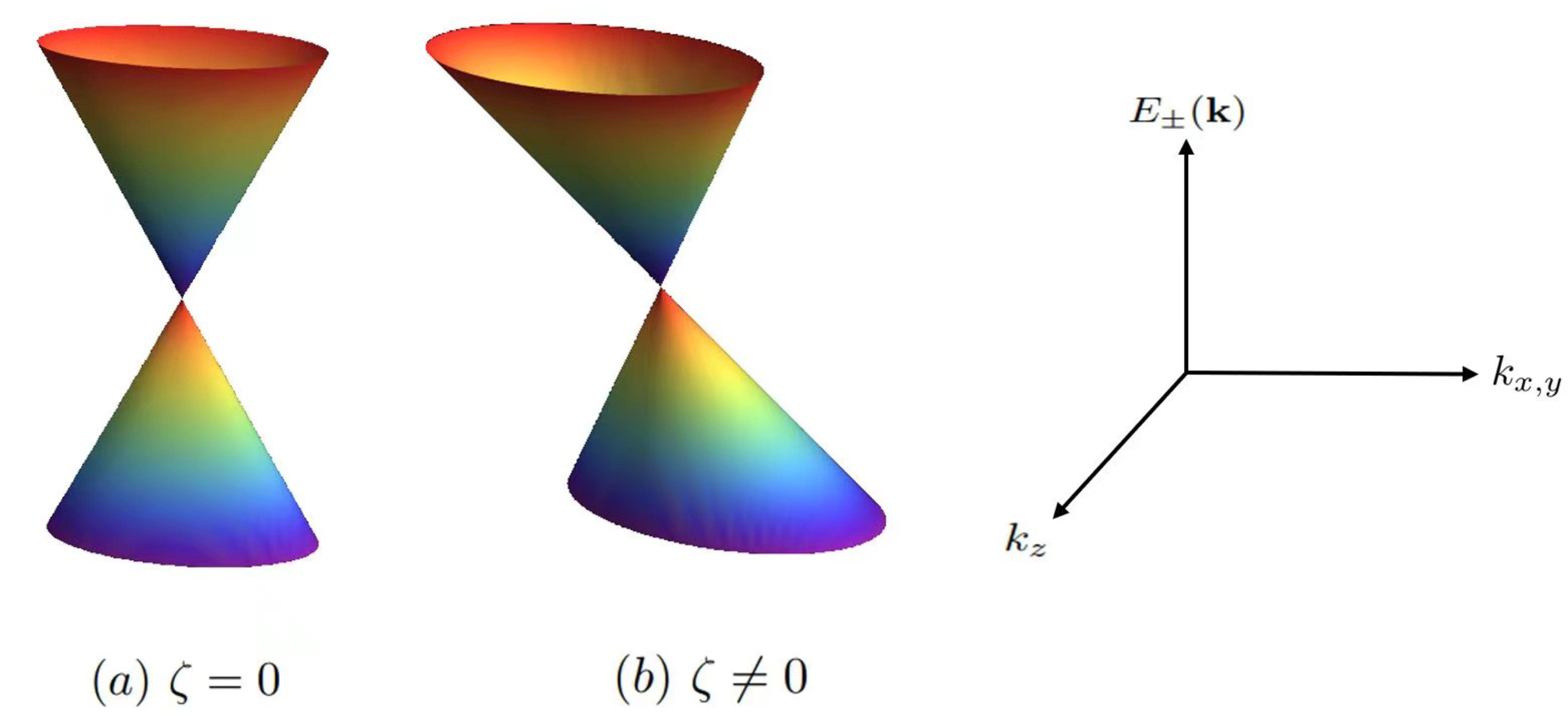}
\vspace{-0.39cm}
\caption{(Color online) Schematic dispersion for (a) the untilted and (b) the tilted 3D DSM.}\label{fig1_schematic-tilted-dispersion}
\end{figure}

To proceed, we hereafter restrict our study to the 3D type-I tDSM, namely $|\zeta|<1$ in Eq.~(\ref{Eq dispersion})
as schematically illustrated in Fig.~\ref{fig1_schematic-tilted-dispersion}. Starting from the Hamiltonian
density in Eq.~(\ref{Eq H_0}), we are left with the following noninteracting fermionic
action,
\begin{eqnarray}
S_{\psi}&=&\sum_{\chi,\alpha}\int_k\psi^{\dag}_{\chi\alpha}(\omega,\mathbf{k})
\{i\omega\sigma_{0}-\zeta v_{z}k_{z}\sigma_{0}
-\chi[v_{z}k_{z}\sigma_{z}\nonumber\\
&&+v(k_{x}\sigma_{x}+k_{y}\sigma_{y})]\}
\psi_{\chi\alpha}(\omega,\mathbf{k}),\label{Eq_S_psi}
\end{eqnarray}
with $\int_k=\int d\omega d^3\mathbf{k}/(2\pi)^4$. Here, the spinors $\psi^{\dag}_{\chi\alpha}(\omega,\mathbf{k})$
and $\psi_{\chi\alpha}(\omega,\mathbf{k})$ with spin degeneracy $\alpha=\pm1$
describe the excited fermionic quasiparticles from the Dirac points
in the first Brillouin zone~\cite{Goerbig2008PRB,Lee2018PRB,Fritz2019arXiv}. As we only consider type-I tDSMs,
the tilting parameter is constrained to $|\zeta|\in(0,1)$ with $\zeta\rightarrow0$
corresponding to normal (untitled) 3D Dirac systems.
It is worth highlighting that the tilted energy dispersion~(\ref{Eq dispersion})
indicates that the tilted Dirac cones are symmetric under the sign change of the tilting parameter.
Without loss of generality, we from now on restrict our study to the situation with a positive $\zeta$.
In addition, one can expect that the velocities are no longer isotropic
but become anisotropic for the tilted direction ($v_z$) and the other
two orientations ($v$) in the presence of the tilted terms.
To proceed, the free fermionic propagator can be readily derived from Eq.~(\ref{Eq_S_psi}).
\begin{eqnarray}
G^{-1}_{0}(k)
&=&(i\omega-\zeta v_{z}k_{z})\sigma_{0}
-\chi[v_{z}k_{z}\sigma_{z}
+v(k_{x}\sigma_{x}\nonumber\\
&&+k_{y}\sigma_{y})].\label{Eq_G_psi}
\end{eqnarray}

In addition to the fermionic excitations~(\ref{Eq H_0}),
we incorporate the contributions from phonons, which signify the potential lattice
distortions. Since the couplings between optical phonons and electrons are marginal but instead irrelevant between
acoustic phonons and electrons~\cite{Mahan1990Book}, this indicates that the optical phonons would
be much more important than acoustic ones upon approaching the potential
instability in the low-energy regime. To simplify the analysis, we can neglect the latter and only focus on
the former~\cite{Ruhman2019PRX}. In principle, there exist two different kinds
of modes corresponding to the transverse phonon and
longitudinal phonon. Accordingly, the relevant phonon ingredients can be expressed
as~\cite{Ruhman2019PRX,Khmelnitskii1971SPSS,Strukov2012Book}
\begin{eqnarray}
S_{\mathrm{u}}\!&=&\!\int \!\!d^{4}x\frac{1}{2}u_{j}(x)\!\left[(-\partial_{0}^{2}+\omega^2_{T,L})\delta_{ji}-C^j_{T}C^i_{T}
(\nabla^{2}\delta_{ji}-\partial_{j}\partial_{i})\right.\nonumber\\
&&\left.-C^j_{L}C^i_{L}\partial_{j}\partial_{i}\right]u_{i}(x)+V^{ij}\int d^{4}x[u_{i}(x)u_{j}(x)]^2,\label{Eq_S_action-u}
\end{eqnarray}
where $u_{i}$ ($u_j$) represents
the phonon field with $i,j=x,y,z$; $\omega_{T,L}$ serve as the phonon mass,
and $C^{i}_{T,L}$ specify the velocities of the transverse
and longitudinal phonons, while the coupling $V^{ij}$ characterizes the self-interactions
among phonons themselves. In principle, the phonon mass is an adjustable parameter, which is closely
related to the critical regime where the fluctuations
are important. Given that the concerned regime is adjacent to
the potential phase transition, we follow the strategy in Ref.~\cite{Ruhman2019PRX}
to tune the mass terms to be zero, which places the system near the ferroelectric critical regime.
This makes the phonon can completely
couple with other degrees of freedom to induce the potential critical behavior in the low-energy regime.
Under this assumption, the free transverse and longitudinal
phonon propagators in the momentum space can be written as,
\begin{eqnarray}
G_{\mathrm{u},ji}^{\mathrm{T}}(q_{0},\mathbf{q})
&=&\frac{\delta_{ji}-\widehat{q}_{j}\widehat{q}_{i}}
{q_{0}^{2}+C_{T}^{2}\mathbf{q}^{2}},\label{Eq_G_u-T}\\
G_{\mathrm{u},ji}^{\mathrm{L}}(q_{0},\mathbf{q})
&=&\frac{\widehat{q}_{j}\widehat{q}_{i}}
{q_{0}^{2}+C_{L}^{2}\mathbf{q}^{2}},\label{Eq_G_u-L}
\end{eqnarray}
with $\widehat{q}_{j}\widehat{q}_{i}\equiv q_{j}q_{i}/\mathbf{q}^2$ for $i,j=x,y,z$ and
$C_{T,L}^{2}\mathbf{q}^{2}\equiv \sum_{i=,x,y,z}(C^i_{T,L})^{2}q^{2}_i$.
Besides their self-interactions in Eq.~(\ref{Eq_S_action-u}), the phonons are expected to couple with
the fermionic excitations, and this coupling can be constructed as follows~\cite{Ruhman2019PRX},
\begin{eqnarray}
S_{\mathrm{u}\psi}=\sum_j\lambda_{j}\int d^4x\psi^{\dag}\sigma_{j}\psi u_{j},\label{Eq_S_u-psi}
\end{eqnarray}
where the coupling $\lambda_{j}$ with $j=x,y,z$ are utilized to measure the
strength between fermions and phonons.

Furthermore, it is of particular significance to take into account the Coulomb interaction among
the low-energy fermionic excitations. For the sake of simplicity, such degrees of freedom
can be effectively established by introducing
an auxiliary bosonic field $\phi$ as~\cite{Lee2018PRB,Ruhman2019PRX,Moon2016PRB,Moon2016SR,Nandkishore2017PRB,Mandal2022PRB}
\begin{eqnarray}
S_{\mathrm{Coul}}
\!\!&=&\!\!\frac{1}{2}\int_{q}\phi(q_{0},\mathbf{q})D_{0}^{-1}(\mathbf{q})\phi(-q_{0},-\mathbf{q})\nonumber\\
&&+ig\!\!\int_{q,k}\!\phi(q_{0},\mathbf{q})
\psi^{\dag}(\omega+q_{0},\mathbf{k}+\mathbf{q})\sigma_{0}\psi(\omega,\mathbf{k}),\label{Eq_S_Coulomb}
\end{eqnarray}
where the free propagator for the auxiliary bosonic field is given by
\begin{eqnarray}
D_{0}(\mathbf{q})=\frac{4\pi}{\epsilon\mathbf{q}^2}.\label{Eq_G_phi}
\end{eqnarray}
Here, the parameter $\epsilon$ serves as the dielectric constant
with $\epsilon\mathbf{q}^2\equiv\sum_{i=x,y,z}\epsilon_iq^2_i$, and $g$ characterizes
the coupling strength between fermion and the auxiliary bosonic field.

\begin{figure}
\centering
\includegraphics[width=3.2in]{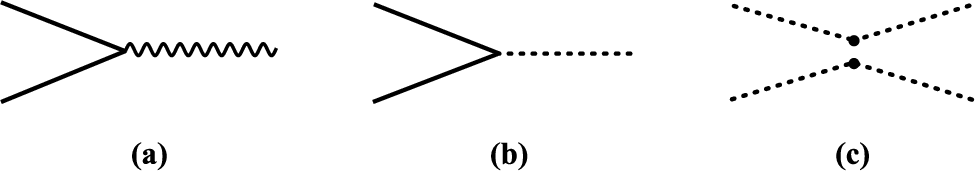}
\vspace{-0.21cm}
\caption{Tree-level vertexes: (a) Coulomb interaction, (b) electron-phonon interaction,
and (c) phonon-phonon interaction, respectively (the solid, wavy, and dashed lines
denote the free fermionic, auxiliary bosonic, and phonon propagators).}\label{fig2_interaction-vertex}
\end{figure}

Based on above presentations, we gather the physical elements, including the
low-energy fermionic excitations~(\ref{Eq_S_psi}) and phonons~(\ref{Eq_S_action-u}), in conjunction with their
entanglements (\ref{Eq_S_u-psi}) and (\ref{Eq_S_Coulomb}), and eventually arrive at our effective theory,
\begin{eqnarray}
S_{\mathrm{eff}}=S_{\psi}+S_{\mathrm{u}}+S_{\mathrm{u}\psi}+S_{\mathrm{Coul}},\label{Eq_eff-action}
\end{eqnarray}
where the corresponding free propagators for fermion, phonon, and the auxiliary bosonic field are presented in
Eqs.~(\ref{Eq_G_psi});  (\ref{Eq_G_u-T})-(\ref{Eq_G_u-L}); and (\ref{Eq_G_phi}), respectively.
Additionally, the associated tree-level vertexes for the interactions
among low-energy fermion and phonons are illustrated in Fig.~\ref{fig2_interaction-vertex}.
With these in hand, it is suitable to construct all the one-loop diagrams contributing
to the interaction parameters, as detailed in Appendix~\ref{appendix-one-loop-corrections}.
Subsequently, we adopt the effective action~(\ref{Eq_eff-action})
as our starting point to examine the critical fates of 3D tDSMs in the low-energy regime,
taking into account the influence of the Coulomb interaction, electron-phonon interaction,
and phonon-phonon interaction.

\section{RG analysis and coupled evolutions}\label{Sec_RGEqs}

Given the distinctive energy dispersions characterized by the tilted Dirac cones in type-I 3D tDSM,
it is appropriate to adopt the energy-shell method for the RG analysis~\cite{Wilson1975RMP,Polchinski9210046,Shankar1994RMP}.
This henceforth requires us to integrate the energy shells out one by one during the RG  transformations~\cite{Shankar1994RMP,Lee2018PRB,Huh2008PRB,She2010PRB,Wang2011PRB,Qiong2019NPJB,Dong2020PRB}.
To this end, we introduce the useful transformations
and then utilize the Jacobian transformation
to parametrize the tilted energy dispersion~(\ref{Eq dispersion}).
This parametrization is expressed as follows~\cite{Lee2018PRB,Lee2019PRB,Goerbig2008PRB,Qiong2019NPJB}
\begin{eqnarray}
v_zk_{z}&=&-\frac{\zeta}{1-\zeta^2}E+\frac{|E|}{1-\zeta^2}\cos\theta
\label{Eq kz},\\
vk_{x}&=&\frac{|E|}{\sqrt{1-\zeta^2}}\sin\theta\cos\varphi\label{Eq kx},\\
vk_{y}&=&\frac{|E|}{\sqrt{1-\zeta^2}}\sin\theta\sin\varphi\label{Eq ky},\\
\!\!\!\!\!\!\int \!\!dk_{x}dk_{y}dk_{z}\!\!
&=&\!\!\!\!\int\!\!dE\!\!\int^{\pi}_0\!\!\!\!d\theta \!\!\int^{2\pi}_0\!\!\!\!\!\!d\varphi
\frac{E^2\sin\theta(\eta_{E}-\zeta\cos\theta)}{v^2v_z(1-\zeta^2)^2,}\label{Eq E}
%\!\!\!\!\!\!\int \!\!dk_{x}dk_{y}dk_{z}\!\!
%&=&\!\!\!\!\int\!\!dE\!\!\int^{\pi}_0\!\!\!\!d\theta \!\!\int^{2\pi}_0\!\!\!\!\!\!d\varphi
%\frac{\blue{\eta_E}E^2\sin\theta(\eta_{E}-\zeta\cos\theta)}{v^2v_z(1-\zeta^2)^2,}\label{Eq E}
\end{eqnarray}
where $E$ denotes the energy scale with the $\eta_E$ collecting its sign, and
$\theta$ and $\varphi$ are two associated angles. In the spirt of the energy-shell framework,
the fast modes of degrees of freedom within the energy shell $\Lambda/b<E<\Lambda$ would
be integrated out, where $\Lambda$ is the energy scale and variable parameter $b$ can be
specified as $b=e^{-l}<1$ with $l>0$.  Then, the renormalized ``slow modes" are obtained, with which
the RG processes can be fulfilled by performing the
RG transformation rescalings~\cite{Wilson1975RMP,Polchinski9210046,Shankar1994RMP}.

\begin{figure}
\centering
\includegraphics[width=3.2in]{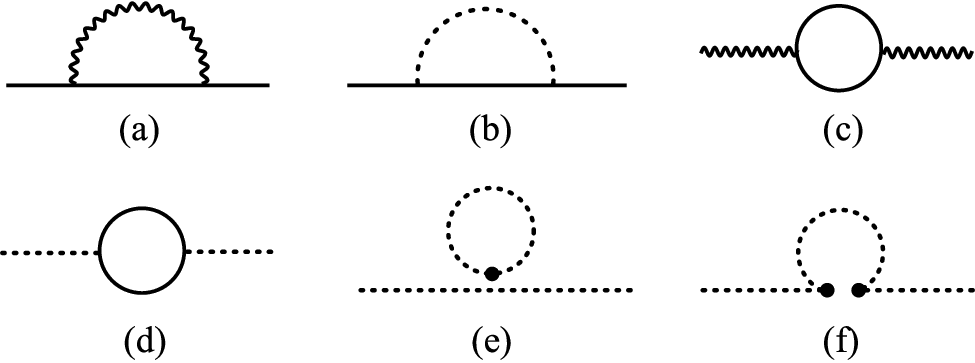}
\vspace{-0.2cm}
\caption{One-loop corrections caused by the Coulomb and electron-phonon as well
as phonon-phonon interactions to (a)-(b) fermionic propagator,
(c) auxiliary bosonic propagator, and (d)-(f) phonon
propagator (the solid, wavy, and dashed lines
denote the free fermionic, auxiliary bosonic, and phonon propagators, respectively).}\label{fig3_propagators}
\end{figure}

To proceed, we are going to derive the RG rescaling transformations
that serve as crucial bridges connecting two successive RG steps. Following
the RG approach~\cite{Wilson1975RMP,Polchinski9210046,Shankar1994RMP},
the free frequency term in the noninteracting action~(\ref{Eq_S_psi}) can be regarded
as the initial fixed point that is invariant during the RG process.
In collaboration with Eqs.~(\ref{Eq kz})-(\ref{Eq E}), it
gives rise to the following RG rescalings~\cite{Shankar1994RMP,Wang2011PRB,
Lee2018PRB,Lee2019PRB,Huh2008PRB,She2010PRB},
\begin{eqnarray}
&&\omega\longrightarrow\omega'e^{-l},\label{Eq_RG-rescaling-omega}\\
&&E\longrightarrow E'e^{-l},\\
&&\psi\longrightarrow\psi'e^{\frac{5}{2}l-\eta_\psi l},\\
&&\phi\longrightarrow\phi'e^{3l-\eta_\phi l},\\
&&u_T\longrightarrow u'_Te^{3l-\eta_{u_T}l}\\
&&u_L\longrightarrow u'_Le^{3l-\eta_{u_L}l},\label{Eq_RG-rescaling-u-L}
\end{eqnarray}
where all the fields $\psi,\phi,u_{T}$, and $u_{L}$ are defined in the frequency and momentum space. In addition,
$\eta_\psi$, $\eta_\phi$, $\eta_{u_T}$, and $\eta_{u_L}$
are the anomalous dimensions of related fields that capture the one-loop
corrections due to all kinds of interactions in our effective
theory~(\ref{Eq_eff-action}). To determine them, we resort to
the one-loop contributions to the free propagators
as shown in Fig.~\ref{fig3_propagators}. The
straightforward calculations gives rise to
\begin{eqnarray}
\Sigma_{f}(i\omega,\mathbf{k})&=&-\{
[i\omega\mathcal{A}_{0}
-\zeta v_{z}k_{z}\mathcal{A}_{3}]\sigma_{0}
-\chi[\mathcal{A}_{2}v_{z}k_{z}\sigma_{z}\nonumber\\
&&+v\mathcal{A}_{1}(k_{x}\sigma_{x}
+k_{y}\sigma_{y})]\}l,\\
\Sigma_{b}(\mathbf{q})
&=&\frac{-[v^2(q_{x}^2+q_{y}^2)+v_{z}^2q_{z}^2]
g^2l}{4\pi^2v^2v_{z}(1-\zeta^2)},\\
\Sigma_{u_{T,L}}(q_0)
&=&\frac{-5\lambda_{j}\lambda_{i}\delta_{ji}q_{0}^2l}{2\pi^2v^2v_{z}},
\end{eqnarray}
for fermionic, auxiliary bosonic, and phonon propagators, respectively.
Here, the coefficients $\mathcal{A}_{0}$, $\mathcal{A}_{1}$, and $\mathcal{A}_{2}$ are designated in
Eqs.~(\ref{Eq_A_1}), (\ref{Eq_A_2}), and (\ref{Rq_appendix-A0}) of Appendix~\ref{appendix-one-loop-corrections}.
Consequently, such one-loop corrections lead to~\cite{Huh2008PRB,She2010PRB,Wang2011PRB}
\begin{eqnarray}
\eta_\psi=-\frac{\mathcal{A}_{0}}{2},\,\,\,\eta_\varphi=0,\,\,\,
\eta_{u^j_L}=\eta_{u^j_T}=-\frac{5\lambda^2_{j}}{8\pi^2v^2v_{z}},\label{Eq_eta}
\end{eqnarray}
with $j=x,y,z$ as aforementioned in Eq.~(\ref{Eq_S_u-psi}).
Subsequently, the detailed calculations of one-loop corrections to the interaction vertices are presented in Appendix~\ref{appendix-one-loop-corrections}. After combining the RG rescaling transformations~(\ref{Eq_RG-rescaling-omega})-(\ref{Eq_RG-rescaling-u-L})
with the anomalous dimensions~(\ref{Eq_eta}) and all one-loop corrections in
Appendix~\ref{appendix-one-loop-corrections}, the coupled RG equations of all
interaction parameters are derived as follows~\cite{Shankar1994RMP,Lee2018PRB,Huh2008PRB,She2010PRB,Wang2011PRB}
\begin{eqnarray}
\frac{dv}{dl}&=&(-\mathcal{A}_{1}-2\eta_\psi)v,\label{Eq_RGEq-v}\\
\frac{dv_z}{dl}&=&(-\mathcal{A}_{2}-2\eta_\psi)v_z,\\
\frac{d\zeta}{dl}&=&(\mathcal{A}_{2}-\mathcal{A}_{3})\zeta,\\
\frac{d \epsilon}{dl}&=&\frac{g^2}{\pi v_{z}(1-\zeta^2)}\epsilon,\\
\frac{d \epsilon_z}{dl}&=&\frac{v_{z}g^2}{\pi  v^2(1-\zeta^2)}\epsilon_z,\\
\frac{d C_{T,L}}{dl}&=&-\eta_{u_x} C_{T,L},\\
\frac{d C^z_{T,L}}{dl}&=&-\eta_{u_z} C^z_{T,L},\\
%\frac{d C_{L}}{dl}&=&-\eta_{u_x} C_{L},\\
%\frac{d C^z_{L}}{dl}&=&-\eta_{u_z} C^z_{L},\\
\frac{dg}{dl}&=&(\mathcal{B}-2\eta_\psi )g,\\
\frac{d\lambda}{dl}&=&(\mathcal{D}-2\eta_\psi -\eta_{u_x})\lambda,\\
\frac{d\lambda_z}{dl}&=&(\mathcal{D}_z-2\eta_\psi -\eta_{u_z})\lambda_z,\\
\frac{dV^{xx}_{T,L}}{dl}&=&(-4\eta_{u_{x}} - \mathcal{F}^{xx}_{T,L})V^{xx}_{T,L},\\
\frac{dV^{zz}_{T,L}}{dl}&=&(-4\eta_{u_{z}} - \mathcal{F}^{zz}_{T,L})V^{zz}_{T,L},\\
\frac{dV^{xy}_{T,L}}{dl}&=&[-(2\eta_{u_x}+2\eta_{u_y}) - \mathcal{F}^{xy}_{T,L}]V^{xy}_{T,L},\\
\frac{dV^{xz}_{T,L}}{dl}&=&[-(2\eta_{u_x}+2\eta_{u_z}) - \mathcal{F}^{xz}_{T,L}]V^{xz}_{T,L}.\label{Eq_RGEq-V_T-L}
%\frac{dV^{xx}_L}{dl}&=&(-4\eta_{u_{x}} - \mathcal{F}^{xx}_L)V^{xx}_L,\\
%\frac{dV^{zz}_L}{dl}&=&(-4\eta_{u_{z}} - \mathcal{F}^{zz}_L)V^{zz}_L,\\
%\frac{dV^{xy}_L}{dl}&=&[-(2\eta_{u_x}+2\eta_{u_y}) - \mathcal{F}^{xy}_L]V^{xy}_L,\\
%\frac{dV^{xz}_L}{dl}&=&[-(2\eta_{u_x}+2\eta_{u_z}) - \mathcal{F}^{xz}_L]V^{xz}_L,
\end{eqnarray}
Here, all the associated coefficients $\mathcal{A}_{1,2,3}$, $\mathcal{B}$,
$\mathcal{D}$, $\mathcal{D}_z$, and $\mathcal{F}_{T,L}$ are designated in Appendix~\ref{appendix-one-loop-corrections}.
It is worth emphasizing that the directions $x$ and $y$ are isotropic as displayed in
Eq.~(\ref{Eq H_0}). To simplify the analysis, we have we introduced the notations
$C^{x}_{T,L}=C^{y}_{T,L}\equiv C_{T,L}$, $\lambda_{x}=\lambda_{y}\equiv\lambda$,
and $\epsilon_x=\epsilon_y\equiv\epsilon$ in above RG equations to
describe the related parameters appearing in Eqs.~(\ref{Eq_S_action-u})-(\ref{Eq_S_u-psi}) and Eq.~(\ref{Eq_G_phi}).
Besides, several flows of parameters, including $V^{yy}_{T,L}$ and $V^{yz}_{T,L}$,
similar to their $xx$ and $xz$-component counterparts, can be neglected to simplify our analysis.

\begin{figure}
\centering
\includegraphics[width=3.2in]{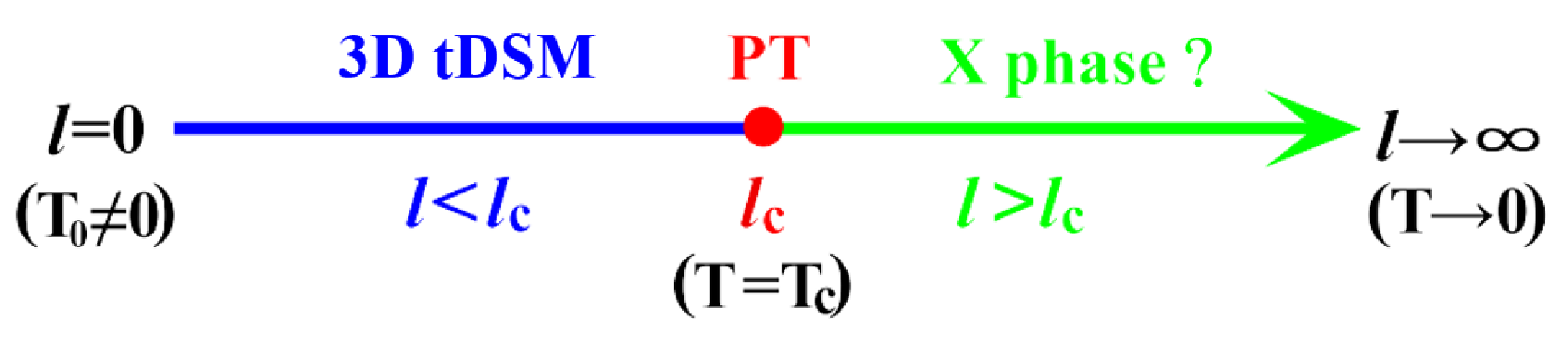}
\vspace{-0.1cm}
\caption{(Color online) A schematic diagram illustrating the tendencies of 3D tDSM and
potential instability as the energy scale is tuned. The label $l_c$ (or $T_c$) specifies the critical
energy scale (or critical temperature) and ``PT" designates the accompanying phase transition
induced by the potential instability from the 3D tDSM to an $X$ phase. A further investigation into
this transition will be conducted in Sec.~\ref{Sec_FP-instab}.}\label{fig4_lc-PT}
\end{figure}

The energy-dependent coupled evolutions~(\ref{Eq_RGEq-v})-(\ref{Eq_RGEq-V_T-L})
encapsulate the low-energy information resulting from the interplay among all interactions
in our effective theory. Deciphering the physics encoded in these equations, Fig.~\ref{fig4_lc-PT}
schematically presents the underlying properties from the initial state to the lowest energy limit.
We realize that the competition among different sorts of interactions results in a
number of unique behavior as the energy scale decreases. In particular, it potentially induces some
instability, which may drive a phase transition from a 3D tDSM to an $X$ phase
as the critical energy scale denoted by $l_c$ is approached.
Within the RG framework, the very scale $l_c$ is adopted to represent the ``critical energy scale",
marking the terminal point of the RG flows where the system approaches a quantum critical
regime~\cite{Shankar1994RMP,Sachdev2011Book}. At $l_c$, the correlation length diverges ($\xi \to \infty$)~\cite{Vojta2003RPP,Sachdev2011Book}, generally accompanying by certain instabilities and
the emergence of symmetry-breaking ordered
phases~\cite{Halboth2000PRL,Maiti2010PRB,Cvetkovic2012PRB,Murray2014PRB,Roy2018PRX,Khodas2016PRX}.

Our primary focus hereafter is on studying these low-energy consequences.
Specifically, we are going to investigate the tendencies of interaction parameters and their related
implications at $l<l_{c}$ in the upcoming Sec.~\ref{Sec_SM-tend-fates}. The examination of potential instability
at $l\rightarrow l_c$ (depicted in Fig.~\ref{fig4_lc-PT}) and critical behavior as $l$ approaches $l_c$
will be discussed in Secs.~\ref{Sec_FP-instab} and \ref{Sec_critical_implications}, respectively.

\begin{figure}
\centering
\includegraphics[width=4.1in]{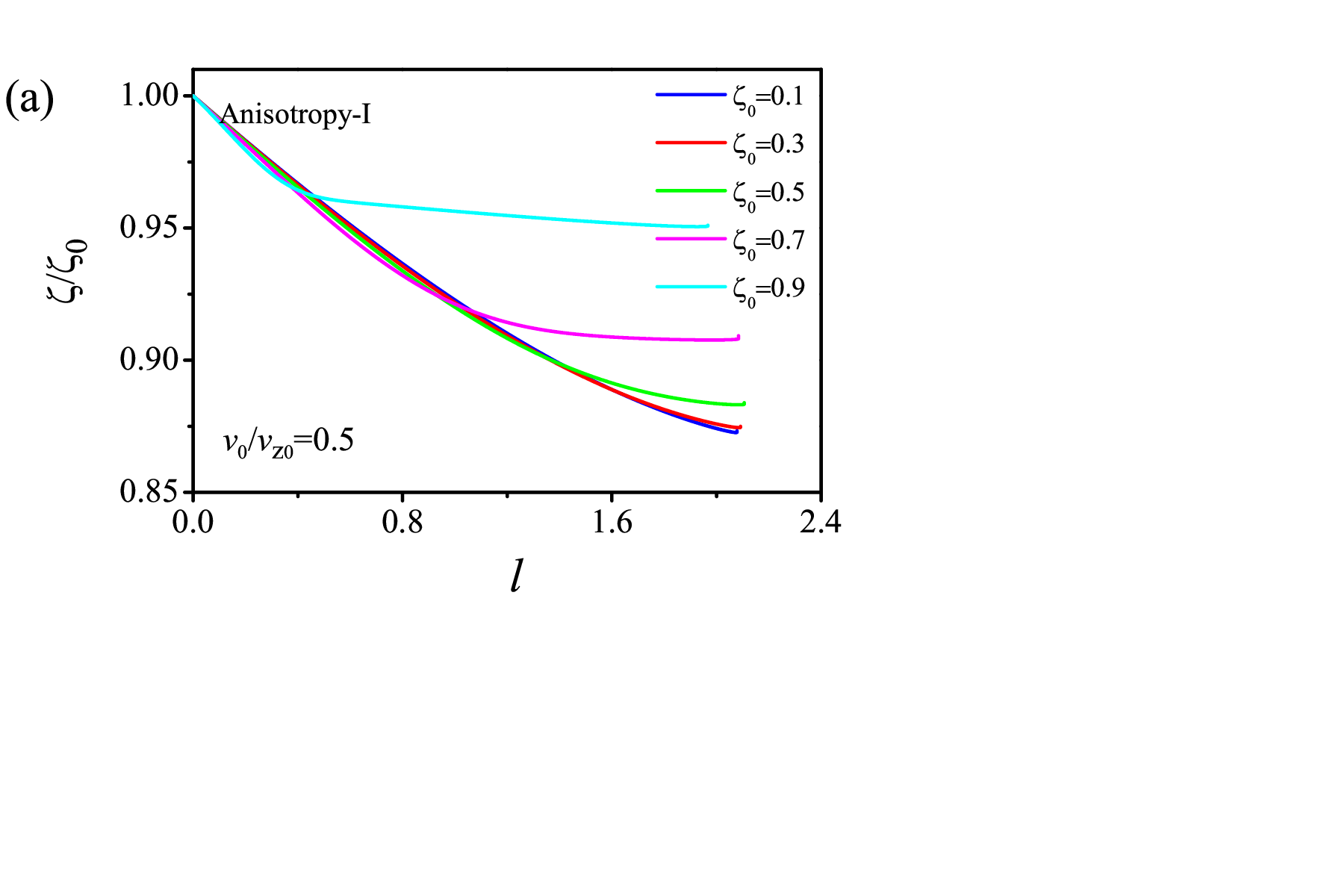}\vspace{-2.7cm}
\includegraphics[width=4.1in]{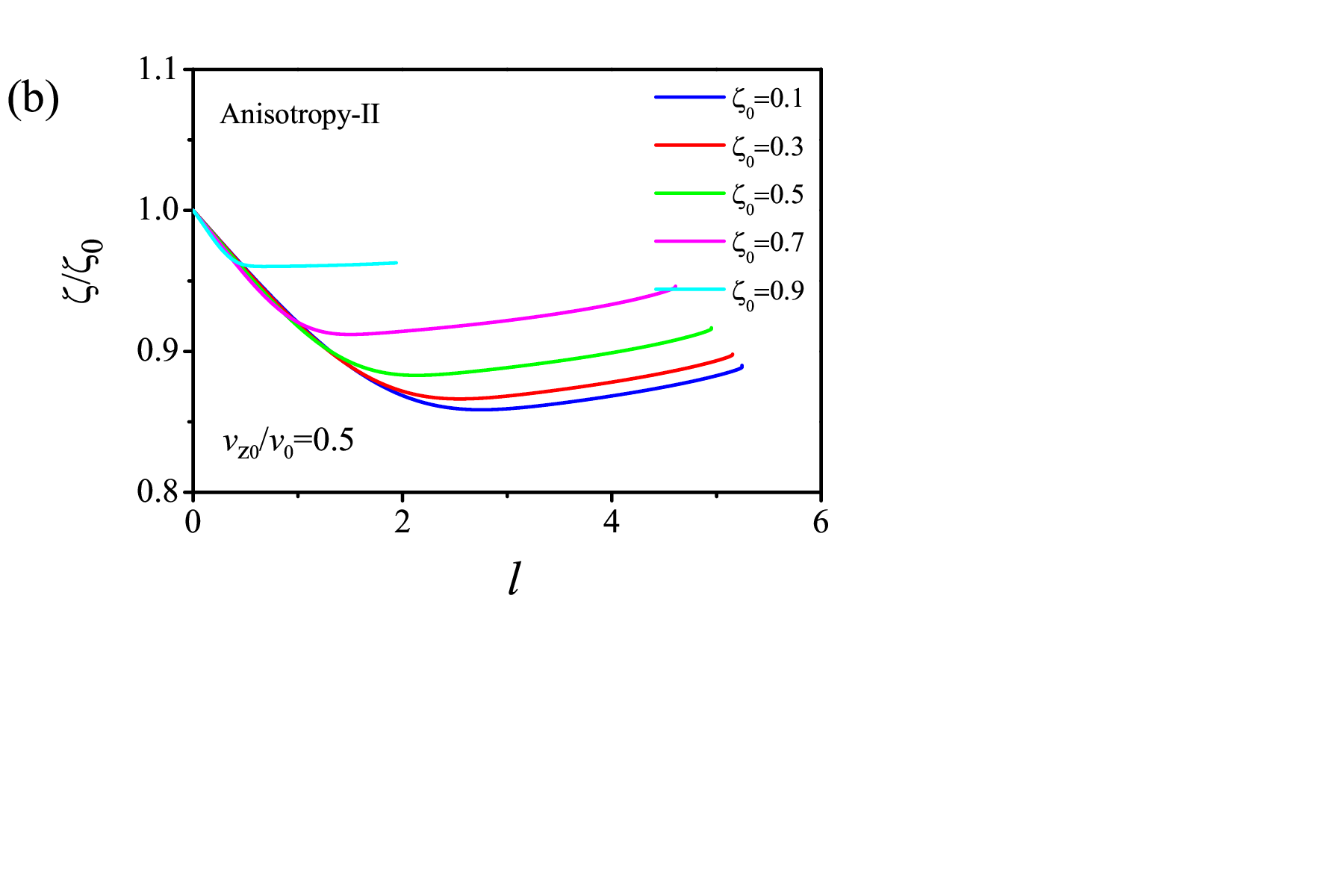}\\
\vspace{-2.5cm}
\caption{(Color online) Energy-dependent evolutions of $\zeta/\zeta_{0}$ as $\zeta_{0}$ varies, starting from the
(a) Anisotropy-I ($v_0/v_{z0}=0.5$) and (b) Anisotropy-II ($v_{z0}/v_0=0.5$) situations.}\label{Fig_zeta-1}
\end{figure}

\section{Tendencies and fates of interaction parameters}\label{Sec_SM-tend-fates}

As aforementioned in Sec.~\ref{Sec_RGEqs}, the coupled RG evolutions encompass all
the low-energy properties of 3D tDSM influenced by the Coulomb interaction as well
as electron-phonon and phonon-phonon interactions. Within this
section, we are going to investigate the energy-dependent coupled RG flow
equations and endeavor to extract the low-energy behavior of all relevant parameters
from the intimate competition of these interactions.

\begin{figure}
\centering
\includegraphics[width=2.8in]{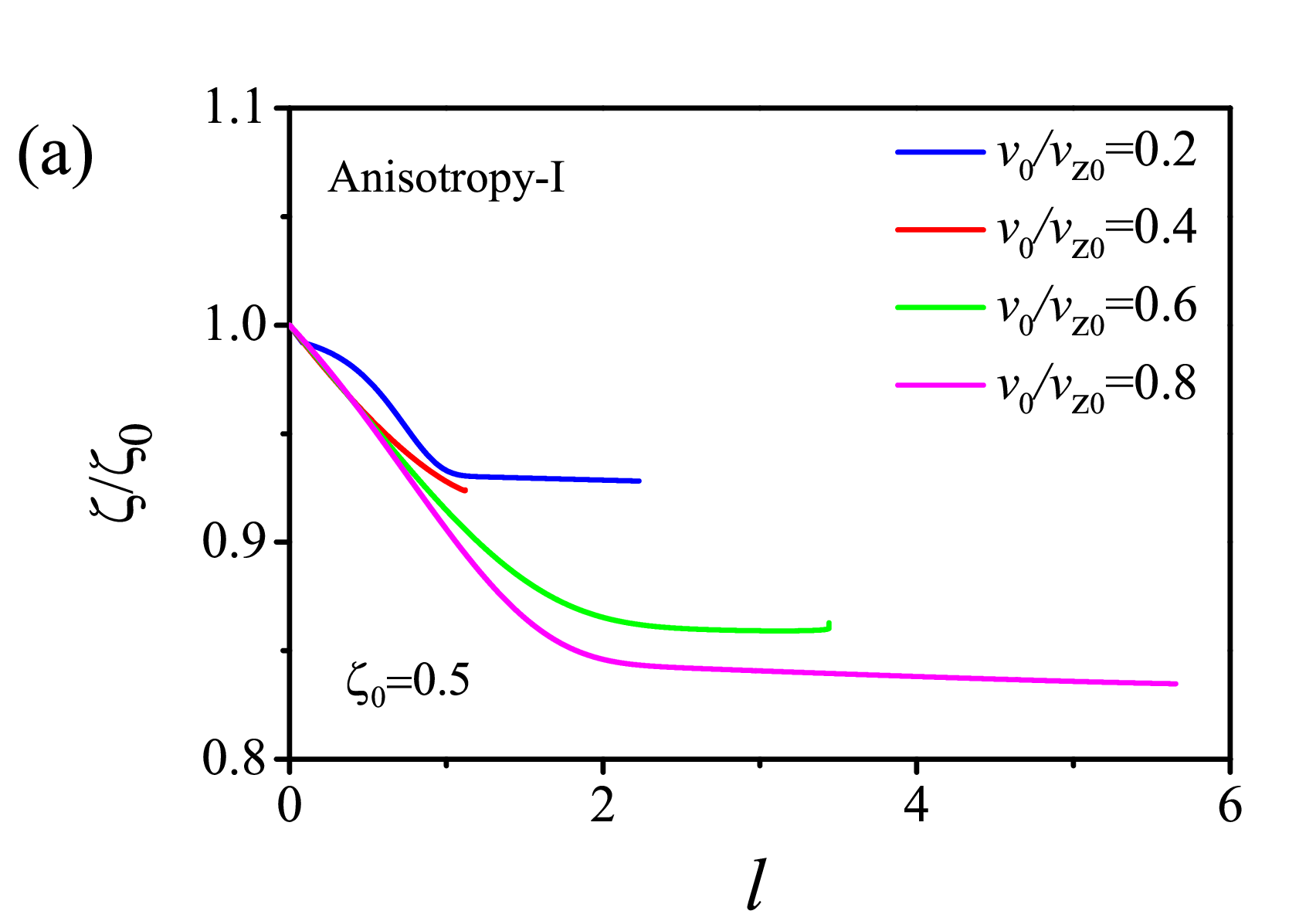} %\vspace{-2.7cm}
\includegraphics[width=2.8in]{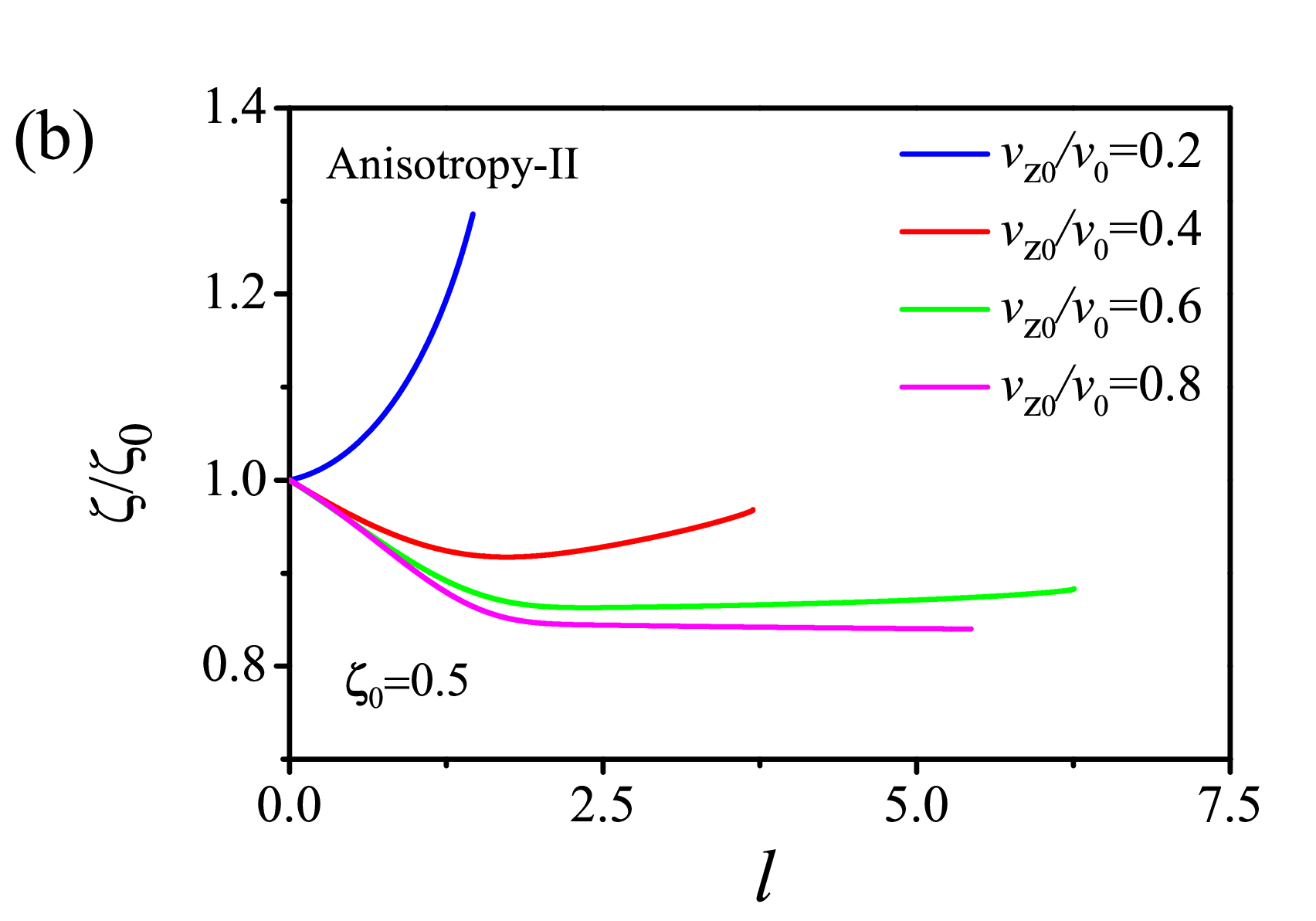}\\ %\vspace{-2.7cm}
\includegraphics[width=2.8in]{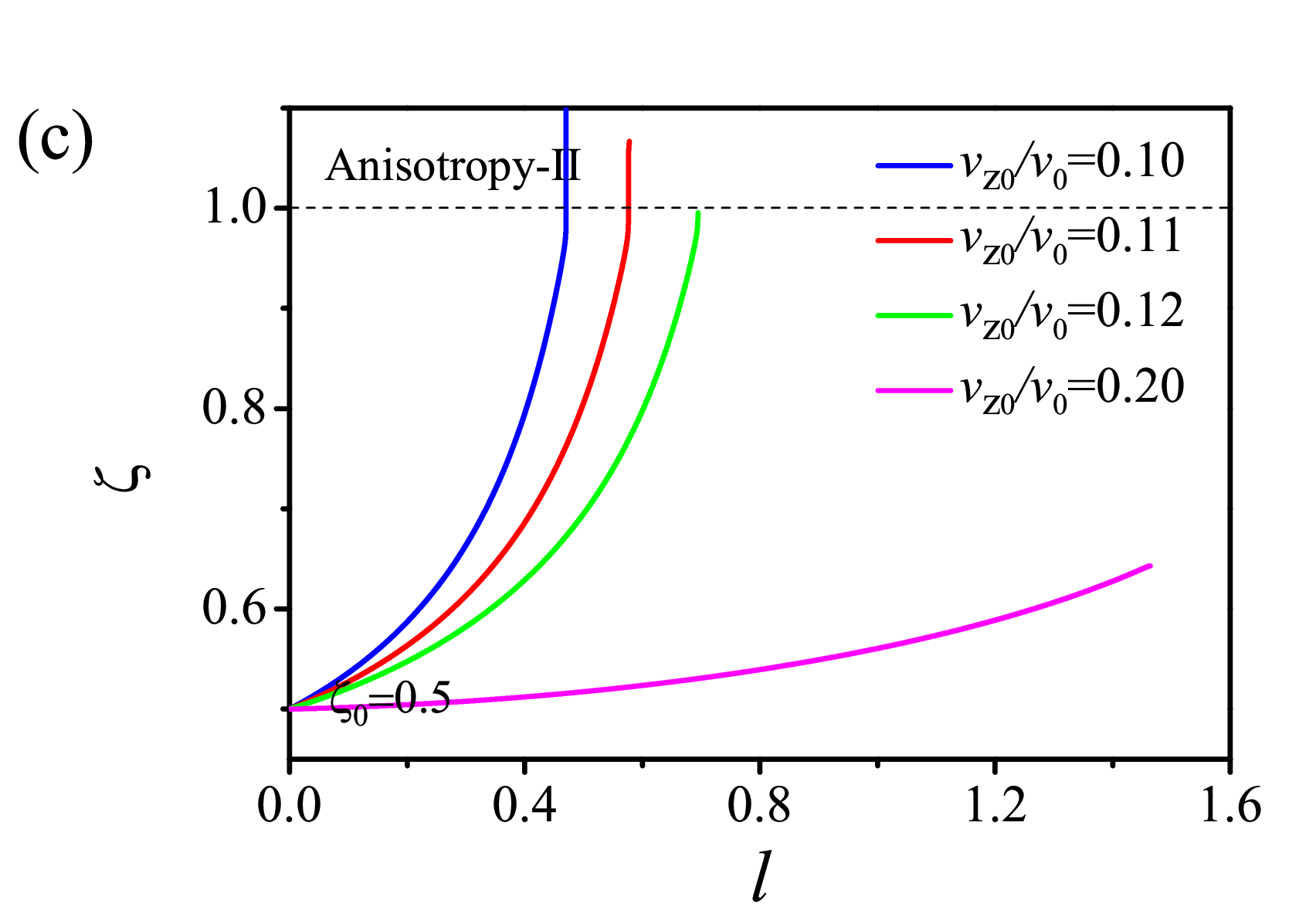}\\
\vspace{-0.1cm}
\caption{(Color online) Energy-dependent evolutions of $\zeta/\zeta_{0}$ with variation of $v_{z0}/v_0$, starting from
(a) Anisotropy-I, (b) Anisotropy-II situations, and (c) of $\zeta$ for the Anisotropy-II case.}\label{Fig_zeta-2}
\end{figure}

\begin{figure}
\centering
\includegraphics[width=2.35in]{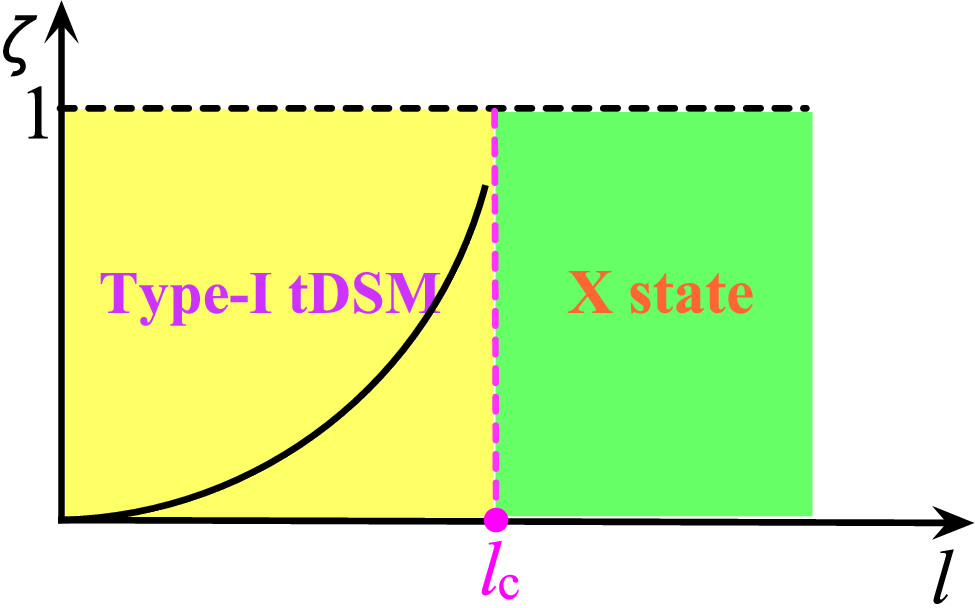}\vspace{0.7cm}
\includegraphics[width=2.35in]{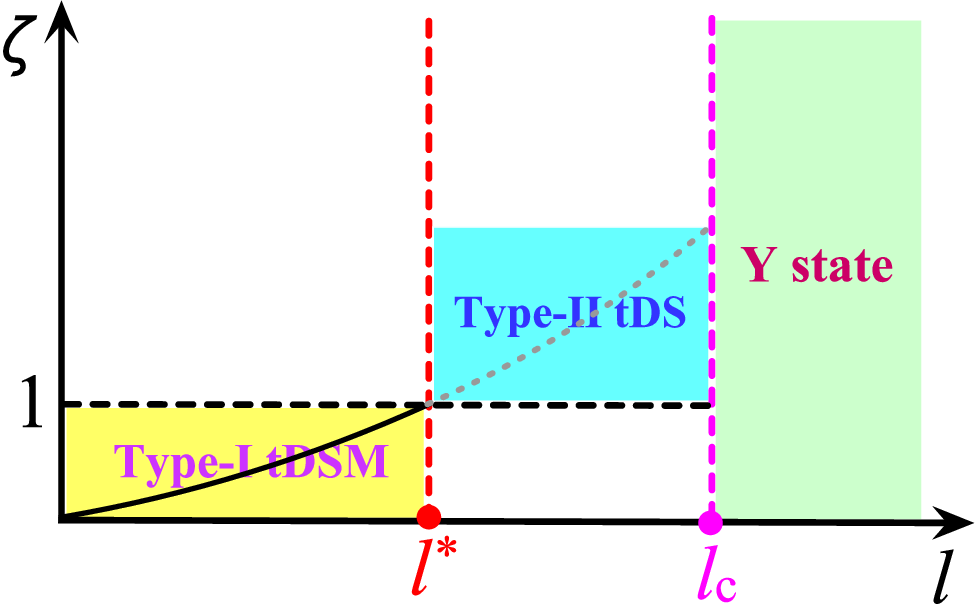}\\
\vspace{0.1cm}
\caption{(Color online) A schematic diagram illustrates two distinct scenarios for the phase transition:
(a) transitioning to the $X$ state characterized by $\zeta<1$, and (b) transitioning to $Y$ state with $\zeta>1$.}\label{Fig_phase}
\end{figure}

\begin{figure*}[htpb]
\centering
\subfigure[]{\includegraphics[width=2.5in]{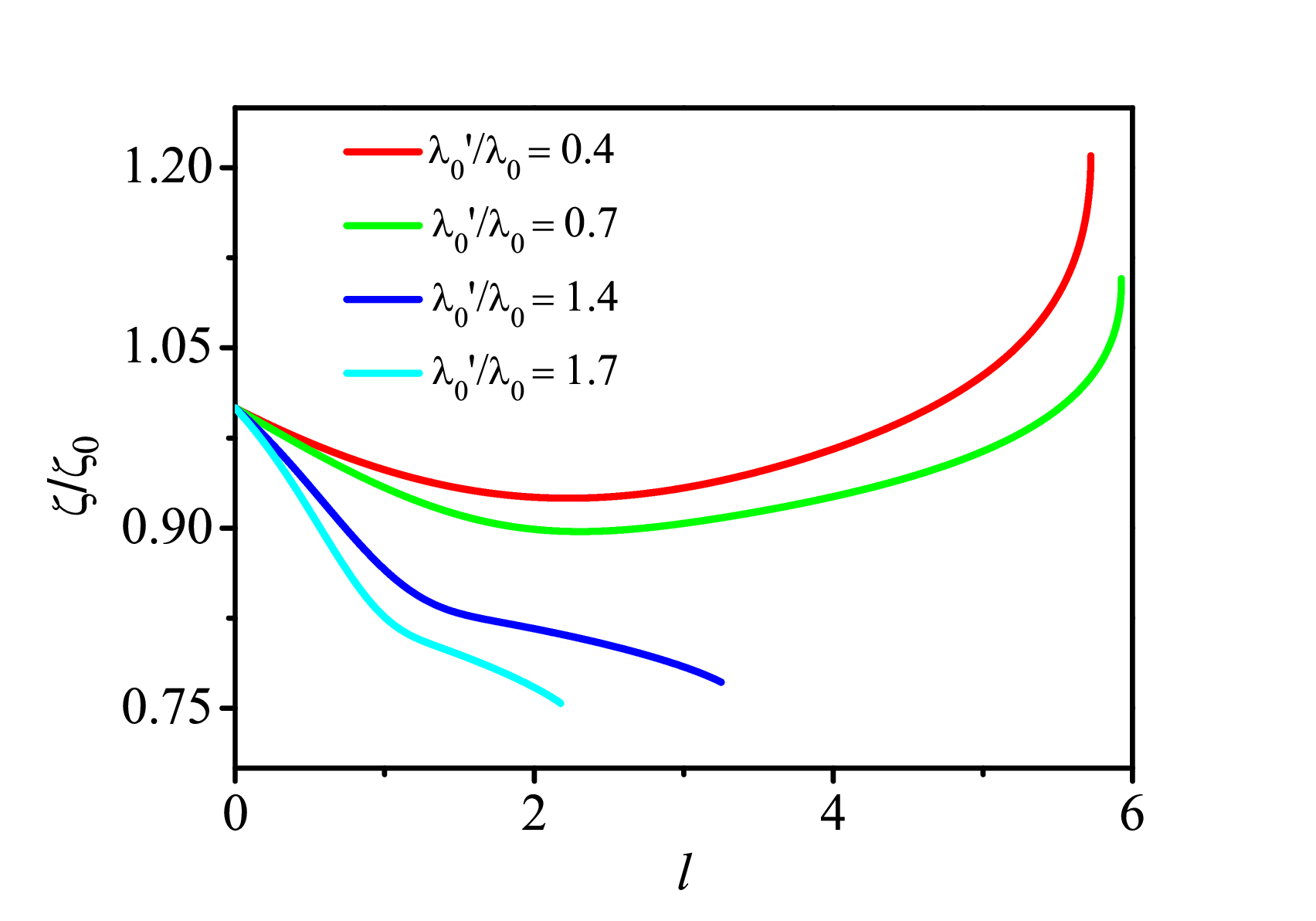}}
\subfigure[]{\includegraphics[width=2.5in]{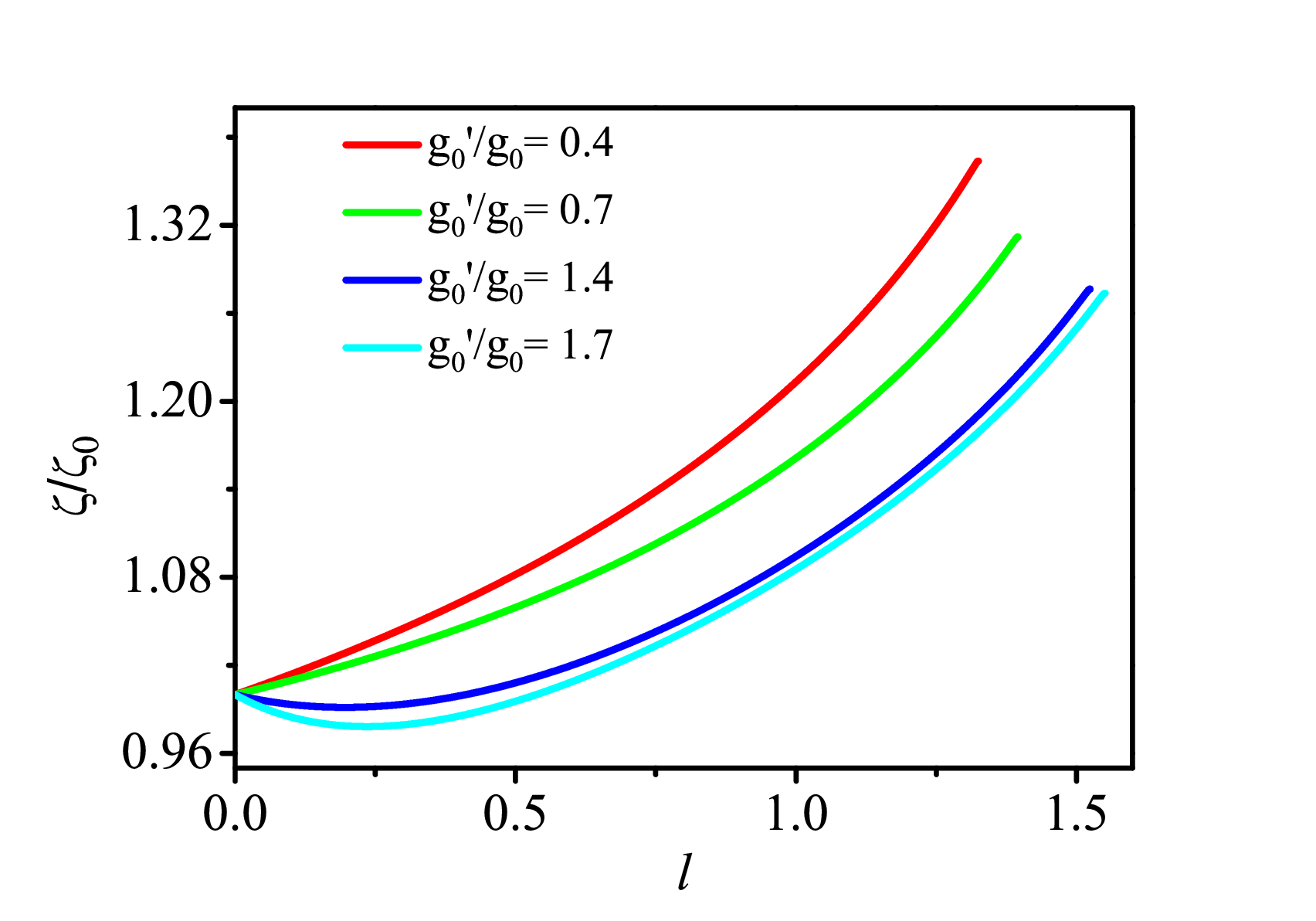}}\\ \vspace{-0.5cm}
\subfigure[]{\includegraphics[width=2.5in]{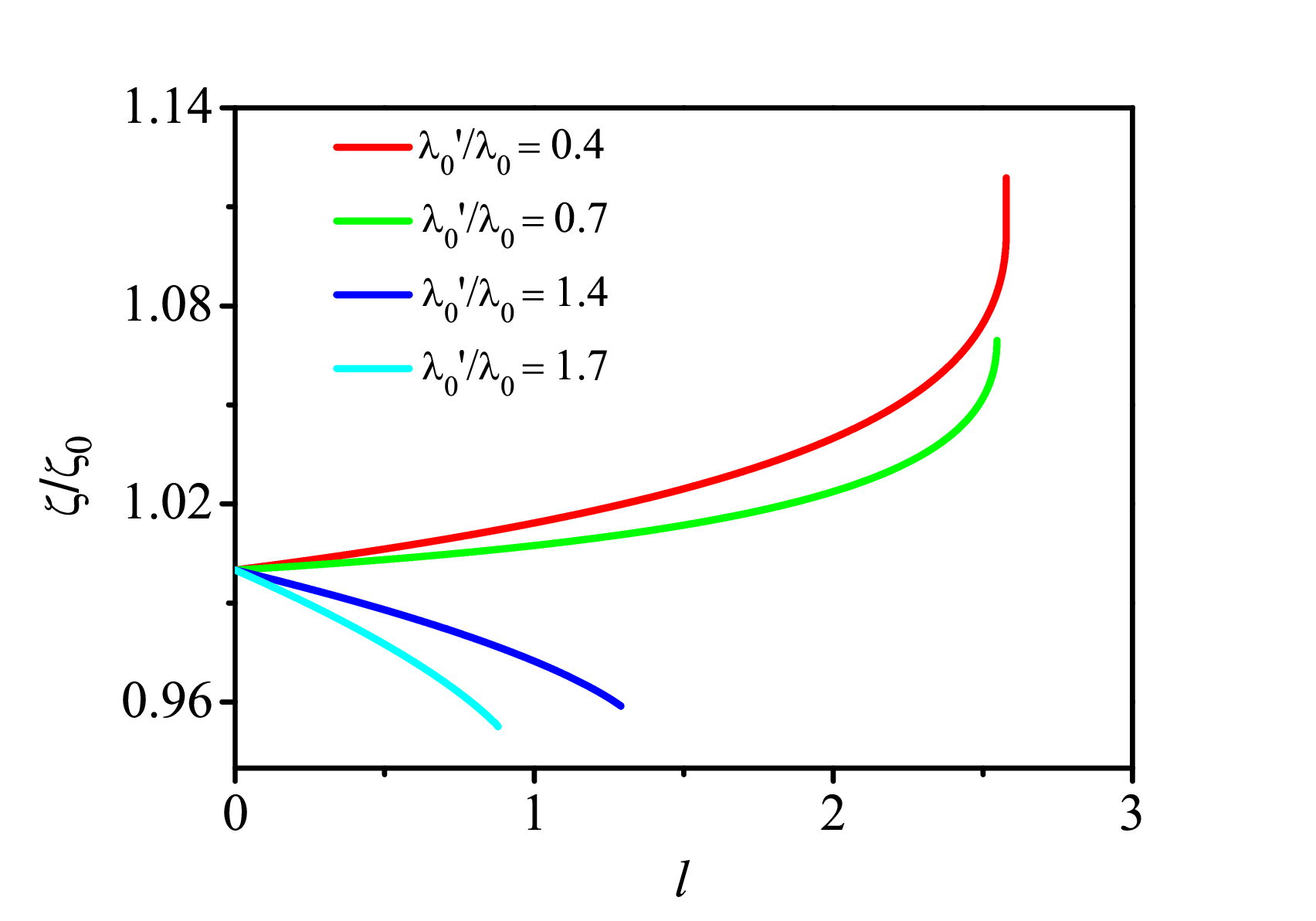}}
\subfigure[]{\includegraphics[width=2.5in]{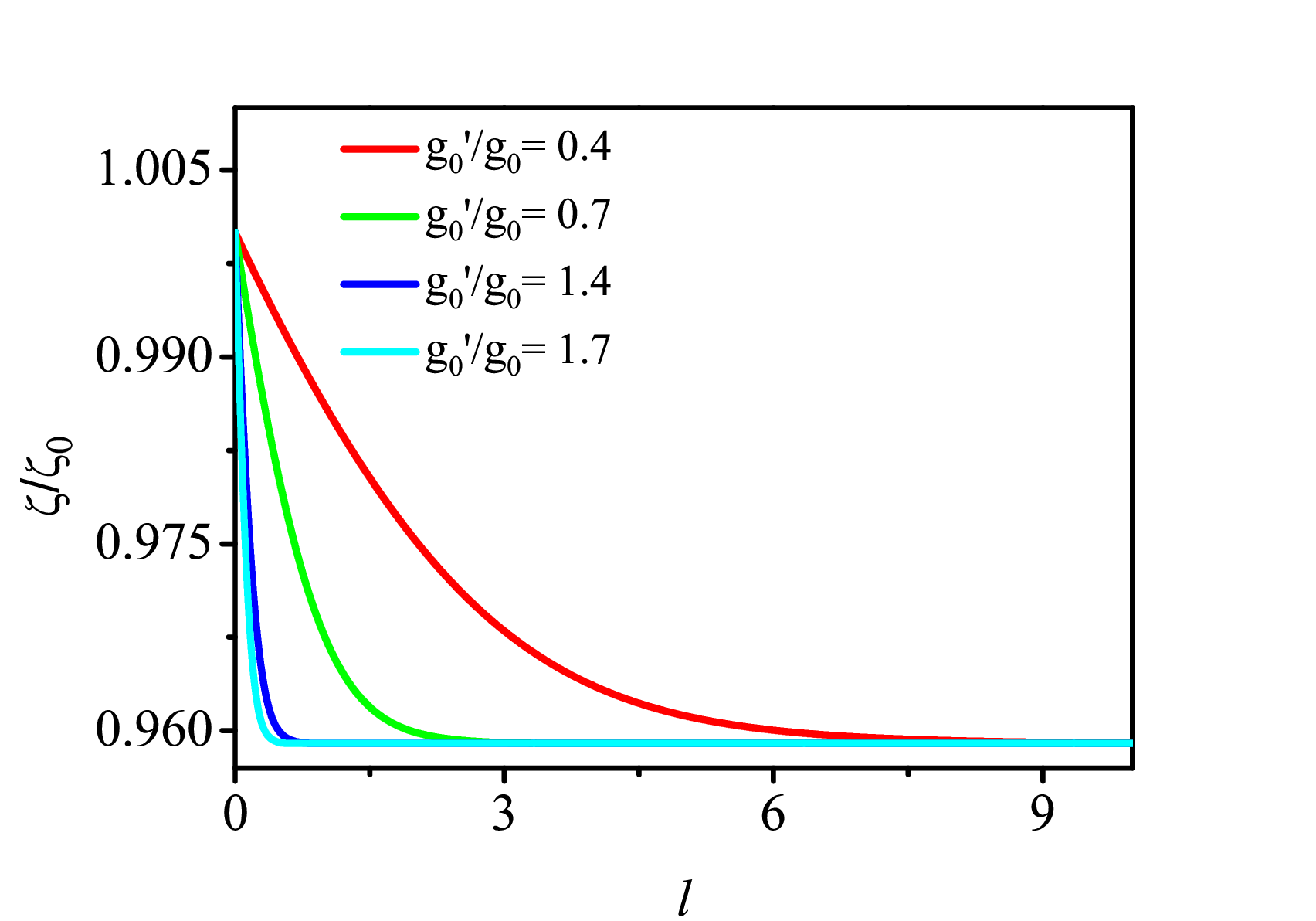}}
\centering
\caption{(Color online) Energy-dependent evolutions of $\zeta/\zeta_0$ with $\zeta_0=0.5$
for several representative initial values of:
(a) the electron-phonon interaction at $v_z(0)/v(0) = 0.6$
and (b) the Coulomb interaction at $v_z(0)/v(0)=0.2$,
as well as (c) the electron-phonon interaction in the absence of the Coulomb interaction $(g=0)$, and
(d) the Coulomb interaction in the absence of the electron-phonon interaction ($\lambda=0$).
The basic results for both Anisotropy I are similar and not shown (hereby $X'_0/X_0$ with $X=V^{xx}_L,g,\lambda$ denotes the
ratios of the initial values with respect to those of Fig.~\ref{Fig_zeta-1}).}
\label{zeta_lam_0_g_mult}
\end{figure*}

\subsection{Fate of $\zeta/\zeta_{0}$ and two unstable scenarios}

Before proceeding further, it is of particular necessity to highlight that the effective theory~(\ref{Eq_eff-action}) is
restricted to the type-I tDSMs with $\zeta\in (0,1)$. To ensure the well-defined RG equations,
we accordingly commence by examining the low-energy behavior of the tilting parameter $\zeta$. For convenience
of reference in future discussions, we designate the system with $v_{0}<v_{z0}$, $v_{0}>v_{z0}$, $v_{0}=v_{z0}$ as
the Anisotropy-I, Anisotropy-II, and Isotropic cases, respectively.

As discussed in Sec.~\ref{Sec_model}, we hereby focus on the $\zeta>0$ case due to
the symmetric consideration. Carrying out the numerical analysis of RG equations~(\ref{Eq_RGEq-v})-(\ref{Eq_RGEq-V_T-L})
gives rise to the basic properties of $\zeta$ in Figs.~\ref{Fig_zeta-1} and \ref{Fig_zeta-2}.
At first, we fix the initial values of fermion velocities and study the evolution of $\zeta$ for
several representative initial values. Figure \ref{Fig_zeta-1}(a) illustrates that, in the Anisotropy-I case, $\zeta/\zeta_{0}$
decreases with lowering energy scales, and finally converges to a finite value
that is always less than $1$ at a certain critical value. In comparison, Fig.~\ref{Fig_zeta-1}(b) showcases
$\zeta/\zeta_{0}$ in the Anisotropy-II initially decreases and then increases in the low energy regime,
yet remains constrained to $\zeta/\zeta_{0}\leq1$. Interestingly, we notice that the decrease in energy
scale has a relatively less pronounced effect on $\zeta/\zeta_{0}$ for a bigger initial tilting system.
The Isotropic case exhibits similar basic results. As a consequence, the type-I tDSM is adequately robust
against the variations in the initial values of fermion velocities, ensuring the validity of the
coupled RG equations.

\begin{figure}
\centering
\includegraphics[width=4.1in]{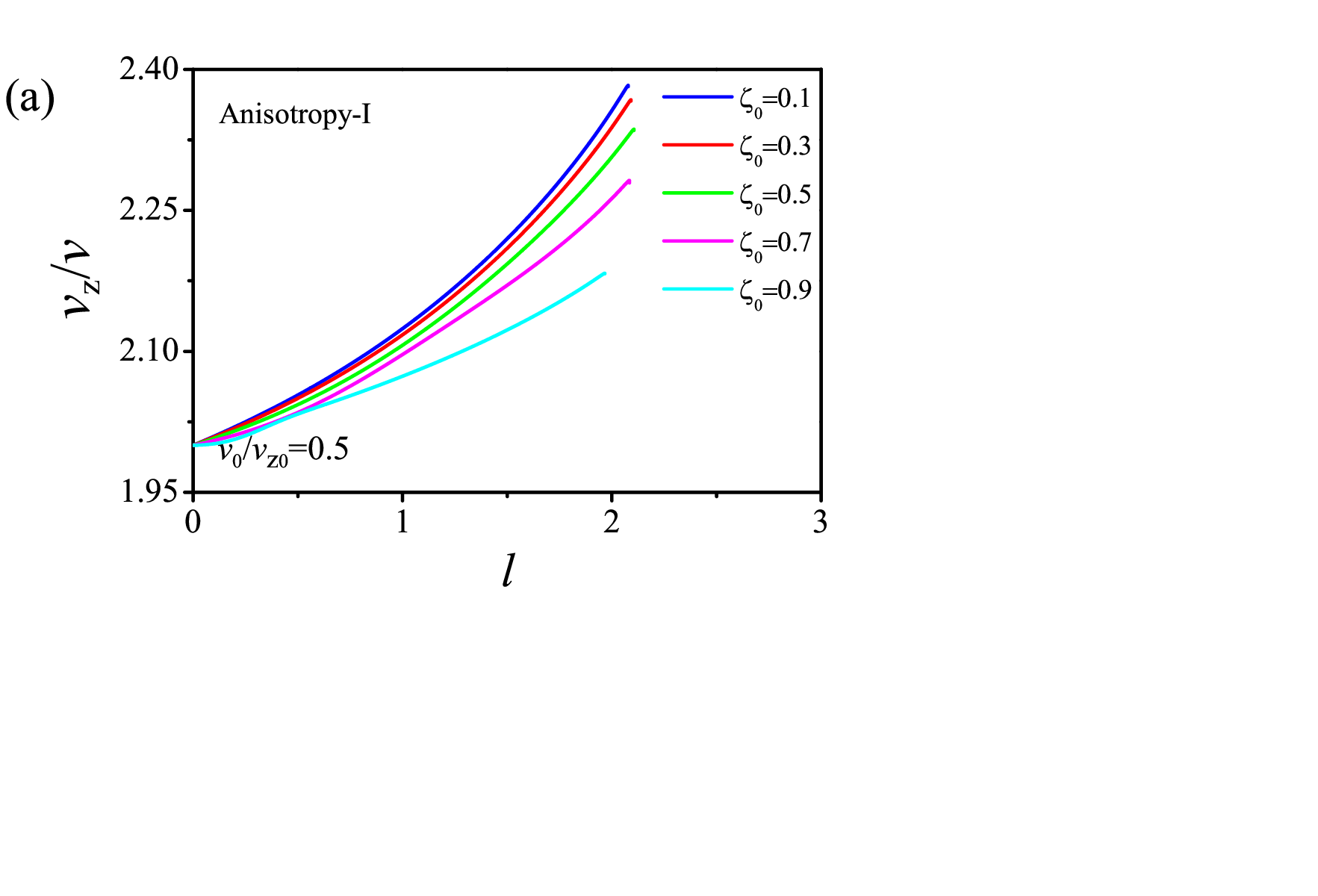}\vspace{-2.7cm}
\includegraphics[width=4.1in]{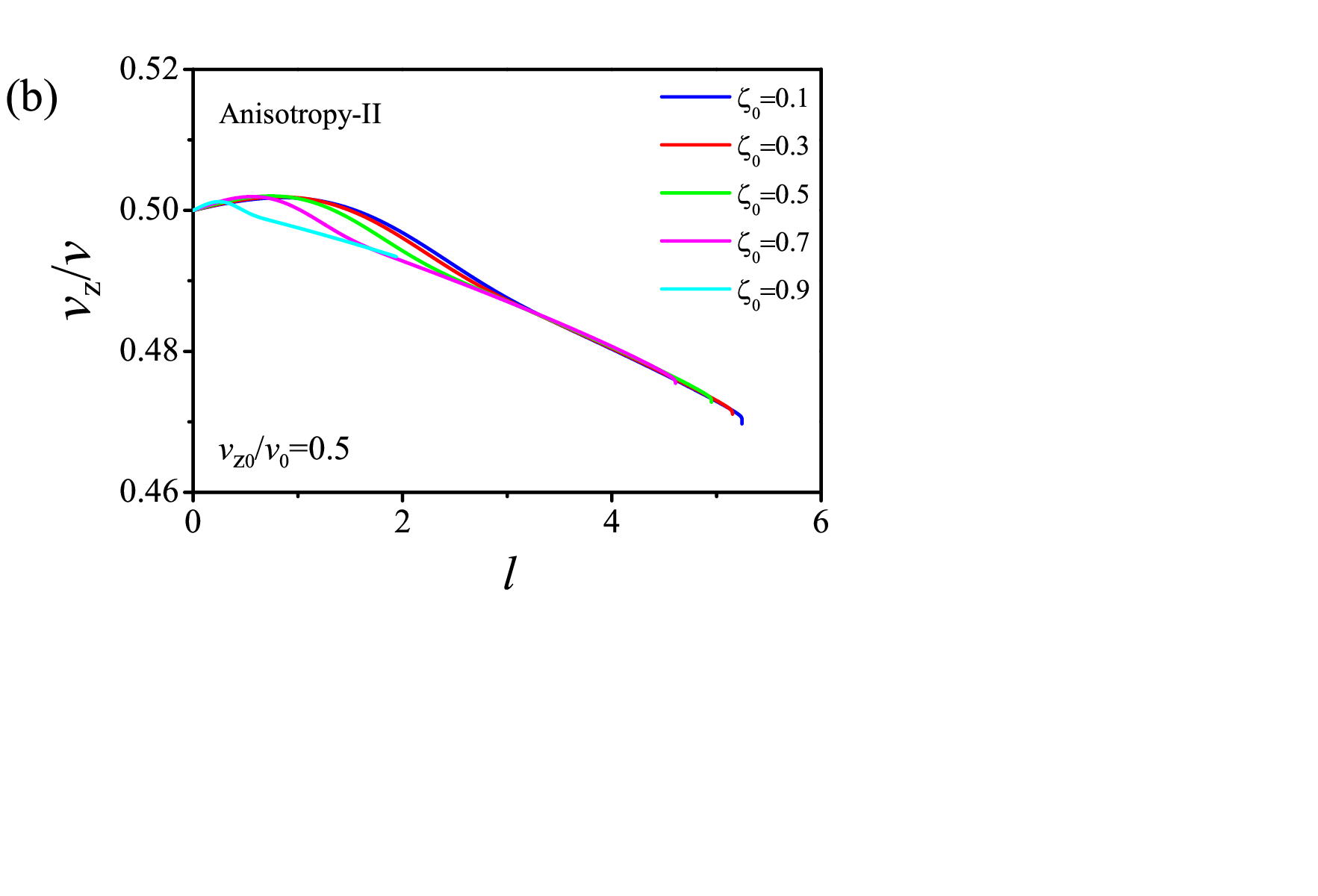}\\
\vspace{-2.7cm}
\caption{(Color online) The energy-dependent evolutions of $v_{z}/v$ are depicted with respect to the variation of $\zeta_0$, starting from
(a) Anisotropy-I and (b) Anisotropy-II situations.}\label{Fig_vz-1}
\end{figure}

Next, we study the stability of the tilting parameter under varying values of fermion velocities.
Given the observed robustness of $\zeta$ to changes in its initial value for both Anisotropy-I and Anisotropy-II scenarios
as shown in Fig.~\ref{Fig_zeta-1}, it is suitable to select a fixed $\zeta_0$ and examine its behavior by tuning the
fermion velocities. Figure \ref{Fig_zeta-2}(a) signals that, in the Anisotropy-I case, $\zeta/\zeta_{0}$ decreases and
gradually reaches a finite value within the type-I tDSM. Considering the Anisotropy-II case, depicted in Fig.~\ref{Fig_zeta-2}(b),
we observe similarities in the behavior of $\zeta/\zeta_{0}$ to that of Anisotropy-I when
$v_{z}/v_{0}>0.4$. However, at $v_{z}/v_{0}=0.4$, there is a tendency for $\zeta/\zeta_{0}$ to increase,
reaching $1.3$ at $v_{0}/v_{z0}=0.2$. Even though $\zeta$ is still less than $1$ with $\zeta/\zeta_{0}=1.3$,
this suggests a potential for $\zeta$ to exceed $1$, indicating a departure from the type-I tDSM.
These findings imply the significance of fermion velocities in influencing the stability and nature of $\zeta$.

To verify this, we provide Fig~\ref{Fig_zeta-2}(c), which includes more initial values for the Anisotropy-II case.
It exhibits that $\zeta$ is indeed capable of exceeding $1$ at $v_{z0}/v_0\approx0.12$. In particular, the tilting
parameter experiences a more rapid increase and attains a bigger value while $v_{z0}/v_0$ is smaller than $0.12$.
In this sense, we infer that the critical ratio of fermion velocities initiates the transition from the type-I tDSM to type-II tDSM
in the vicinity of $v_{z0}/v_0\approx0.12$. Schematically depicted in Fig.~\ref{Fig_phase}, there exist
two possible scenarios for such a transition. In the first scenario, the type-I tDSM undergoes a direct
transition to an $X$ state at $\emph{l}_{c}$ as illustrated in Fig.~\ref{Fig_phase}(a),
driven by the competition among Coulomb interaction and electron-phonon coupling as well as phonon-phonon interaction.
In the second scenario, shown in Fig.~\ref{Fig_phase}(b), the type-I tDSM first transitions to type-II tDSM at $l_*<l_c$
before entering into a $Y$ state.

Furthermore, Fig.~\ref{zeta_lam_0_g_mult} presents the effects of the electron-phonon
and Coulomb interactions on the tilting parameter with $\zeta_0=0.5$ for the distinct kinds of initial conditions.
By tuning the initial value of the electron-phonon interaction, Fig.~\ref{zeta_lam_0_g_mult}(a) shows
that, for the weak electron-phonon interactions,
$\zeta/\zeta_0$ decreases and then increases as the energy scale decreases, but for a bigger electron-phonon interactions,
it monotonically decreases. In comparison, with variation of the initial value of Coulomb interaction
as shown in Fig.~\ref{zeta_lam_0_g_mult}(b), the tilting parameter increases with decreasing energy scales
and the basic tendency would be stable and only qualitatively modified. It is worth highlighting that the tilting parameter $\zeta$
remains less than 1 (i.e., $\zeta/\zeta_0<2$) and thus the system is restricted to the first scenario
in the low-energy regime. Besides, Fig.~\ref{zeta_lam_0_g_mult}(c) for the absence of Coulomb interaction
indicates that the tendency of the tilting parameter is qualitatively consistent with
that depicted in Fig.~\ref{zeta_lam_0_g_mult}(a). In contrast,
switching off the electron-phonon interaction shown in Fig.~\ref{zeta_lam_0_g_mult}(d),
the tilting parameter gradually decreases upon lowering the energy scales and saturates
at a certain value. This implies that the electron-phonon interaction
contributes more to the tilting parameter than the Coulomb interaction.

To recapitulate, the tilting parameter $\zeta$ is robust with respect to its initial value, but relatively sensitive to
the ratio of fermion velocities. In particular, we identify that a critical value of the ratio, $v_{z0}/v_0\approx 0.12$, below which the type-I tDSM becomes unstable and can potentially transition to type-II tDSMs. However, our effective theory is confined to the Type-I tDSM, and hence the coupled RG equations are only well-defined within this context. Consequently, from now on we
only consider the first scenario as displayed in Fig.~\ref{Fig_phase}(a) and
investigate the behavior of all other related parameters in the remainder of this section.
Furthermore, we judge the candidate phase for $X$ state and explore the physical implications in the next two sections.

\subsection{Fates of fermion velocities}

In the context of ype-I tDSM, we begin with studying the impacts of interactions on fermion velocities.
Figure \ref{Fig_vz-1} illustrates the basic tendencies of the anisotropy
of fermion velocities $v_{z}/v$. Starting from the Anisotropy-I case, one can read from Fig.~\ref{Fig_vz-1}(a)
that, as the energy scale decreases, $v_{z}/v$ progressively increases and the weaker tilting parameter
is preferable to support this increase. In sharp contrast, as illustrated in Fig.~\ref{Fig_vz-1}(b), the ratio $v_{z}/v$
receives a slight increase initially when deviating from the Anisotropy-II, but it subsequently decreases to a certain value
as the energy scale diminishes. Besides, both its evolution and the final value of this ratio are fairly insusceptible to the tilting parameter, which are distinct from its Anisotropy-I counterpart.
This implies that the qualitative behavior of anisotropy of fermion velocities heavily relies on its beginning value
in comparison with the strength of the tilting parameter.

\begin{figure}
\centering
\includegraphics[width=4.1in]{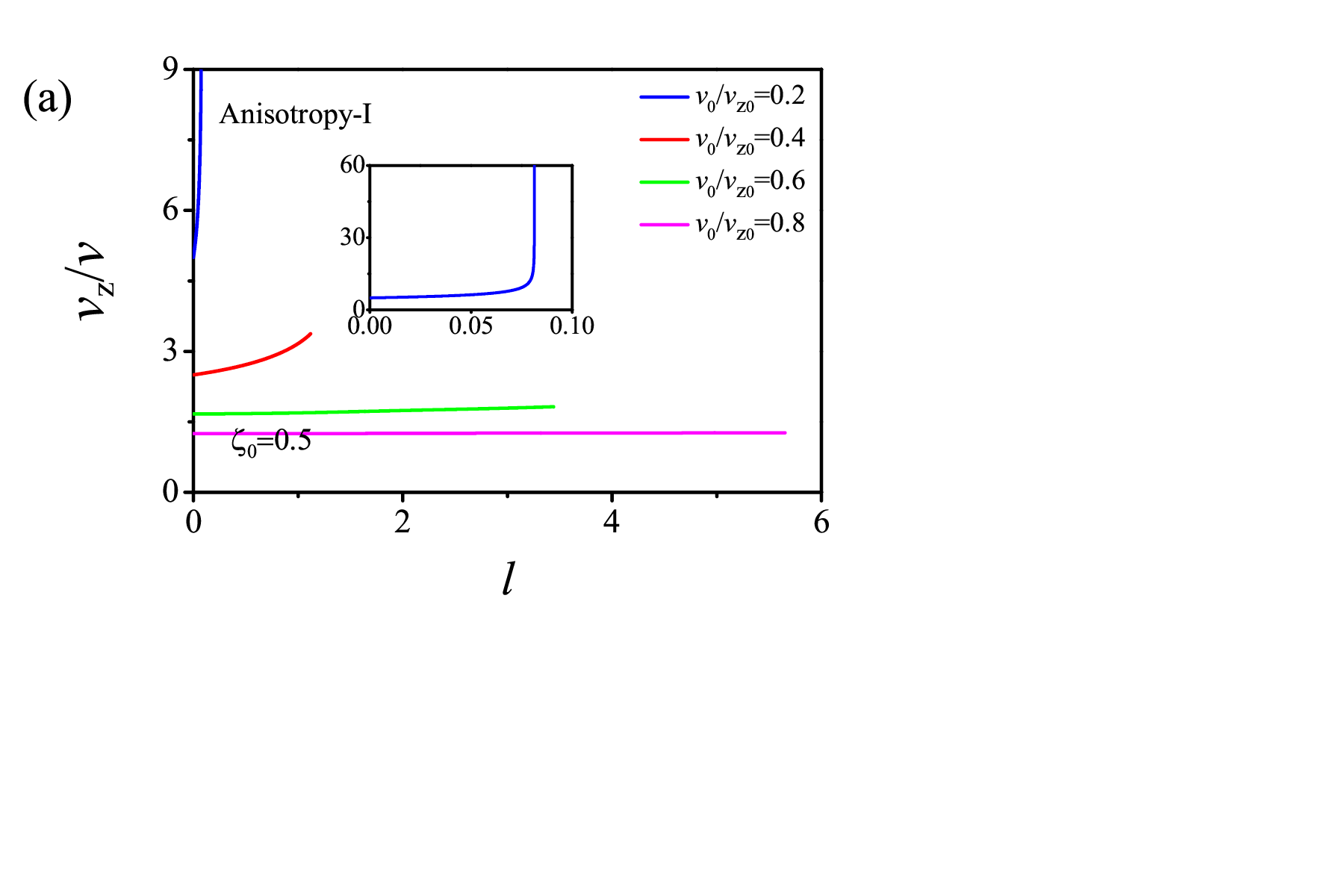}\vspace{-2.7cm}
\includegraphics[width=4.1in]{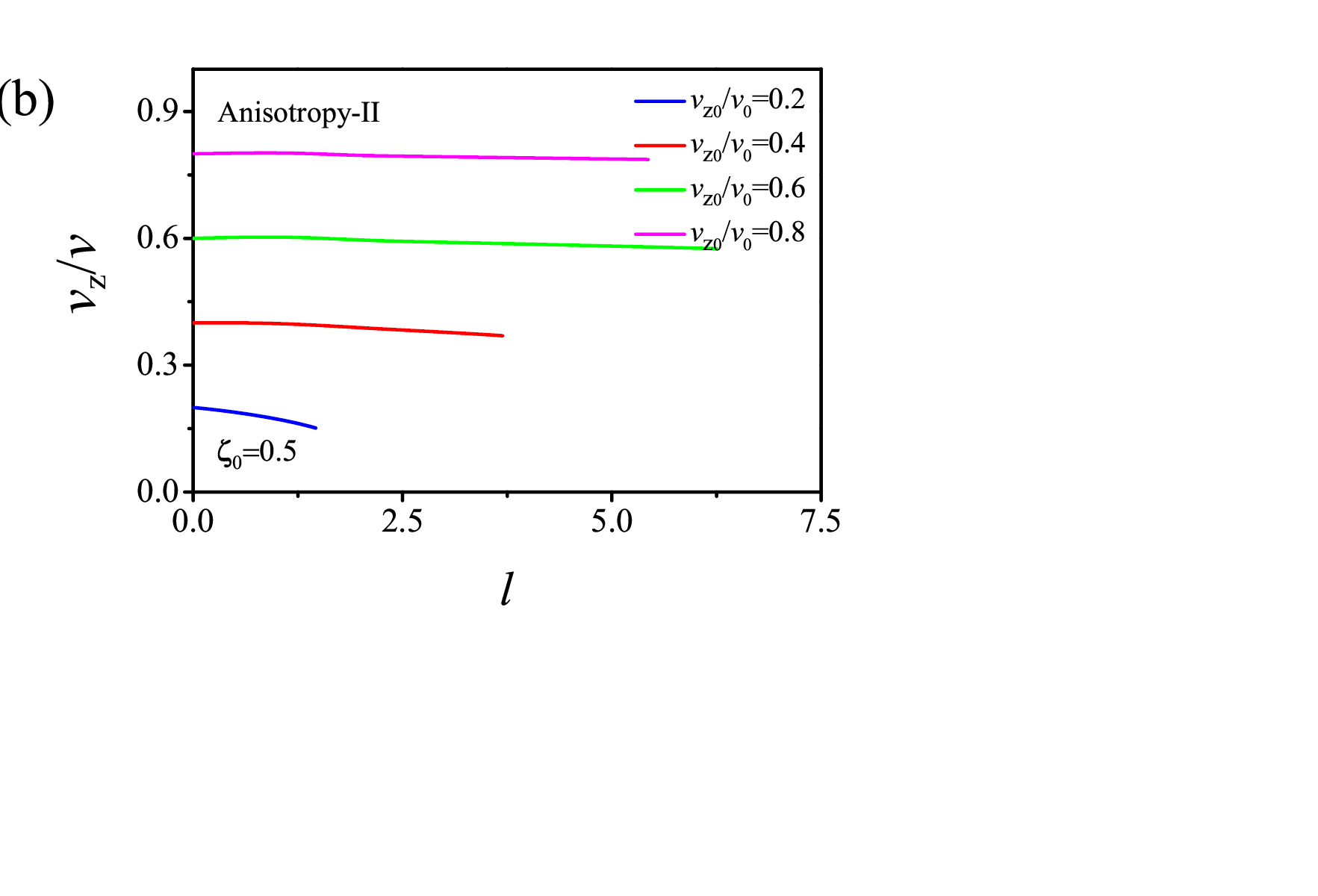}\\
\vspace{-2.7cm}
\caption{(Color online) The energy-dependent evolutions of $v_{z}/v$ are depicted with
respect to the variations of initial fermion velocities,
starting from (a) Anisotropy-I and (b) Anisotropy-II situations.}\label{Fig_vz-2}
\end{figure}

It is therefore necessary to further investigate the influence of the initial condition of anisotropy
on the low-energy fate of fermion velocities. As shown in Fig.~\ref{Fig_vz-2} with several initial values of $v/v_z$, we
adopt a representative tilting parameter $\zeta_{0}=0.5$ to show the energy-dependent
anisotropy of fermion velocities for both Anisotropy-I and Anisotropy-II.
Specifically, for Anisotropy-I, the anisotropy of fermion velocities $v_{z}/v$
is primarily dependent on its initial anisotropy. As displayed in Fig.~\ref{Fig_vz-2}(a),
$v_{z}/v$ is insusceptible to the energy scales, remaining relatively stable in the presence
of a weak starting anisotropy. However, with a strong initial anisotropy, it becomes sensitive and experiences rapid growth.
In contrast, Fig.~\ref{Fig_vz-2}(b) indicates that $v_{z}/v$, departing from Anisotropy-II, can only receive slight negative corrections and even becomes saturated when the starting anisotropy is adequately weak. Therefore, we infer that the fate of $v_{z}/v$ predominantly hinges upon the initial anisotropy. In addition, it is sensitive to the value of $v_0/v_{z0}$ and $\zeta_0$ when starting from Anisotropy-I and Anisotropy-II, respectively.

\begin{figure}[htpb]
\centering
\subfigure[]{\includegraphics[width=3in]{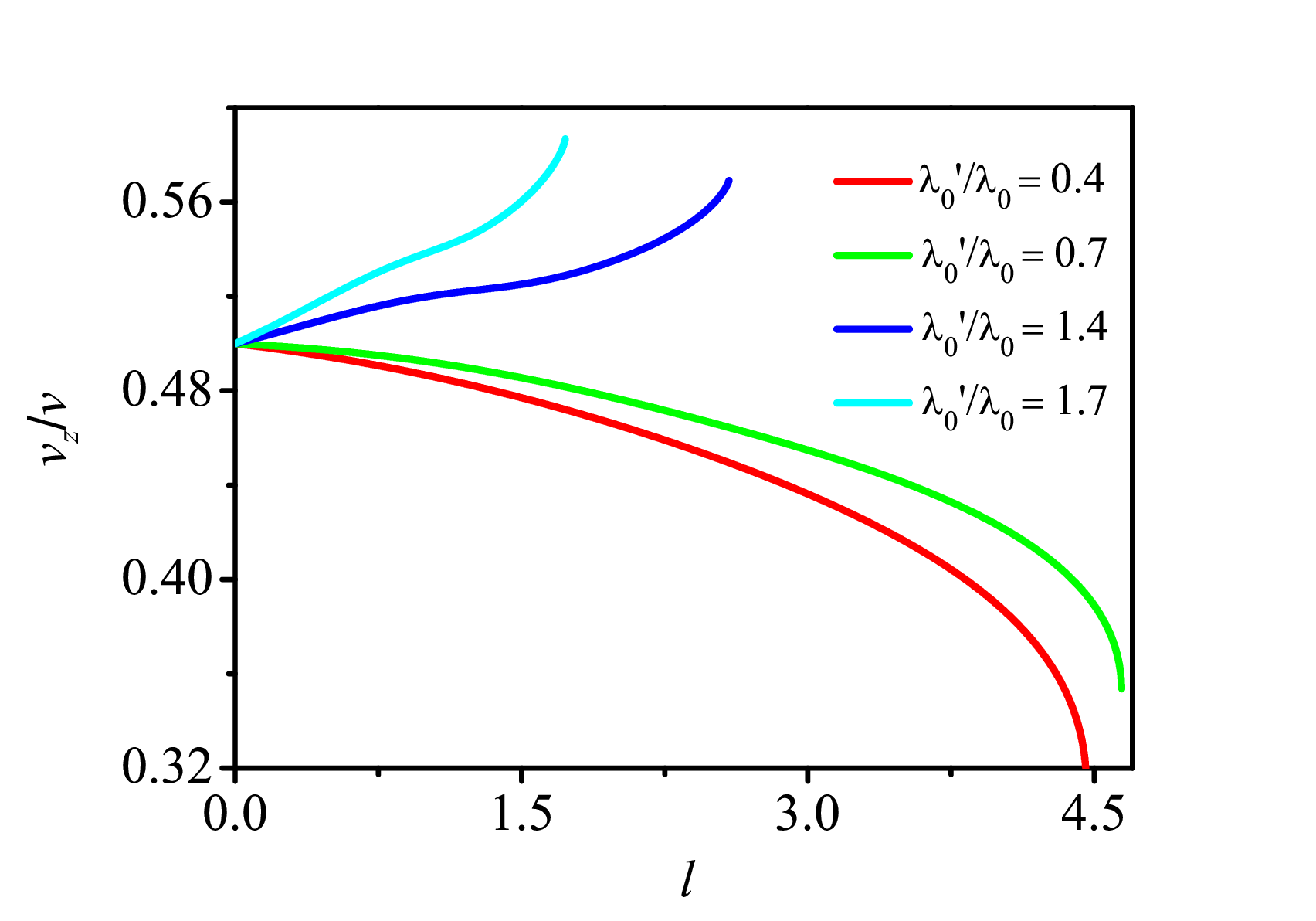}}\\\vspace{-0.6cm}
\subfigure[]{\includegraphics[width=3in]{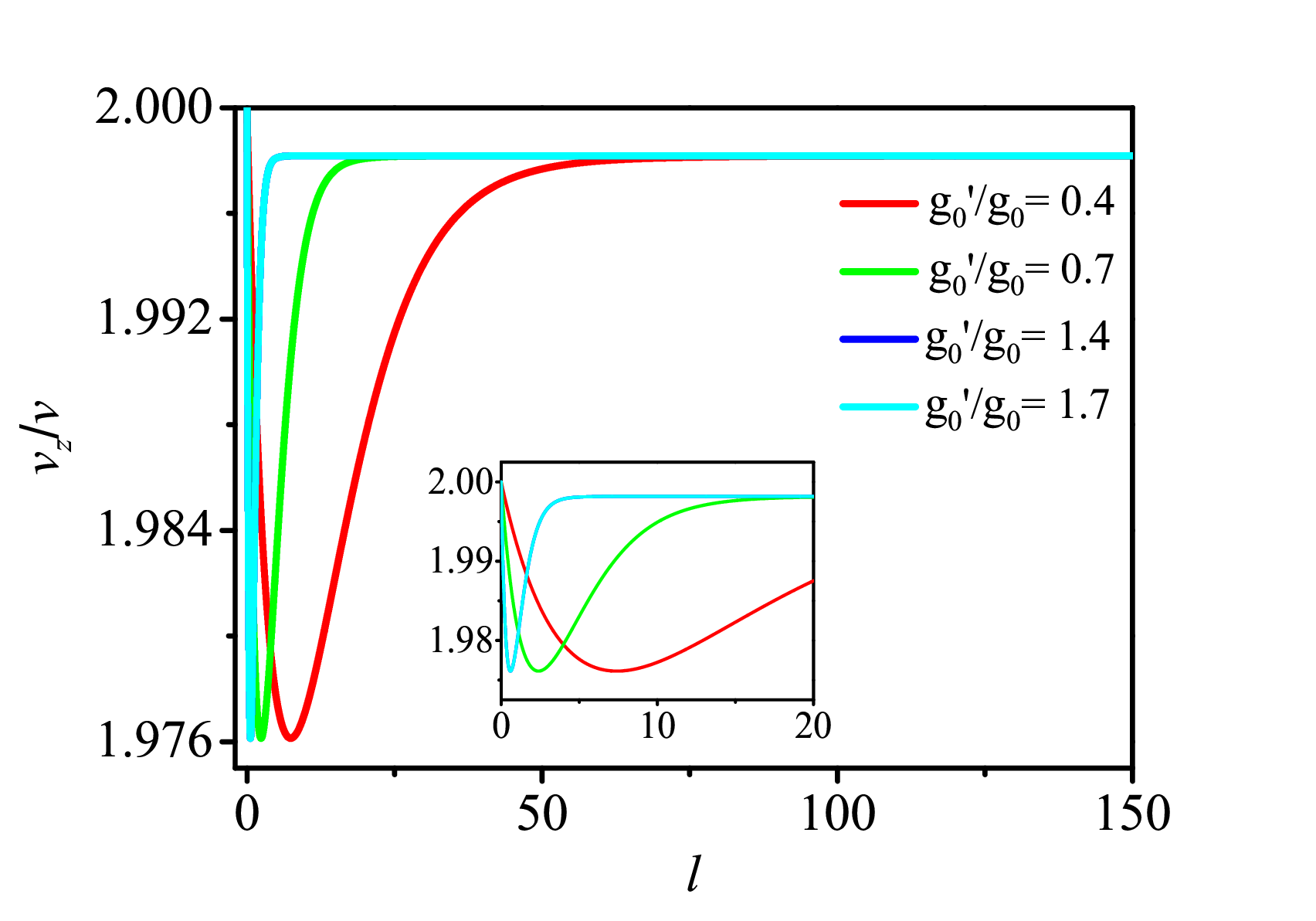}}
\caption{(Color online) Energy-dependent evolutions of $v_z/v$ with $\zeta_0=0.5$
for several representative initial values of: (a) the electron-phonon interaction in the presence of
Coulomb interaction at $v_z(0)/v(0) = 0.5$ and (b) the Coulomb interaction in the absence of the
electron-phonon interaction ($\lambda=0$) at $v_z(0)/v(0)=2.0$. The basic results for both the other initial conditions
are similar and not shown (hereby $X'_0/X_0$ with $X=V^{xx}_L,g,\lambda$ denotes the
ratios of the initial values with respect to those of Fig.~\ref{Fig_zeta-1}).}\label{vzv_lam_g_mult}
\end{figure}

\begin{figure}
\centering
\includegraphics[width=4.1in]{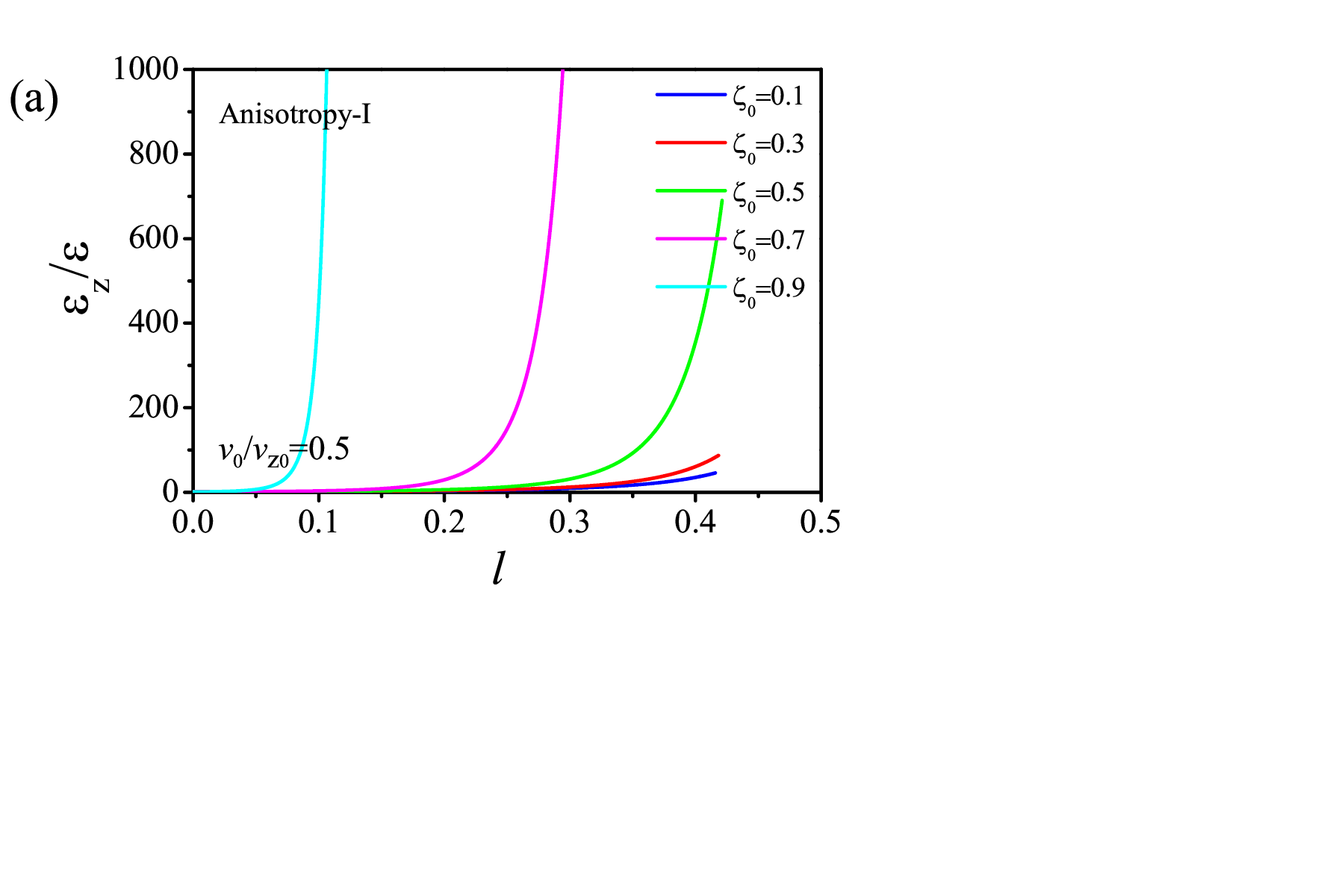}\vspace{-2.7cm}
\includegraphics[width=4.1in]{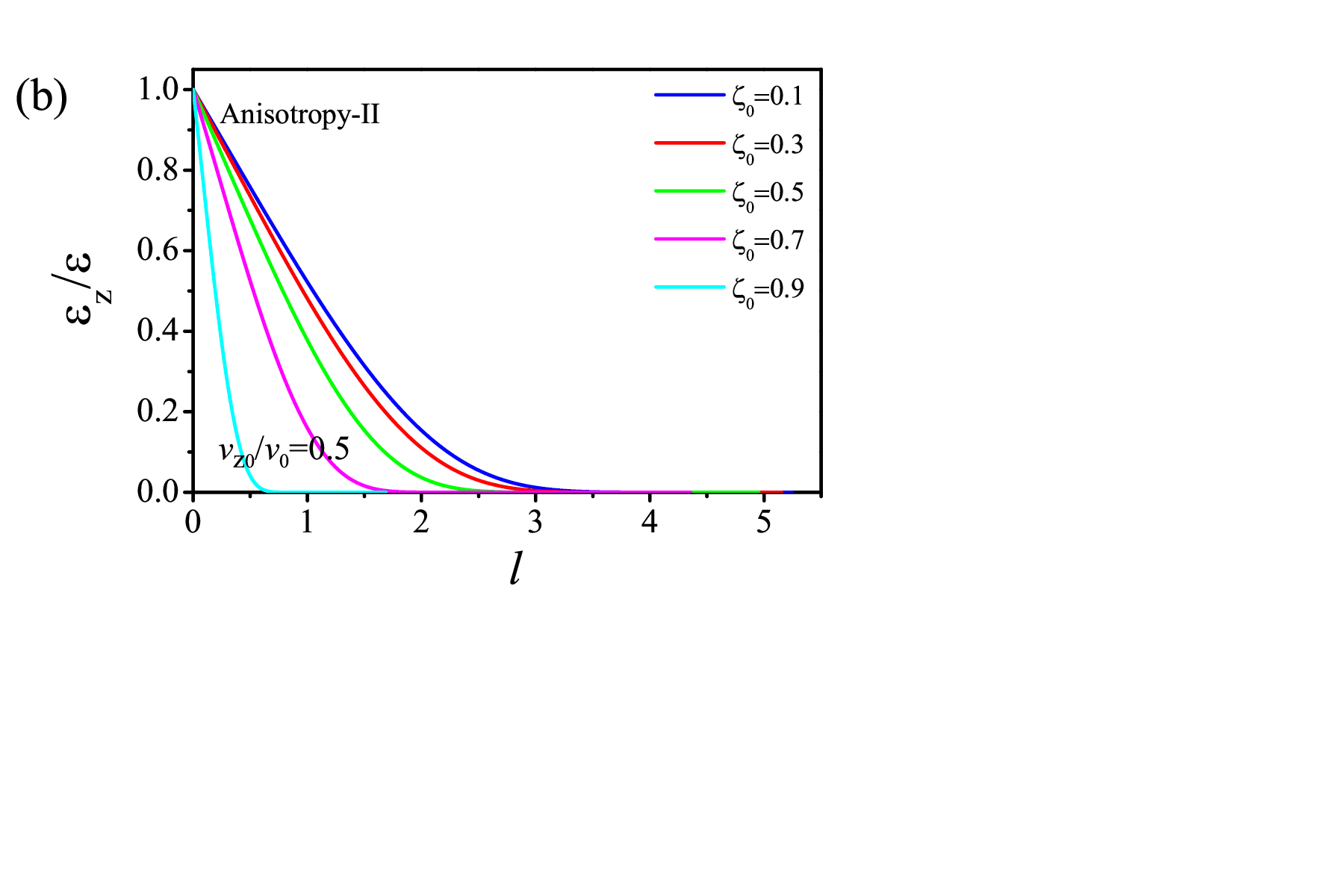}\\
\vspace{-2.7cm}
\caption{(Color online) The energy-dependent evolutions of $\epsilon_{z}/\epsilon$ are
depicted with respect to the variations of $\zeta_{0}$,
starting from (a) Anisotropy-I and (b) Anisotropy-II situations.}\label{Fig_epsilon-1}
\end{figure}

\begin{figure}
\centering
\includegraphics[width=4.1in]{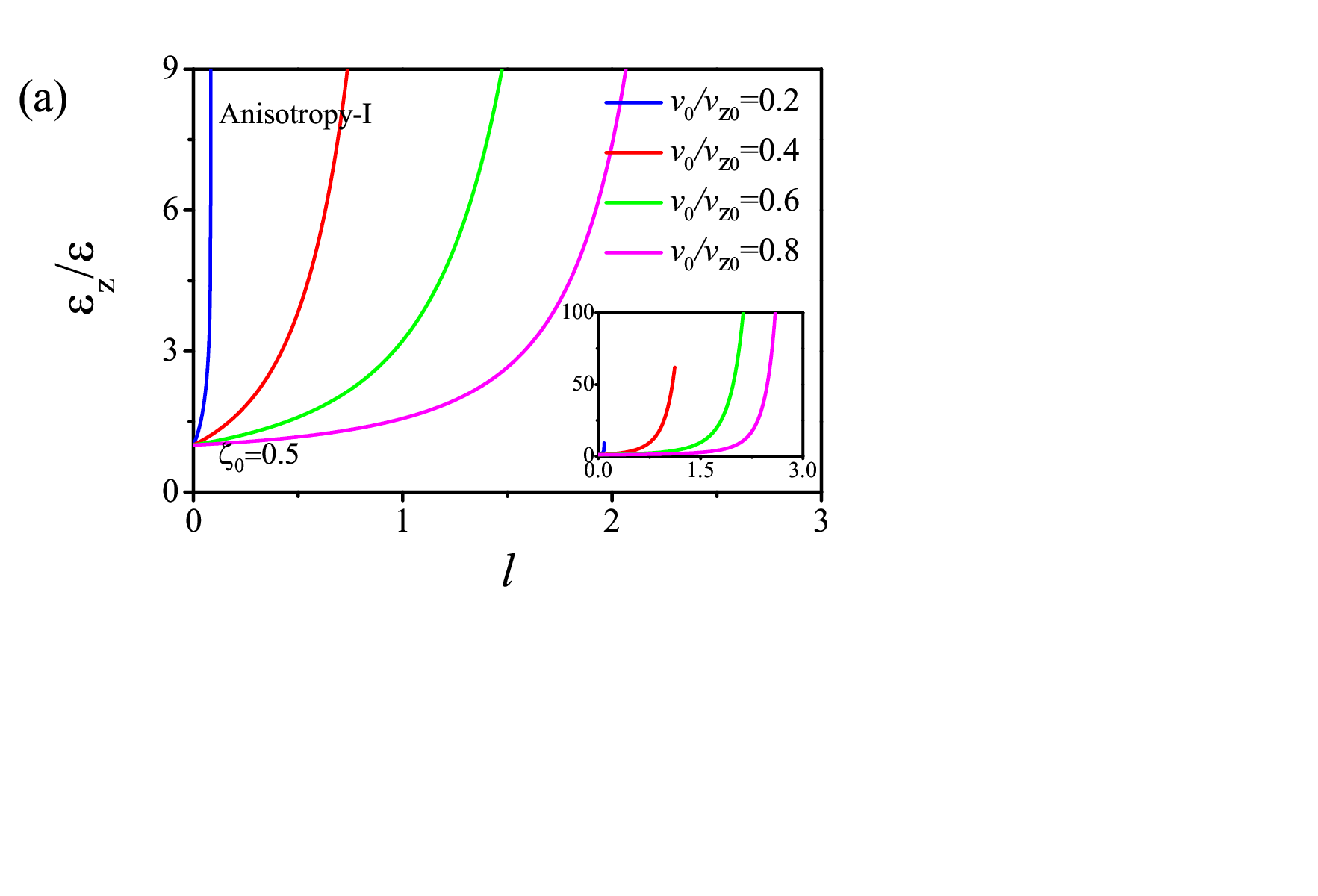}\vspace{-2.7cm}
\includegraphics[width=4.1in]{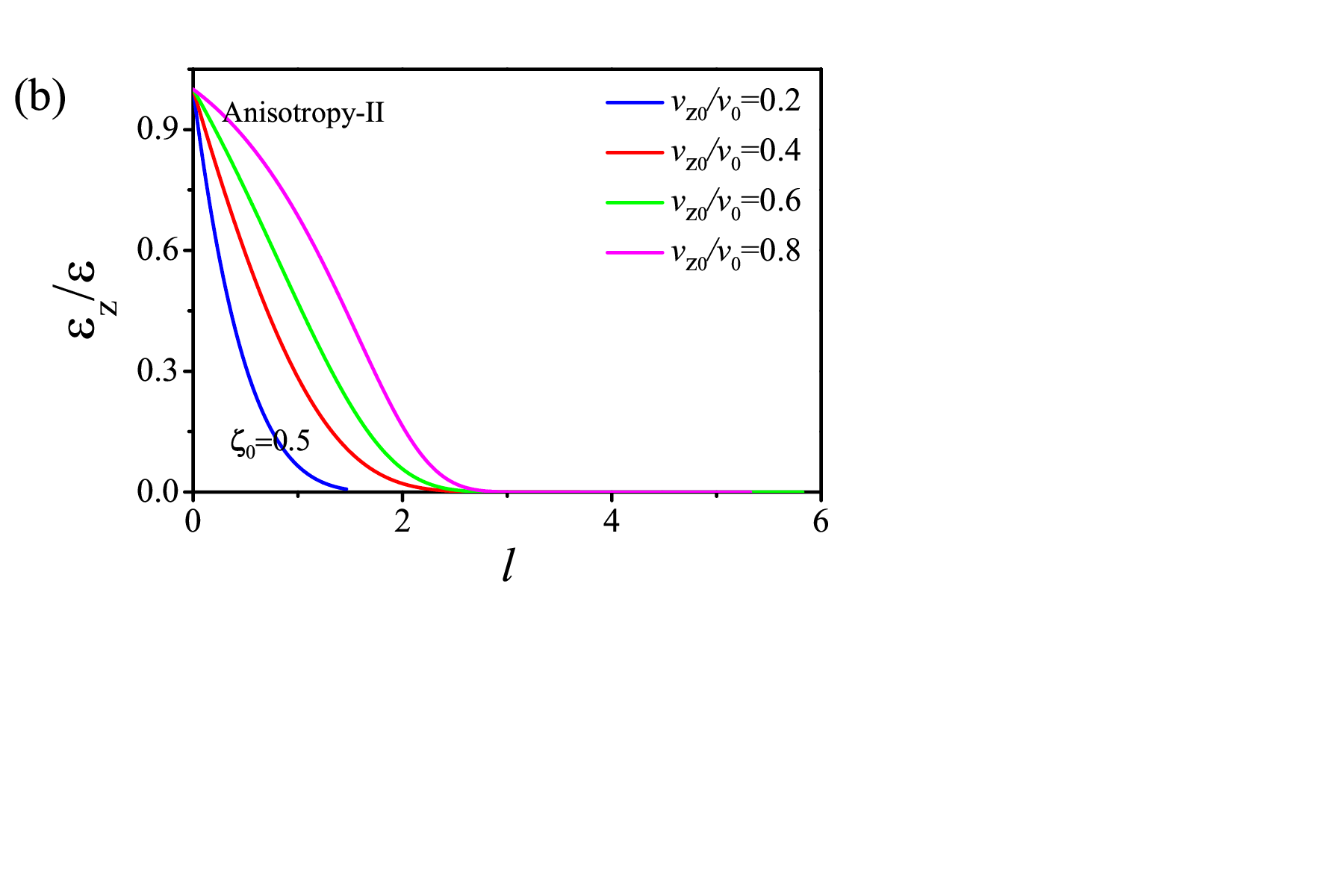}\\
\vspace{-2.7cm}
\caption{(Color online) The energy-dependent evolutions of $\epsilon_{z}/\epsilon$ are depicted
with respect to the variations of initial fermion velocities,
starting from (a) Anisotropy-I and (b) Anisotropy-II situations.}\label{Fig_epsilon-2}
\end{figure}

Furthermore, we parallel the strategy adopted in Fig.~\ref{zeta_lam_0_g_mult} to examine the effects of the electron-phonon
and Coulomb interactions on the anisotropy of fermion velocities. Figure \ref{vzv_lam_g_mult}(a) shows that
the anisotropy of fermion velocities can either be increased or decreased with tuning the initial value of electron-phonon interaction. In comparison, the Coulomb interaction only quantitatively renormalizes
the evolution of anisotropy of fermion velocities but does not changes its basic tendencies presented in
Fig.~\ref{Fig_vz-1} and Fig.~\ref{Fig_vz-2}.
Additionally, switching off the electron-phonon interaction shown in Fig.~\ref{vzv_lam_g_mult}(b),
we find that the anisotropy of fermion velocities evolves and saturates at a
certain value in the low-energy regime.
In comparison, analogous to the behavior of tilting parameter in Fig.~\ref{zeta_lam_0_g_mult}(c) in the absence of the Coulomb interaction,
the electron-phonon interaction can largely reshape the anisotropy of fermion velocities. Consequently, the electron-phonon interaction plays a more important role in the anisotropy of fermion velocities
than the Coulomb interaction.

\begin{figure}
\centering
\includegraphics[width=2.66in]{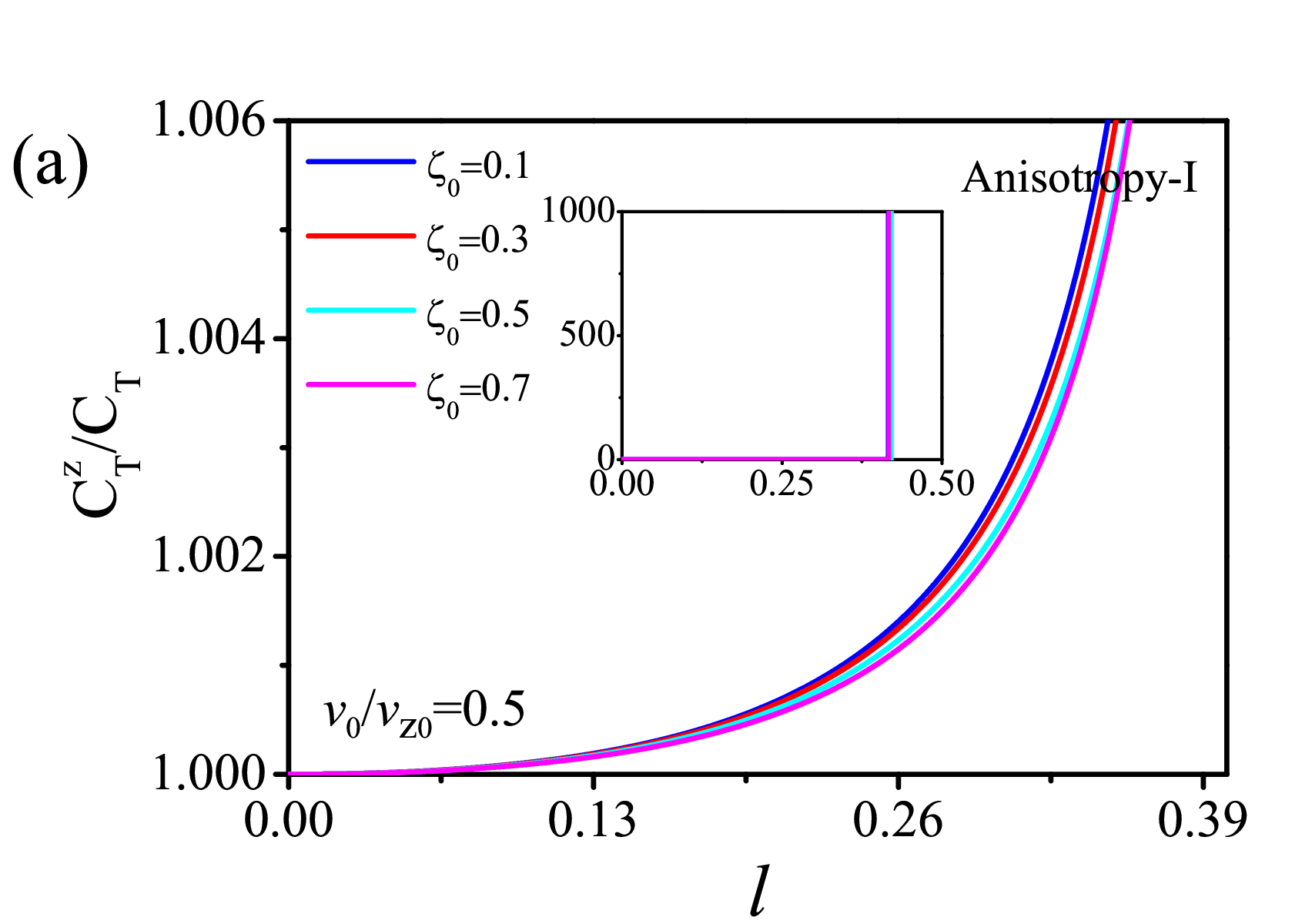}\vspace{-0.3cm}
\includegraphics[width=2.66in]{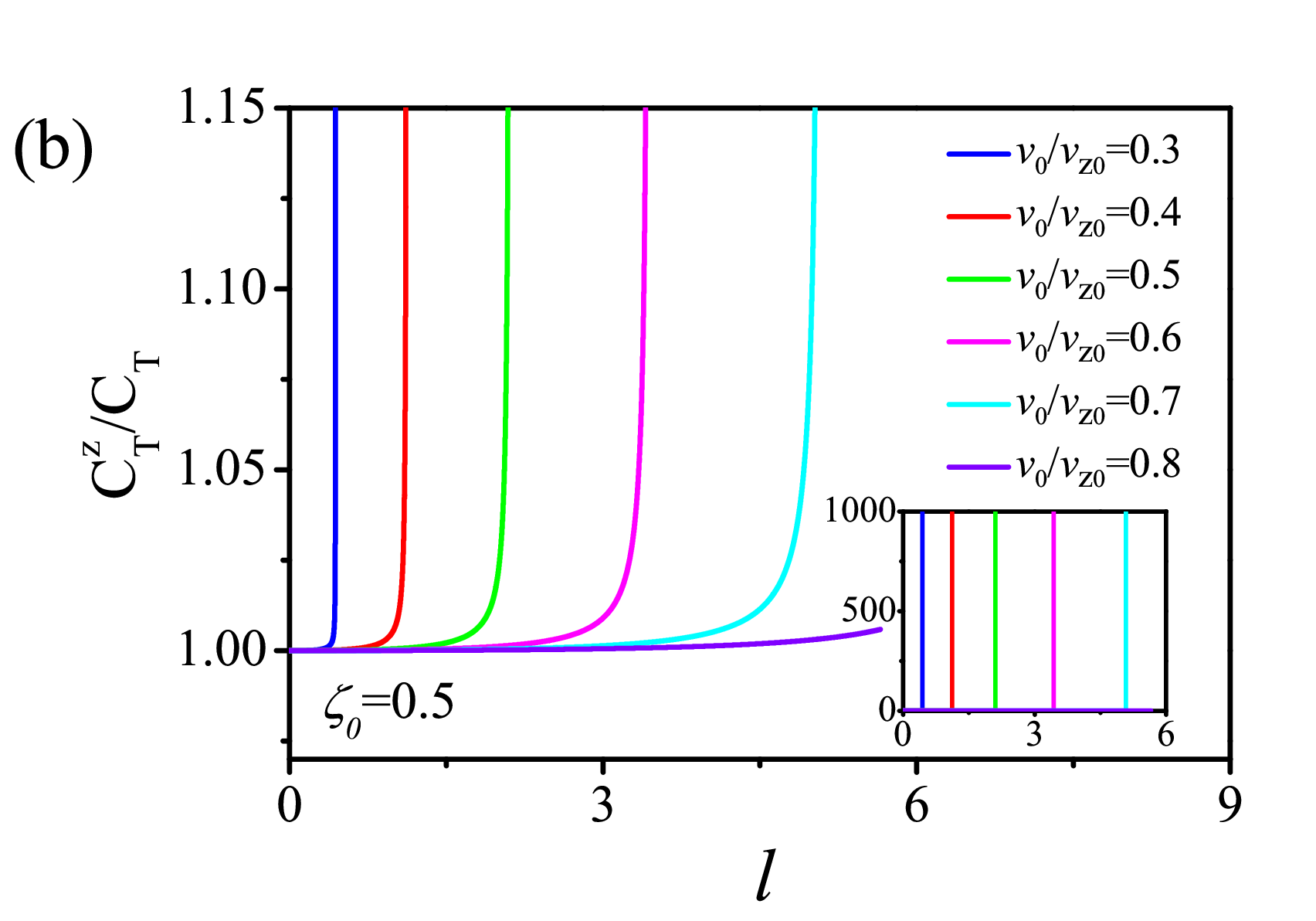}\\
\vspace{-0.3cm}
\caption{(Color online) The energy-dependent evolutions of $C_{T}^{z}/C_{T}$ starting from Anisotropy-I
are depicted with respect to the variations of (a) $\zeta_0$ and (b) $v_0/v_{z0}$.}\label{Fig_CTz-1}
\end{figure}

\subsection{Fates of $\epsilon_z/\epsilon$ and $g/g_0$}

Subsequently, we move to investigate the impacts of coupled interactions on the dielectric
constant, which is an important quantity to measure the strength of Coulomb interactions.
Given the focus on a tilted 3D tDSM, we are more interested in examining the ratio
of the dielectric constant between different orientations.
By selecting a specific initial ratio of fermion velocities, Fig.~\ref{Fig_epsilon-1}
clearly displays that, in the case of Anisotropy-I ($v_{0}/v_{z0}=0.5$), $\epsilon_{z}/\epsilon$
quickly increases. Conversely, it decreases rapidly to zero in Anisotropy-II ($v_{z0}/v_{0}=0.5$).
Despite these results being robust under the variations of the initial value
of tilting parameter $\zeta_{0}$, it is worth pointing out that a bigger $\zeta_0$ is much more helpful
to enhance the critical value of $\epsilon_{z}/\epsilon$ in Anisotropy-I and increase the critical
energy scale that is inversely proportional to $l_c$ in Anisotropy-II, respectively.

\begin{figure*}[htpb]
\centering
%\hspace{-0.8cm}
\subfigure[]
{\includegraphics[width=1.9in]{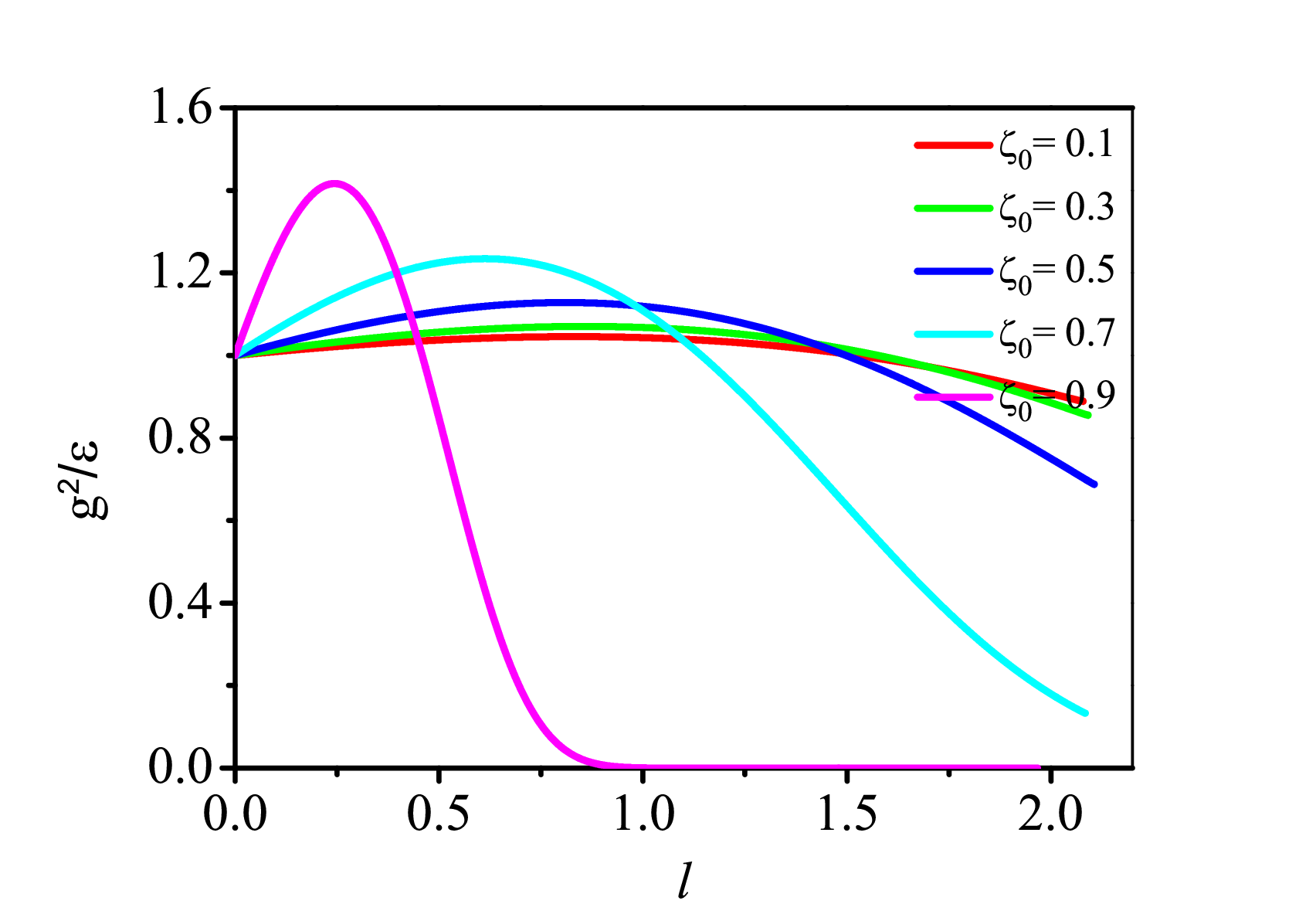}}
\hspace{-0.8cm}
\subfigure[]
{\includegraphics[width=1.9in]{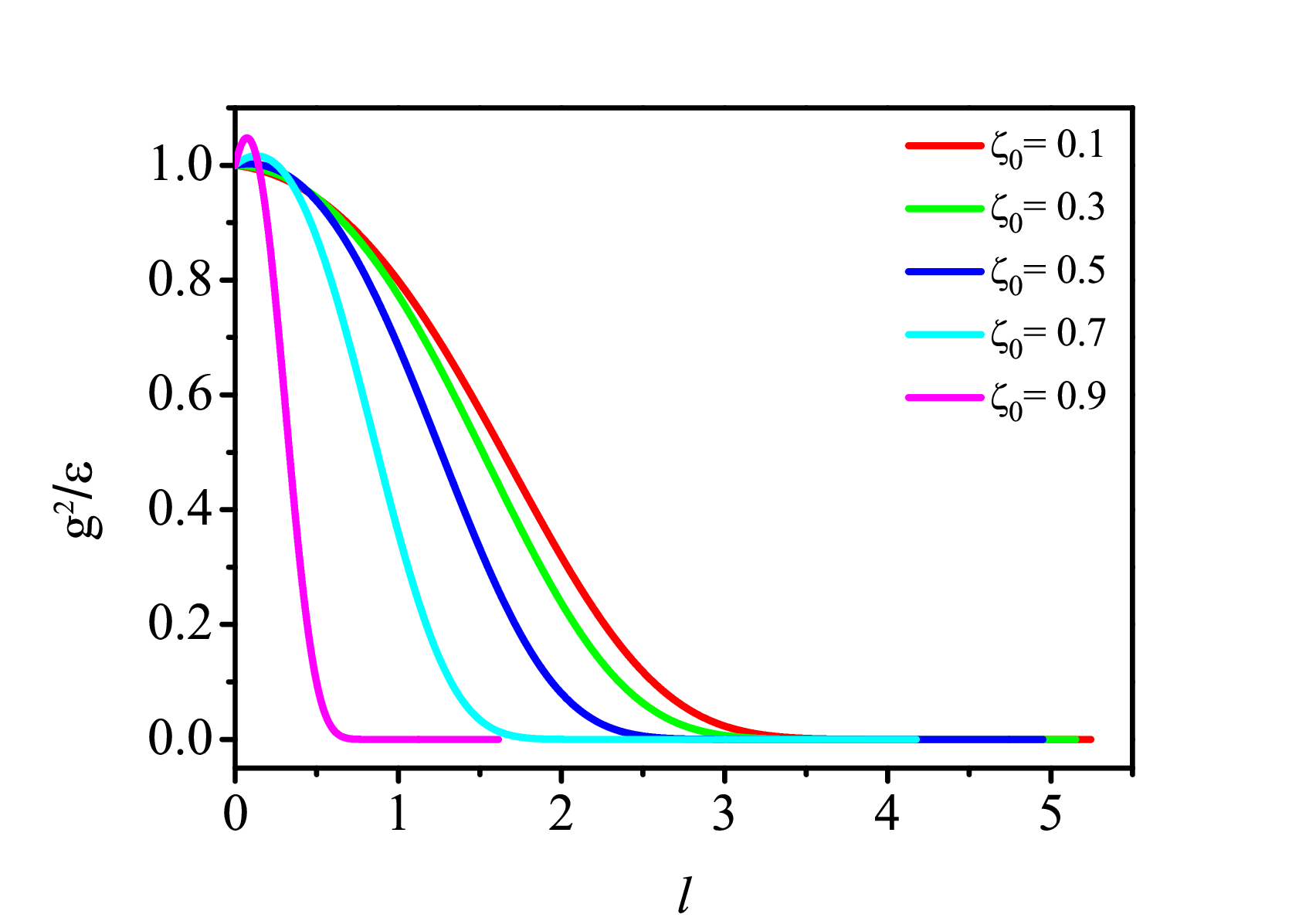}}
\hspace{-0.8cm}
\subfigure[]
{\includegraphics[width=1.9in]{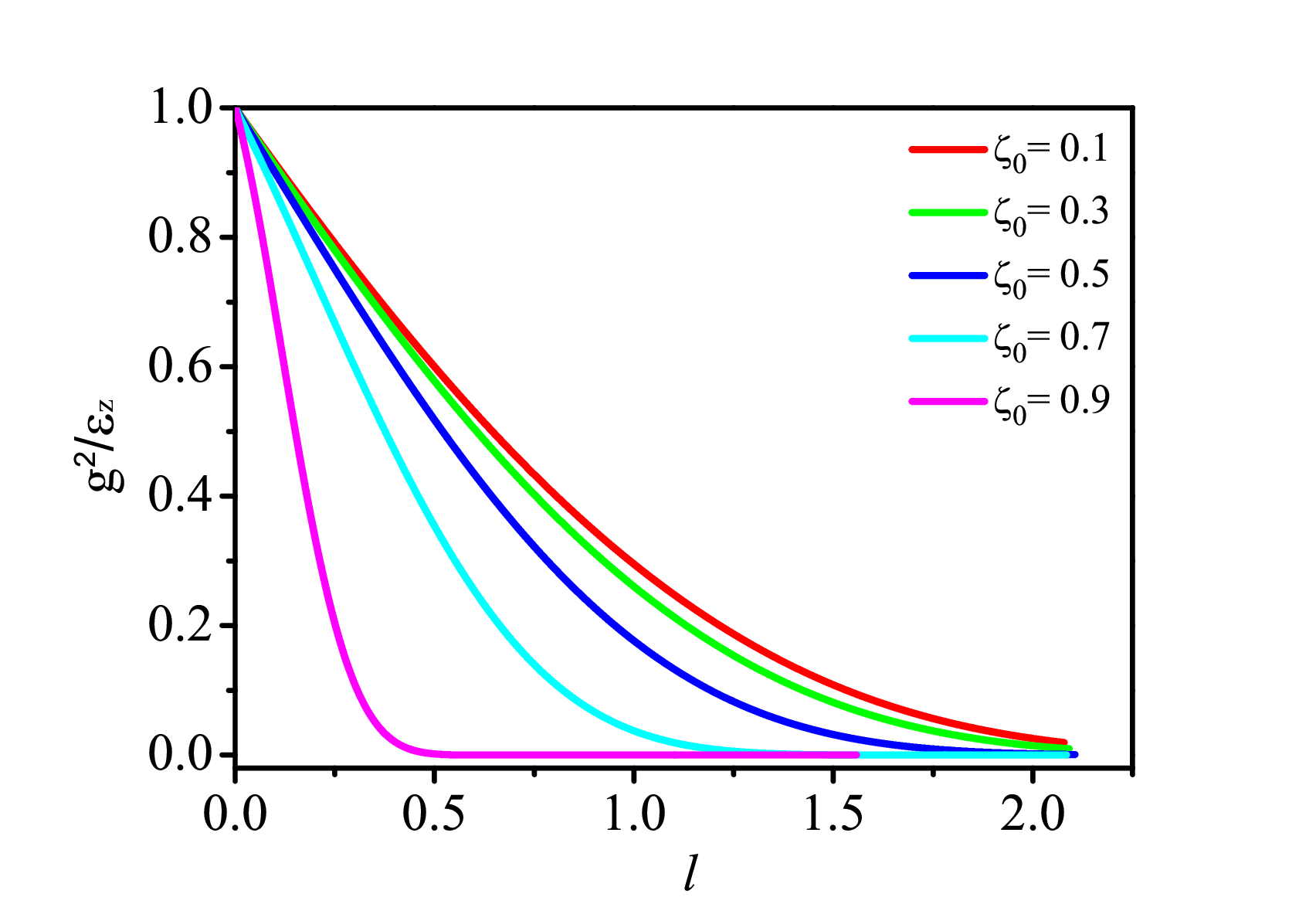}}
\hspace{-0.8cm}
\subfigure[]
{\includegraphics[width=1.9in]{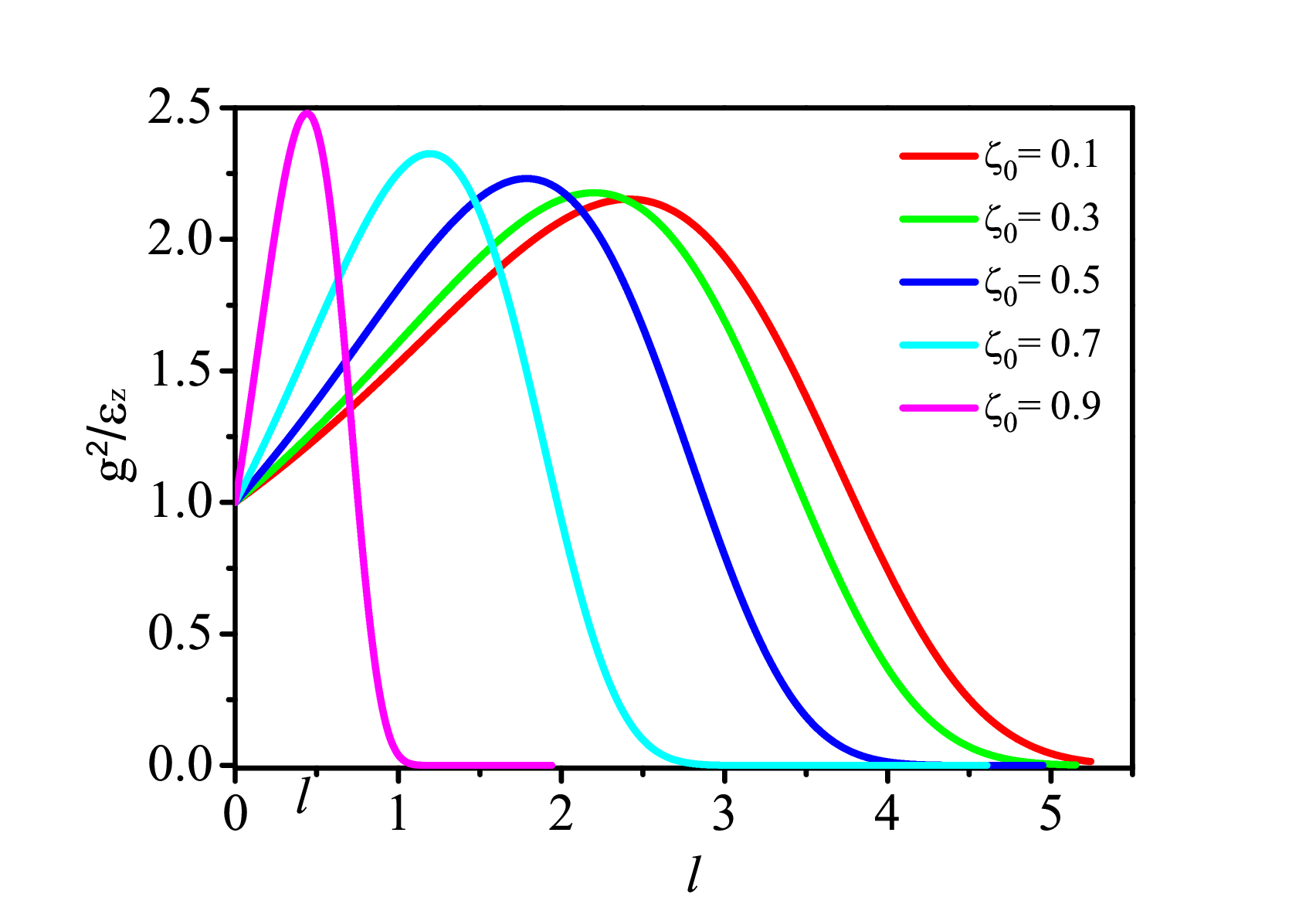}}
\vspace{-0.3cm}
\caption{(Color online) Energy-dependent tendencies of $g^2/\epsilon$ and $g^2/\epsilon_z$ for (a) and (c)
Anisotropy~I at $v_0/v_{z0}=0.5$, and for (b) and (d)
Anisotropy~II at $v_{z0}/v_0=0.5$, with variations of the tilting parameter $\zeta$.}
\label{g2epsilon}
\end{figure*}

In addition to checking the variation of the tilting parameter, we also examine the stability of $\epsilon_{z}/\epsilon$
by tuning the initial anisotropy of fermion velocities. We notice from Fig.~\ref{Fig_epsilon-2} that
$\epsilon_{z}/\epsilon$ exhibits a similar trend under the influence of
$v_0/v_{z0}$ as it does under the effects of $\zeta_0$ in Fig.~\ref{Fig_epsilon-1}. In other words,
as the energy scale decreases, $\epsilon_{z}/\epsilon$ increases quickly and progressively vanishes while starting from
Anisotropy-I and Anisotropy-II, respectively. Besides, a stronger anisotropy of fermion
velocities leads to a quicker increase or decrease.

Furthermore, it is of particular importance to examine the low-energy evolutions of both $g^2/\epsilon$ and $g^2/\epsilon_z$ for
the $o-xy$ plane and $z$ direction, which characterize the effective interaction strength of the Coulomb interaction.
As shown in Fig.~\ref{g2epsilon}(a) for Anisotropy-I, the ratio $g^2/\epsilon$ initially increases but subsequently decreases as energy is lowered. However, it gradually decreases for Anisotropy-II as shown in Fig.~\ref{g2epsilon}(b).
In comparison, $g^2/\epsilon_z$ displays a similar rise-and-fall behavior with increasing scaling parameter $l$,
as shown in Fig.~\ref{g2epsilon}(d), and Fig.~\ref{g2epsilon}(c) shows that it
monotonically decreases ultimately approaching zero in the low-energy regime.
for Anisotropy-II. This indicates that the Coulomb interaction is generally
screened by the interplay of distinct interactions in the low-energy regime. In particular, it is more screened
for Anisotropy~II and eventually vanishes at the lowest-energy limit.

To wrap up, the dielectric constant evolves towards the strong anisotropy at low-energy
due to the competition among various kinds of interactions. Depending on the departure from Anisotropy-I or
Anisotropy-II, it can either be driven to $\epsilon_z/\epsilon\gg1$
or $\epsilon_z/\epsilon\ll1$.  Since the strength of the Coulomb interaction is inversely
proportional to the dielectric constant, this indicates that the Coulomb interaction in direction-$z$ or direction-$x,y$ would be
considerably screened in the low-energy regime. Before going further, it is worth emphasizing that the
basic behavior of $g/g_0$, which characterizes the coupling strength between fermion and auxiliary bosonic field,
is analogous to that of $\epsilon_z/\epsilon$ and hence not shown for brevity.

\begin{figure*}
\hspace{0.6cm}
\includegraphics[width=3.7in,height=2.7in]{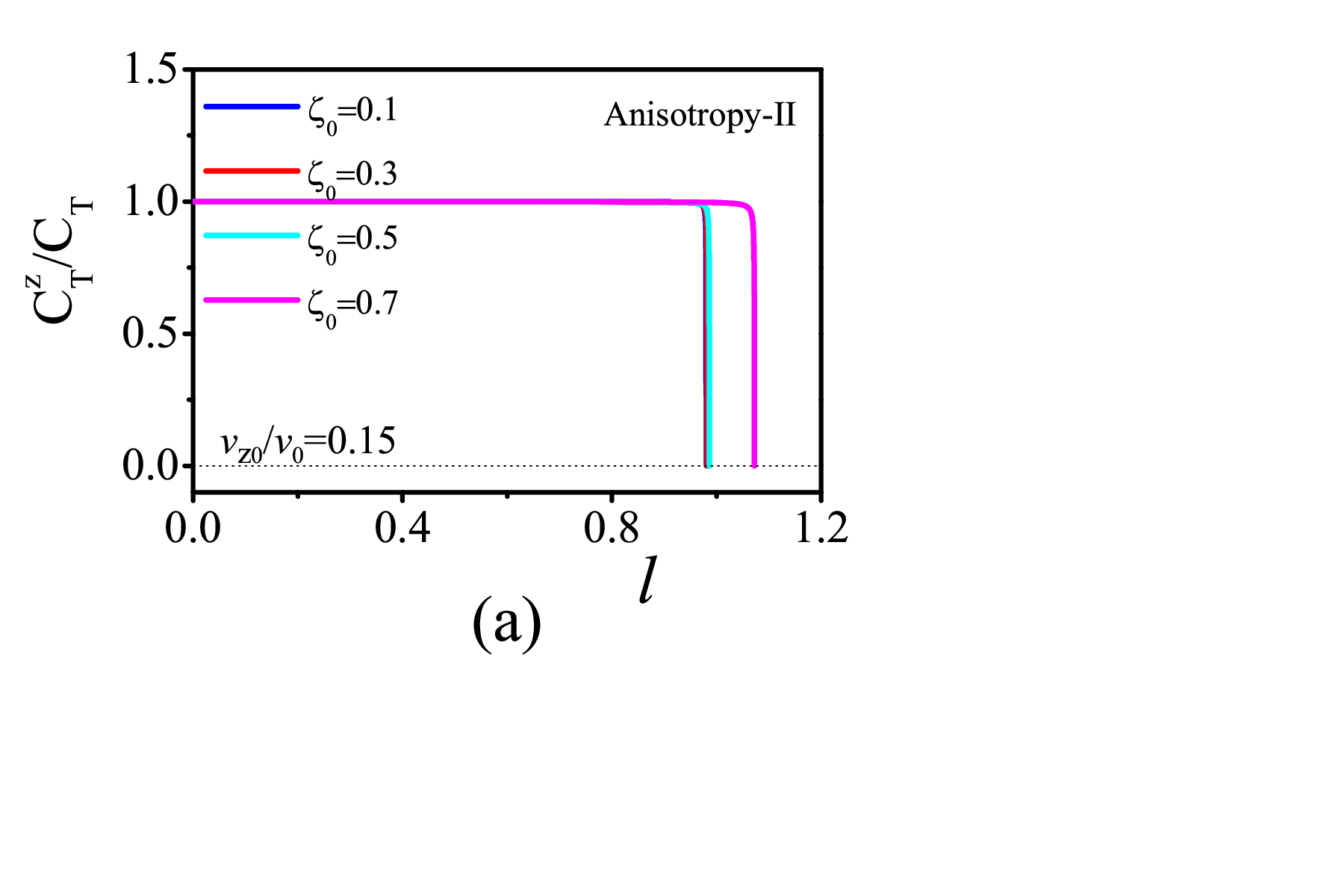}\hspace{-2.0cm}
\includegraphics[width=3.7in,height=2.7in]{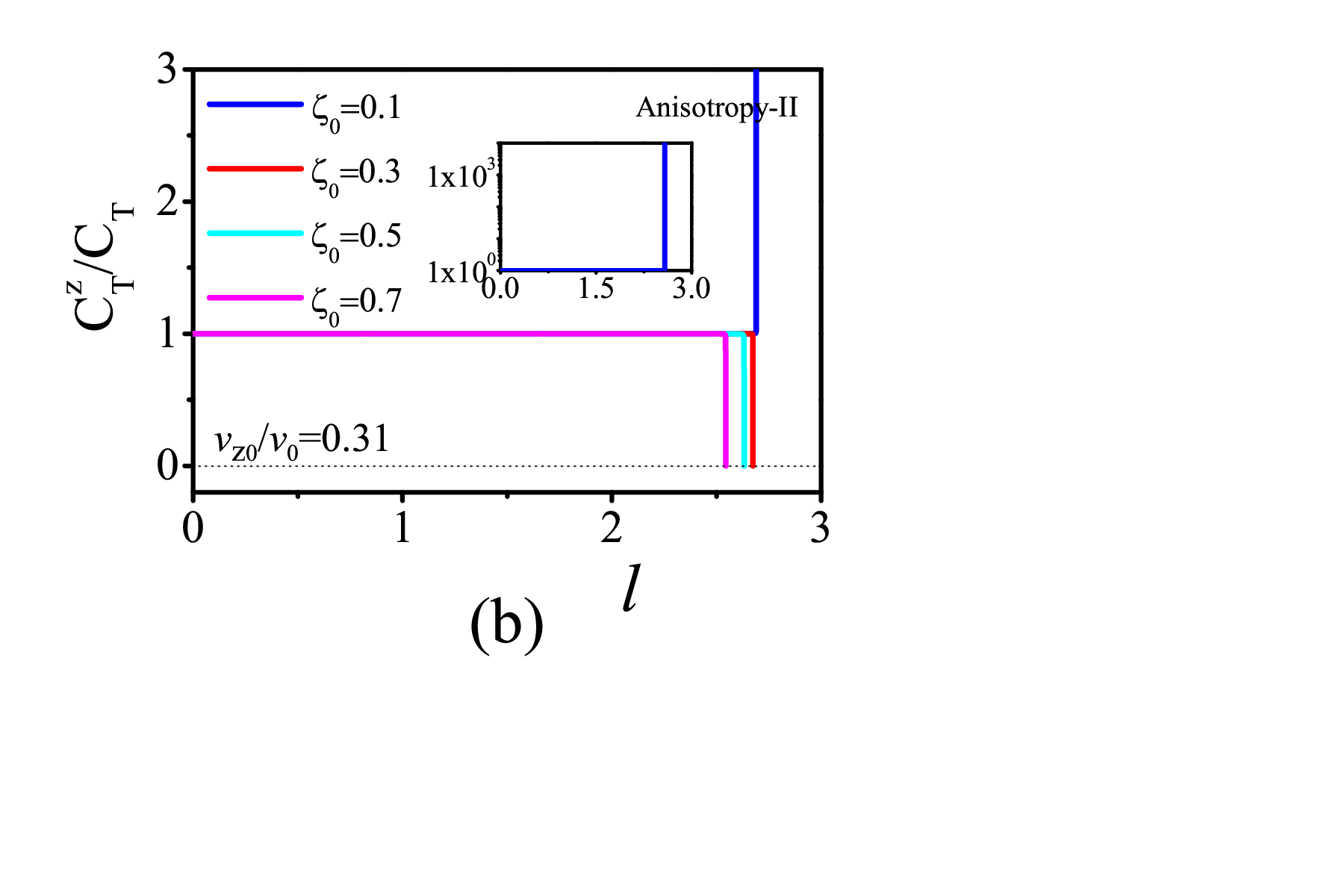}
\\
\vspace{-1.8cm}
\hspace{-2.0cm}
\includegraphics[width=3.7in,height=2.7in]{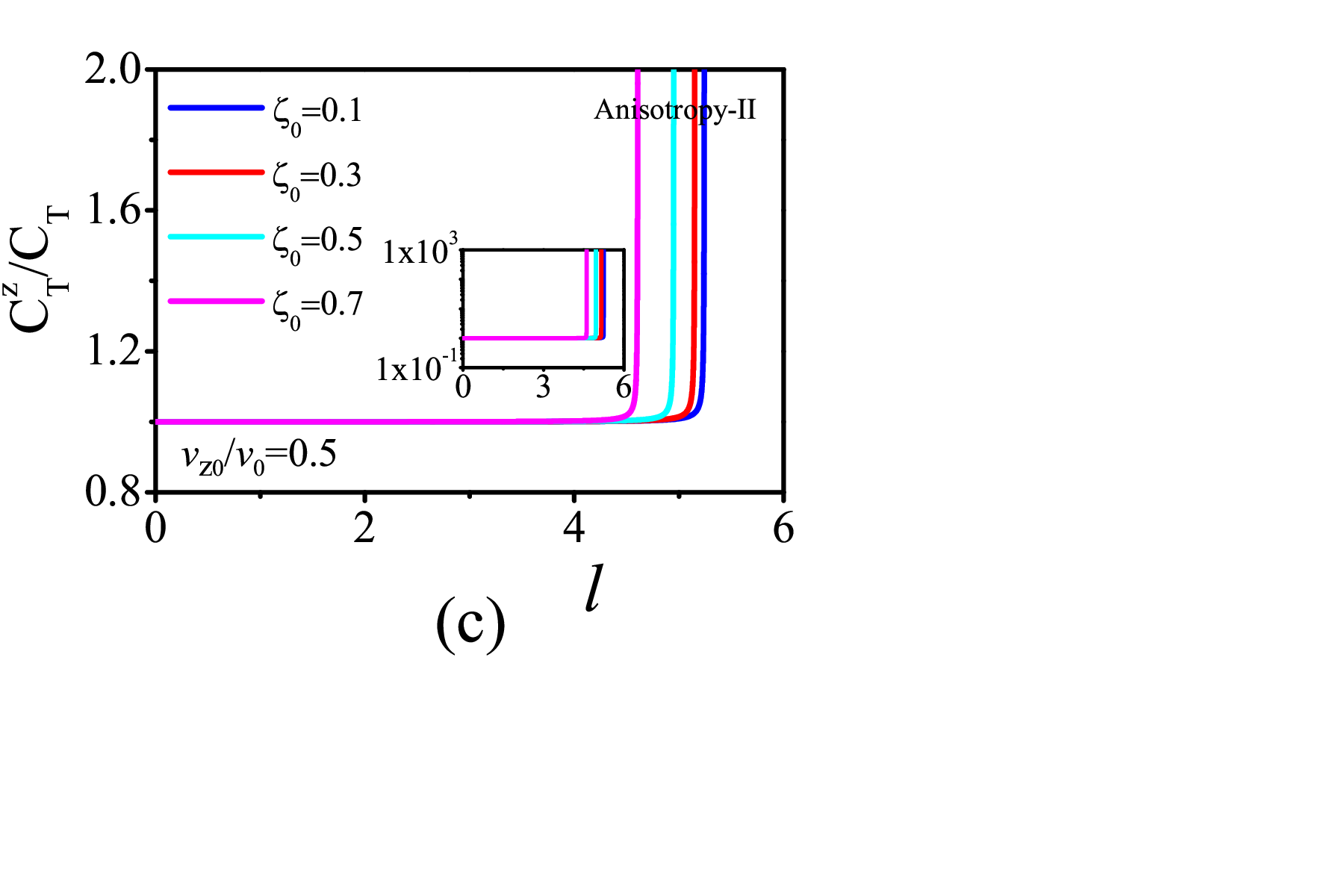}\hspace{-5.5cm}
\includegraphics[width=3.7in,height=2.7in]{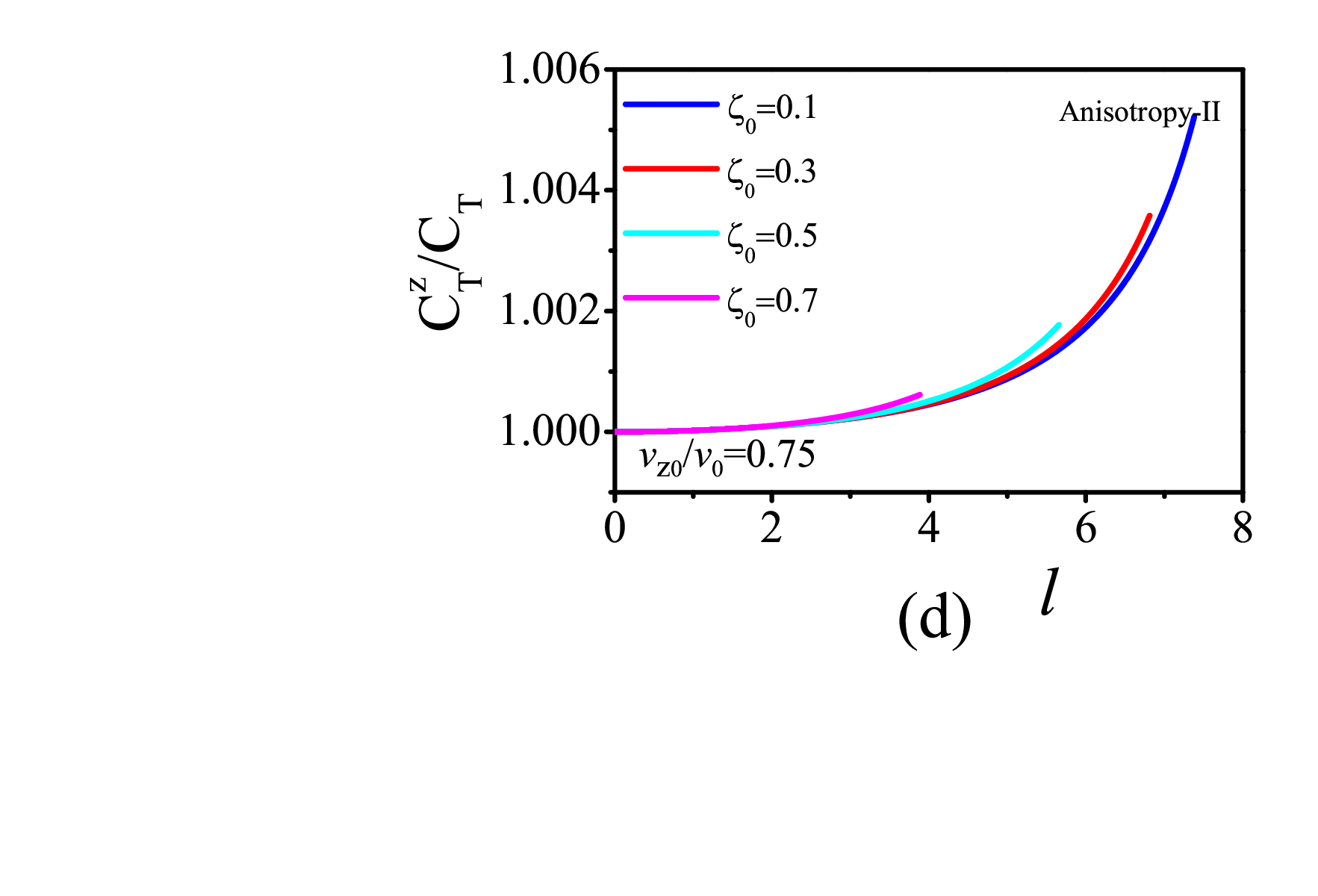}
\\
\vspace{-1.8cm}
\caption{(Color online) The energy-dependent evolutions of $C_{T}^{z}/C_{T}$ starting from Anisotropy-II
are depicted with respect to the variations of $\zeta_0$ for fixing the fermion velocities:
(a) $v_{z0}/v_0=0.15$, (b) $v_{z0}/v_0=0.31$, (c) $v_{z0}/v_0=0.5$, and
(d) $v_{z0}/v_0=0.75$.}\label{Fig_CTz-2}
\end{figure*}

\begin{figure}
\includegraphics[width=4.1in]{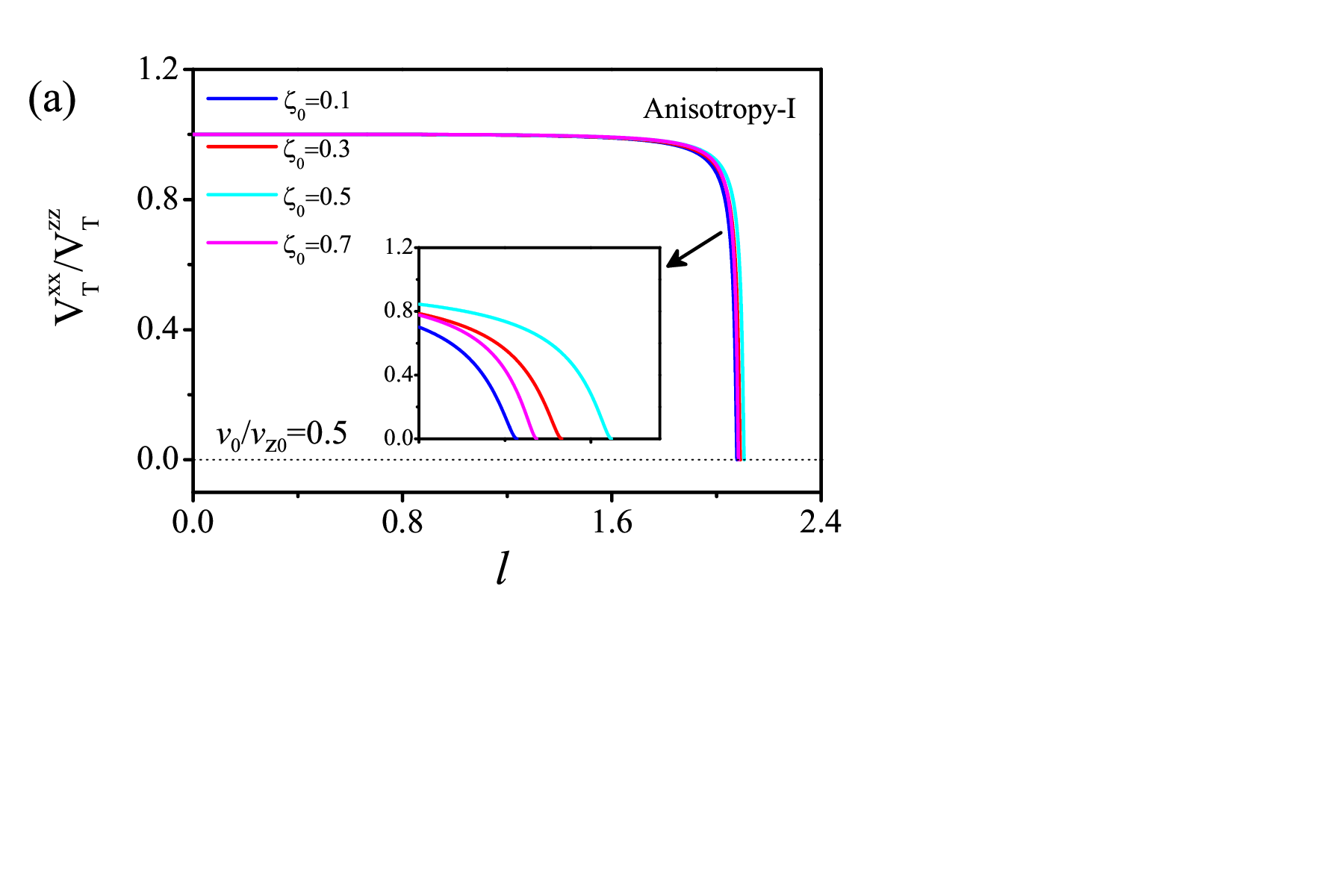}\vspace{-2.8cm}
\includegraphics[width=4.255in]{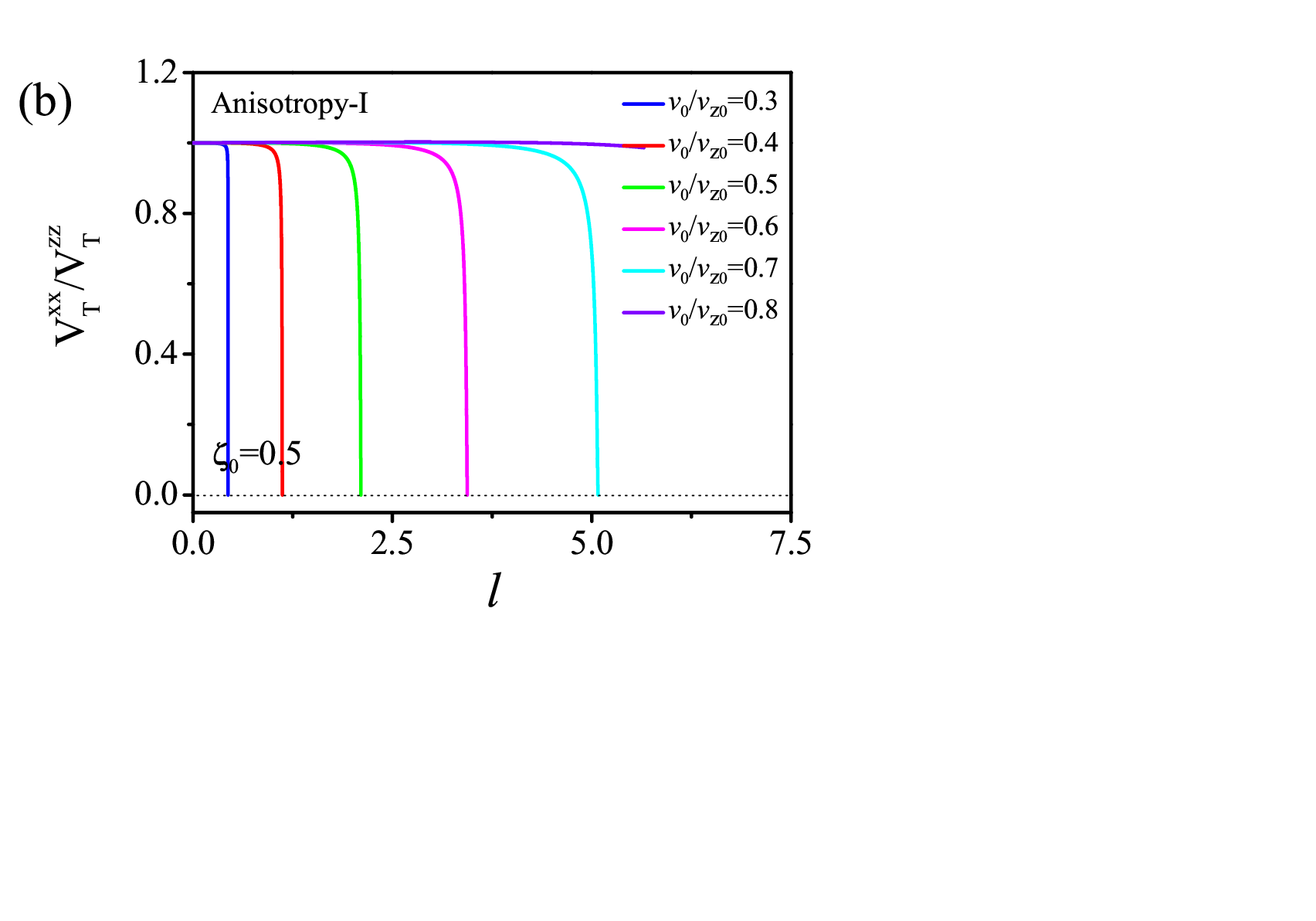}\\
\vspace{-2.8cm}
\caption{(Color online) The energy-dependent evolutions of $V_T^{xx}/V_T^{zz}$ starting from Anisotropy-I
are depicted with respect to the variations of (a) $\zeta_0$ and (b) $v_0/v_{z0}$.}\label{Fig_VTxxVTzz-1}
\end{figure}

\begin{figure*}
\hspace{0.6cm}
\includegraphics[width=3.7in,height=2.7in]{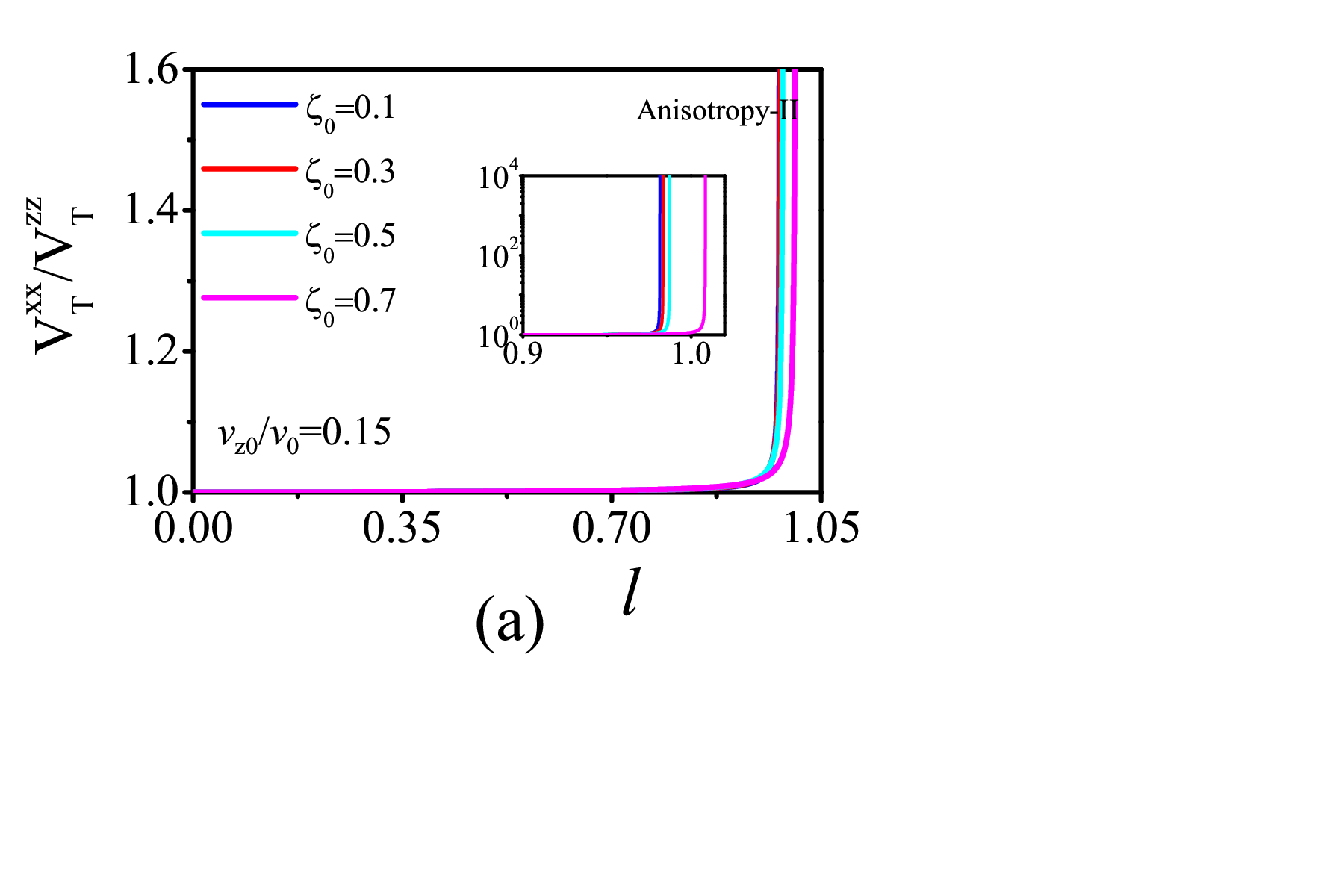}\hspace{-2.0cm}
\includegraphics[width=3.7in,height=2.7in]{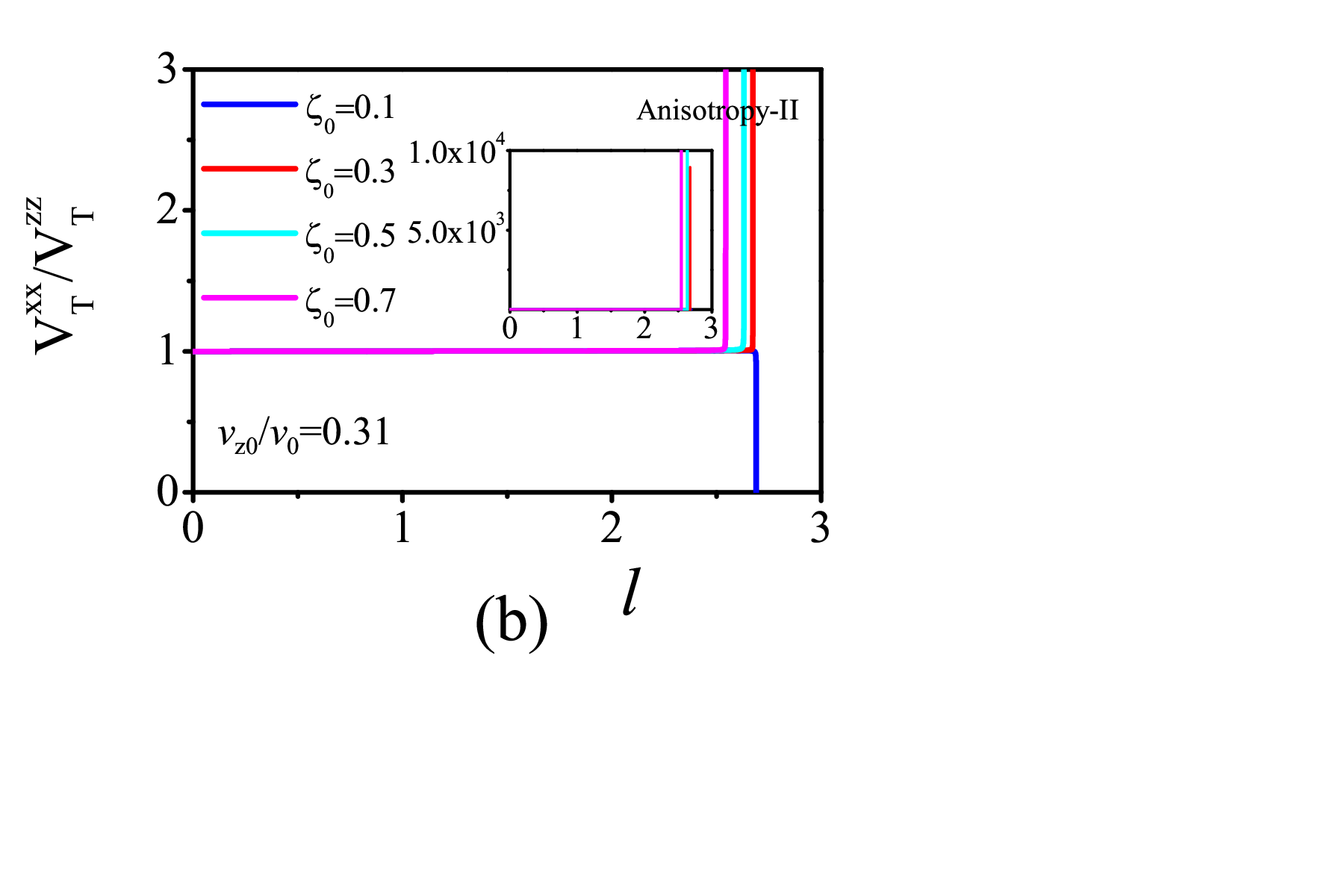}
\\
\vspace{-1.8cm}
\hspace{-2.5cm}
\includegraphics[width=3.7in,height=2.7in]{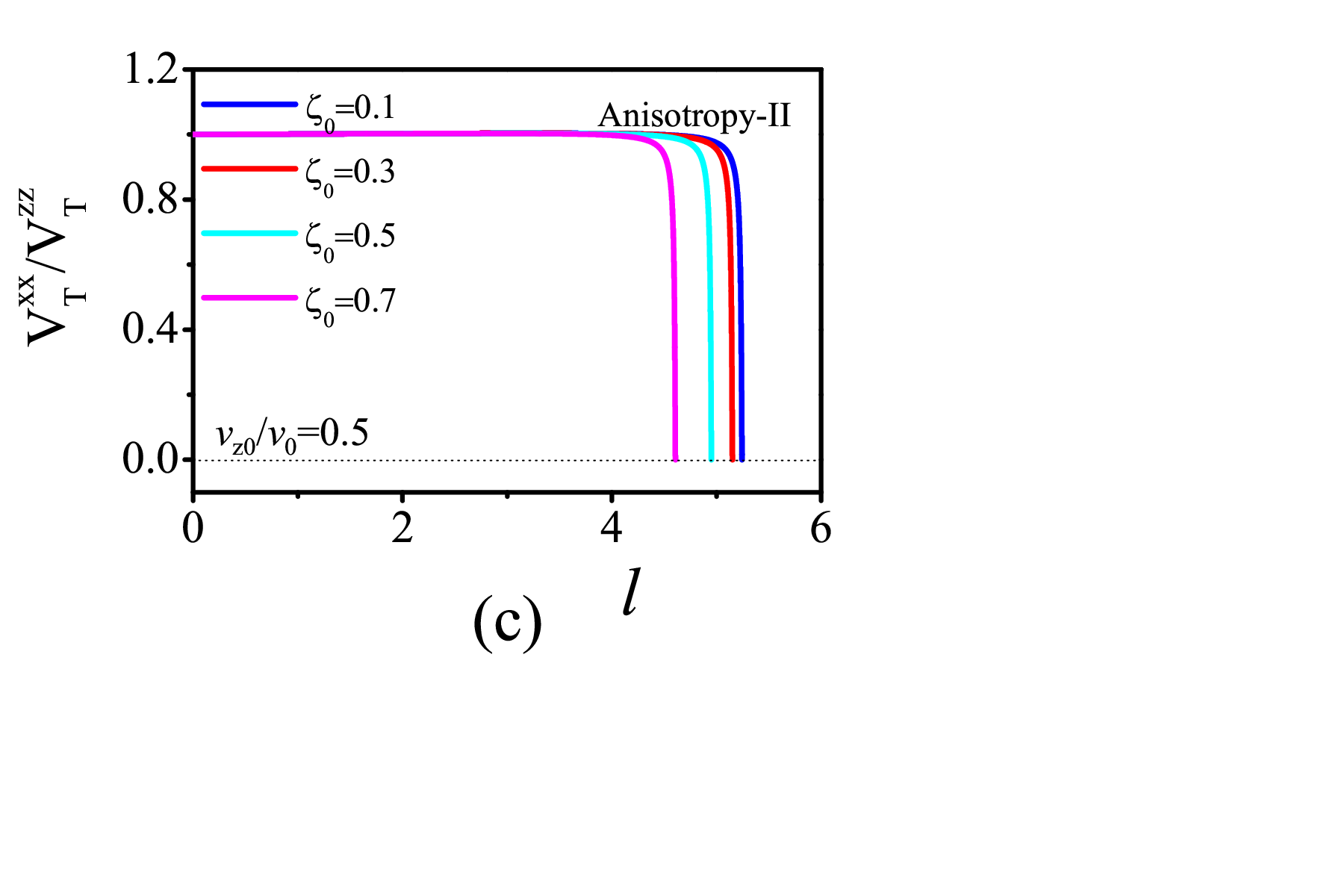}\hspace{-5.0cm}
\includegraphics[width=3.7in,height=2.7in]{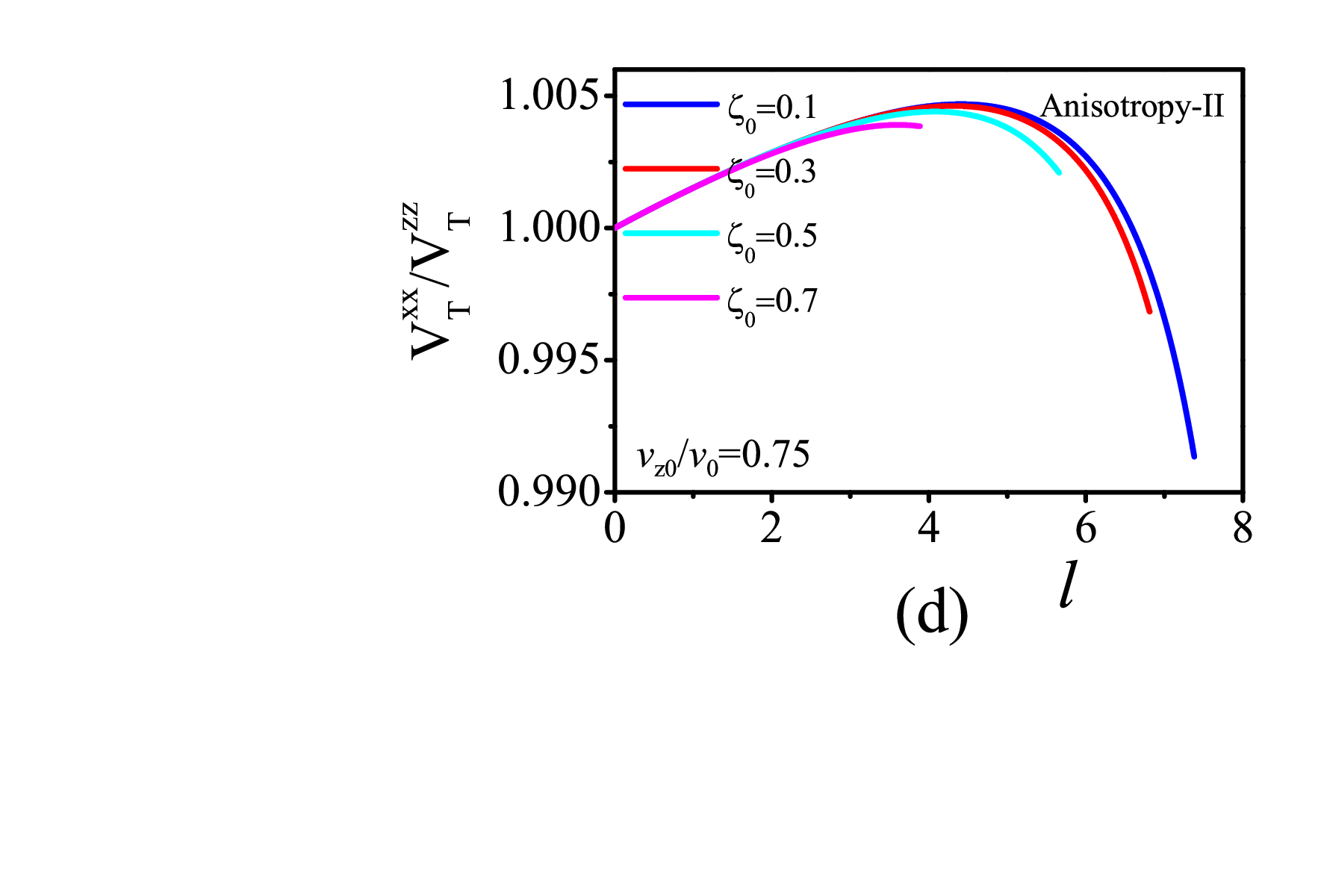}
\\
\vspace{-1.8cm}
\caption{(Color online) The energy-dependent evolutions of $V_T^{xx}/V_T^{zz}$ starting from Anisotropy-II
are depicted with respect to the variations of $\zeta_0$ for fixing the fermion velocities:
(a) $v_{z0}/v_0=0.15$, (b) $v_{z0}/v_0=0.31$, (c) $v_{z0}/v_0=0.5$,
and (d) $v_{z0}/v_0=0.75$.}\label{Fig_VTxxVTzz-2}
\end{figure*}

\subsection{Fates of phonon velocities and $\lambda_z/\lambda$}

Next, we examine the impact of coupled interactions on phonon velocities.
For the sake of simplicity, the focus is put on the velocities of transverse phonons as
their longitudinal counterparts show similar behavior owing to the analogous structures of
RG equations addressed in Sec.~\ref{Sec_RGEqs}.

With lowering the energy scale, Fig.~\ref{Fig_CTz-1} with fixed fermion velocities
presents the basic tendency of $C_{T}^{z}/C_{T}$ for the departure from Anisotropy-I.
One can notice from Fig.~\ref{Fig_CTz-1}(a) that $C_{T}^{z}/C_{T}$ gradually increases and goes towards the divergence as the system approaches a critical point. Notably, these qualitative results are relatively independent of the initial
value of the tilting parameter. Accordingly, we choose a specific tilting parameter $\zeta_0=0.5$ to
examine the impact of the variation of initial anisotropy of fermion velocities as
displayed in Fig.~\ref{Fig_CTz-1}(b). We notice that, as long as the initial ratio of fermion velocities
remains below a critical value around $0.8$, the ratio $C_{T}^{z}/C_{T}$ unambiguously tends towards divergence at a critical energy scale
that is increased by tuning down $v_0/v_{z0}$. Beyond this threshold, it is more favorable for $C_{T}^{z}/C_{T}$
to converge to a finite value, slightly deviating from the isotropic case. This signals that the $C_{T}^{z}/C_{T}$ prefer
to display an extreme anisotropy, with the $z$ component of the phonon velocity playing a dominant role,
in contrast with the $x$ or $y$ component.

Compared with the case starting from Anisotropy-I, the behavior of phonon velocities in Anisotropy-II,
illustrated in Fig.~\ref{Fig_CTz-2}, exhibits considerably more interesting behavior,
which heavily depend on both the initial values of fermion velocities and the tilting parameter.
From Fig.~\ref{Fig_CTz-2}, it is evident that the low-energy fate of $C_T^z/C_T$ is predominantly dictated
by the ferocious competition between $v_{z0}/v_0$ and $\zeta_0$, which is rooted in the coupled RG equations.

To be specific, when $v_{z0}/v_0$ is smaller than a
critical value ($\approx0.31$), we can find from Fig.~\ref{Fig_CTz-2}(a) that it plays a leading role in driving $C_T^z/C_T$ towards an extreme anisotropy ($C_T^z/C_T\rightarrow0$) at the lowest-energy limit. This behavior proves robust against variations
in the value of $\zeta_0$. In sharp contrast, Fig.~\ref{Fig_CTz-2}(c) shows that, at a moderate initial value of $v_{z0}/v_0$,
another form of extreme anisotropy ($C_T^z/C_T\rightarrow\infty$) would be induced upon reaching the critical energy scale.
Additionally, when $v_{z0}/v_0$ falls within $v_{z0}/v_0\in[0.30,0.31]$ as presented
in Fig.~\ref{Fig_CTz-2}(b), we notice that the influence of $v_{z0}/v_0$ is subordinate to
the tilting parameter. This indicates that $v_{z0}/v_0$ can either drive $C_T^z/C_T\rightarrow0$
or $C_T^z/C_T\rightarrow\infty$ by varying the value of $\zeta_0$. It is noteworthy that, for larger
values of $v_{z0}/v_0>0.75$, $C_T^z/C_T$ only experiences a slight deviation from isotropy due to the
concomitant effects of both $v_{z0}/v_0$ and $\zeta_0$.

To reiterate, the coupled RG equations, in tandem with
the intimate competition between initial values of the fermion velocities and the tilting parameter,
can drive the system towards an extreme anisotropy, where $C_T^z/C_T$ tends towards either zero or infinity,
or towards near-isotropy, where $C_T^z/C_T$ approximately converges to one. In consequence, this signals that, as the
energy scale decreases, phonons can exhibit distinct behavior, such as playing a dominant role
in the $z$ direction or $xy$ plane as well as being nearly equivalent in all directions. In addition, we find that
the qualitative behavior of $\lambda_z/\lambda$ bears the similarity to that of $C_T^z/C_T$ and
hence not shown for brevity.

\begin{figure}
\centering
\includegraphics[width=4.1in]{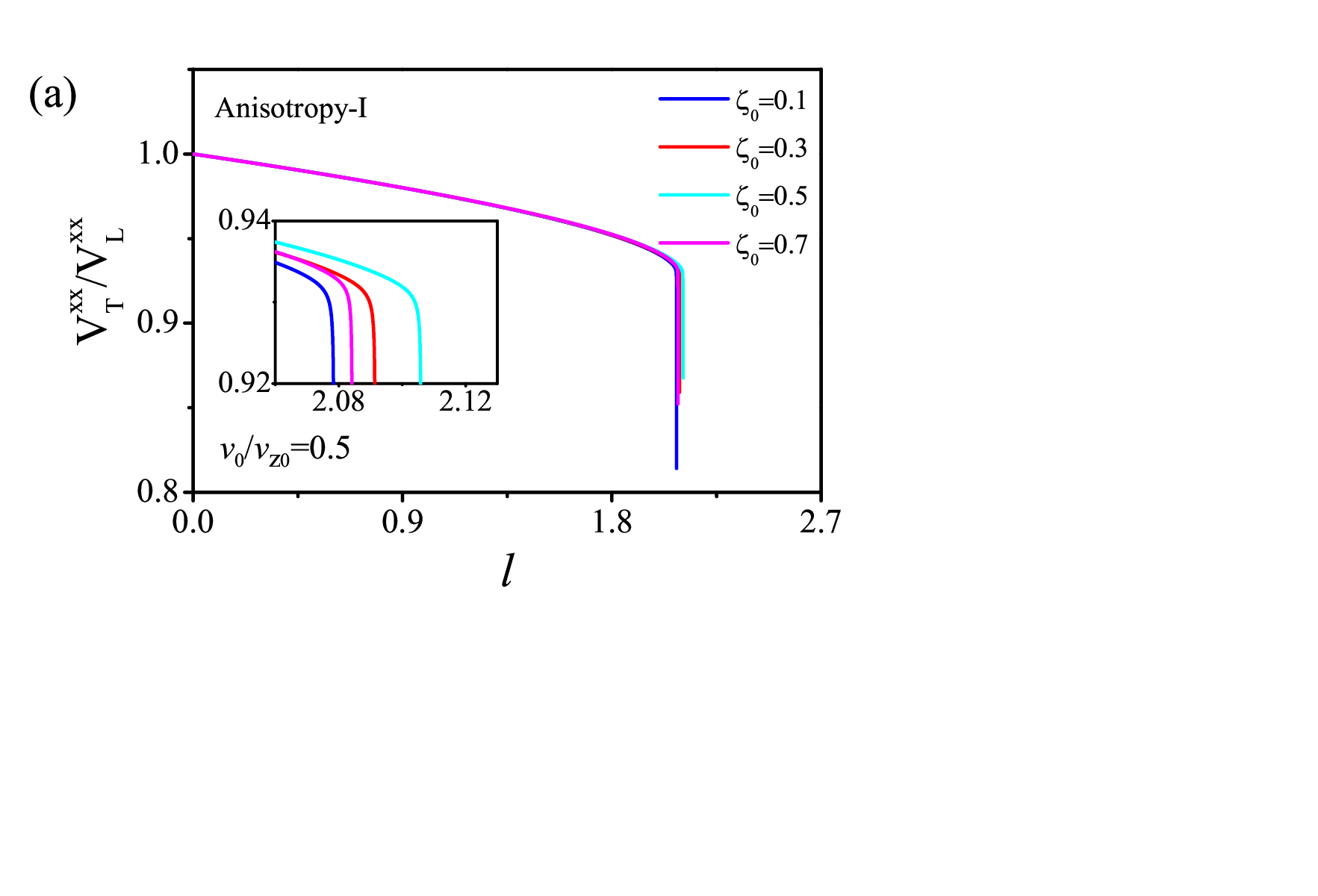}\vspace{-2.7cm}
\includegraphics[width=4.1in]{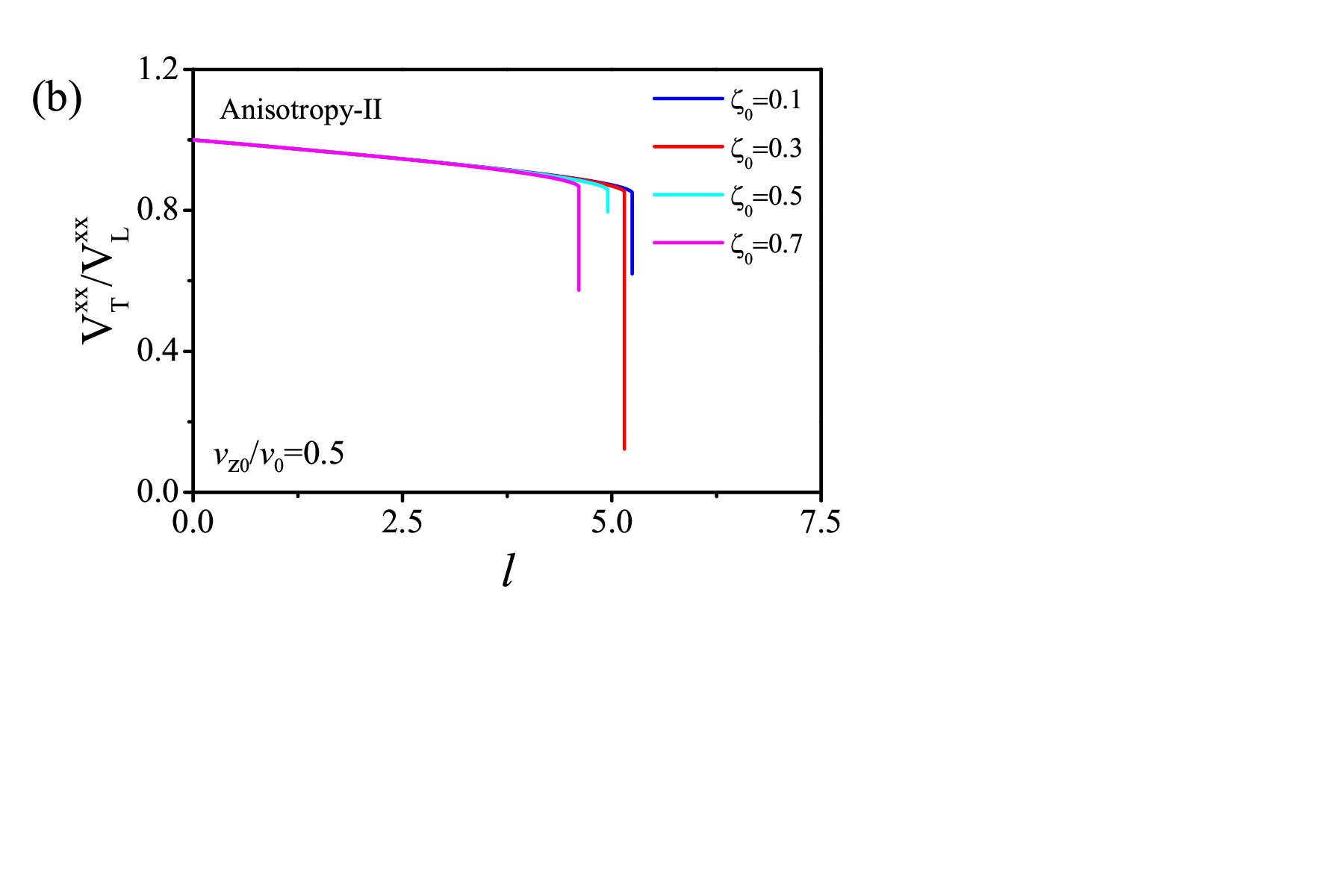}\\
\vspace{-2.7cm}
\caption{(Color online) The energy-dependent evolutions of $V_T^{xx}/V_L^{xx}$ are depicted with respect to the variations of $\zeta_{0}$,
starting from (a) Anisotropy-I and (b) Anisotropy-II situations.}\label{Fig_VTxxVLxx-1}
\end{figure}

\subsection{Fates of phonon-phonon interactions}

At last, we move our attention to the low-energy tendencies of phonon-phonon interactions under the
influence of all interactions in our theory.

To gain a comprehensive understanding of phonon-phonon interactions in this system,
we principally need to analyze the behavior of various quantities derived from
all components of the phonon-phonon couplings~(\ref{Eq_S_action-u}). These include
$V_T^{xy}/V_T^{zz}$, $V_T^{xz}/V_T^{zz}$, $V_L^{xx}/V_L^{zz}$, $V_L^{xy}/V_L^{zz}$,
$V_L^{xz}/V_L^{zz}$, $V_T^{xx}/V_L^{xx}$, $V_T^{xy}/V_L^{xy}$, $V_T^{xz}/V_L^{xz}$, and $V_T^{zz}/V_L^{zz}$,
where the subscripts $T$ and $L$ denote the transverse and longitudinal phonons, respectively.
To simplify our analysis, we can categorize them into three groups: the transverse-phonon components ($V_T^{xx}/V_T^{zz}$,
$V_T^{xy}/V_T^{zz}$, $V_T^{xz}/V_T^{zz}$), the longitudinal-phonon components ($V_L^{xx}/V_L^{zz}$,
$V_L^{xy}/V_L^{zz}$, $V_L^{xz}/V_L^{zz}$ ), and the mixed components ($V_T^{xy}/V_L^{xy}$,
$V_T^{xz}/V_L^{xz}$, $V_T^{zz}/V_L^{zz}$). Fortunately, the numerical analysis shows that
the members within each category exhibit analogous tendencies
as approaching the critical energy scales. Accordingly, we can effectively represent each category
with a single representative behavior

Concerning the transverse-phonon components, we choose to consider the behavior of $V_T^{xx}/V_T^{zz}$.
Starting from Anisotropy-I of fermion velocities, Fig.~\ref{Fig_VTxxVTzz-1}(a) manifestly
shows that $V_T^{xx}/V_T^{zz}$ slowly decreases as $l$ increases, but then rapidly
approaches zero near the critical energy. This indicates that the $z$-component
phonon interaction becomes dominant over that of the $x,y$ components. In comparison
with the initial value of the tilting parameter $\zeta_0$, which contributes only minor corrections to
the basic tendencies, the critical energy scales can be considerably lowered by increasing the value of $v_0/v_{z0}$.
In this context, $V_T^{xx}/V_T^{zz}$ cannot be driven to the extreme anisotropy $V_T^{xx}/V_T^{zz}\rightarrow0$. Instead,
it eventually converges to a slight anisotropy, as shown in Fig.~\ref{Fig_VTxxVTzz-1}(b) at $v_0/v_{z0}=0.8$.
This suggests that the anisotropy of the fermion velocities plays
a more pivotal crucial role in pinning down the fates of
phonon interactions than the tilting parameter.

In comparison, beginning with the Anisotropy-II of fermion velocities,
one can find from Fig.~\ref{Fig_VTxxVTzz-2} that $V_T^{xx}/V_T^{zz}$ bears similarities to
the behavior of phonon velocities discussed in the previous section. If $v_{z0}/v_0$ is
below a critical value ($\approx0.31$) or takes a moderate initial value, it wins against
$\zeta_0$, leading the system to an extreme anisotropy ($V_T^{xx}/V_T^{zz}\rightarrow\infty$ in the former case
and $V_T^{xx}/V_T^{zz}\rightarrow0$ in the latter case, respectively). In addition, an adequate big value of $v_{z0}/v_0$ renders
$V_T^{xx}/V_T^{zz}$ nearly isotropic as shown in Fig.~\ref{Fig_VTxxVTzz-2}(d).
Otherwise, Fig.~\ref{Fig_VTxxVTzz-2}(b)illustrates that the influence of $v_{z0}/v_0$ is subordinate to $\zeta_0$
around the critical value $v_{z0}/v_0\approx0.31$. This drives $V_T^{xx}/V_T^{zz}$ to either
an extreme anisotropy $V_T^{xx}/V_T^{zz}\rightarrow\infty$ or another extreme anisotropy $V_T^{xx}/V_T^{zz}\rightarrow0$.

Since the longitudinal-phonon parts exhibit fates similar to that of their transverse-phonon counterparts, they are not presented here for the sake of brevity. Subsequently, our focus shifts to the mixed parts. As they share similar basic tendencies,
we choose to present the behavior of $V_T^{xx}/V_L^{xx}$, as depicted in Fig.~\ref{Fig_VTxxVLxx-1}.
%and Fig.~\ref{Fig_VTxxVLxx-2}.

From Fig.~\ref{Fig_VTxxVLxx-1}(a), with a fixed value of $v_0/v_{z0}=0.5$,
we observe that, in the scenario initiated from Anisotropy-I, $V_T^{xx}/V_L^{xx}$ displays a slight deviation
from the isotropy as $l$ increases and reaches a finite value at the critical energy scale. This implies
that transverse-phonon and longitudinal-phonon interactions nearly contribute
equally. These basic results are similar upon varying the initial conditions.
In contrast, when beginning with the Anisotropy-II,
we observe in Fig.~\ref{Fig_VTxxVLxx-1}(b) for $v_{z0}/v_0=0.5$ that $V_T^{xx}/V_L^{xx}$
is heavily dependent on the initial conditions. It decreases more significantly than its Anisotropy-I counterpart,
experiencing a substantial drop at some optional $\zeta_0\approx0.3$ or
$v_{z0}/v_0\approx0.6$, where it is driven to a state of strong anisotropy. Consequently, once
we start from Anisotropy-II, the longitudinal-phonon interactions play a more
significant role than transverse-phonon interactions. Additionally, we have verified that
these results are relatively stable against the change of the initial ratio of fermion velocities.

To wrap up, we figured out that the initial anisotropy of the fermion velocities plays a
more crucial role in determining the fates of phonon interactions. The transverse-phonon contributions
are subordinate to their longitudinal-phonon counterparts for the Anisotropy-II case, and
as for the Anisotropy-I situation, both transverse- and longitudinal-phonon contributions are nearly equivalent.

\section{Potential instabilities}\label{Sec_FP-instab}

As presented in the preceding section, the interaction
parameters exhibit a number of interesting behavior dictated by
the RG equations~(\ref{Eq_RGEq-v})-(\ref{Eq_RGEq-V_T-L}), which span from
$l=0$ to $l\rightarrow l_c$ as illustrated in Figs.~\ref{fig4_lc-PT} and
\ref{Fig_phase}(a). At $l>l_c$, the effective theory is invalid and nonphysical behavior may appear.
Thus, we need to stop the RG flows at $l=l_c$. The values of all these parameters at $l_c$ construct the fixed point (FP)
in the phase space~\cite{Shankar1994RMP}. Hereby, let us present several clarifications on the FP.
On one hand, in the parameter space, all the couplings are
vividly attracted by the points with $l=l_c$ as lowering the energy scales~\cite{Shankar1994RMP}.
On the other hand, the FP is a relative concept to describe the destinations of couplings in the parameter space.
Its coordinates are equivalent to $\{x_i(l_c)/ \mathrm{max}\{x_i\}\}$ with the
variable $x_i$ standing for all the interaction parameters~\cite{Murray2014PRB,Roy2018PRX,Cvetkovic2012PRB,Maiti2010PRB,Efremov2017PRB},
which yields fixed-point parameters consisting of several small values.

The FP generally marks a critical juncture
where the potential instability from the 3D tDSM to an $X$ phase may be induced.
After examining the behavior of parameters shown in Sec.~\ref{Sec_SM-tend-fates},
we find that the fates of FPs are heavily dependent upon the tendencies of
$\zeta/\zeta_{0}$, $ v_{z}/v $, $\varepsilon_{z}/\varepsilon $, $ C_{T}^{Z}/C_{T} $,
and $V_{T}^{XX}/V_{T}^{ZZ}$, which qualitatively clusters into three distinct types of FPs
as presented in Table~\ref{Table_classify_FPs}. To simplify our analysis, we subsequently focus
exclusively on these three types of FPs and aim to identify the most favorable instabilities associated with them.

\begin{table}
\caption{Collections of three distinct types of FPs are established based on the tendencies exhibited by interaction parameters.}
\vspace{0.3cm}
\centering{
\renewcommand\arraystretch{2}
\begin{tabular}{c|c|c|c|c|c}
\hline
\hline
 Types & $\zeta/\zeta_{0}$ & $ v_{z}/v $ & $\varepsilon_{z}/\varepsilon $ & $ C_{T}^{Z}/C_{T} $ & $V_{T}^{XX}/V_{T}^{ZZ}$ \\
\hline
 Type-I & $\red{\uparrow}$ & $\blue{\downarrow}$ & $\blue{\downarrow}$ & $\red{\uparrow}$ & $\blue{\downarrow}$ \\
\hline
 Type-II & $\red{\uparrow}$ & $\blue{\downarrow}$ & $\blue{\downarrow}$ & $\blue{\downarrow}$ & $\red{\uparrow}$ \\
\hline
 Type-III & $\blue{\downarrow}$ & $\red{\uparrow}$ & $\red{\uparrow}$ & $\red{\uparrow}$ & $\blue{\downarrow}$ \\
\hline
\hline
\end{tabular}}\label{Table_classify_FPs}
\end{table}

\begin{figure*}
\centering
\includegraphics[width=2.5in]{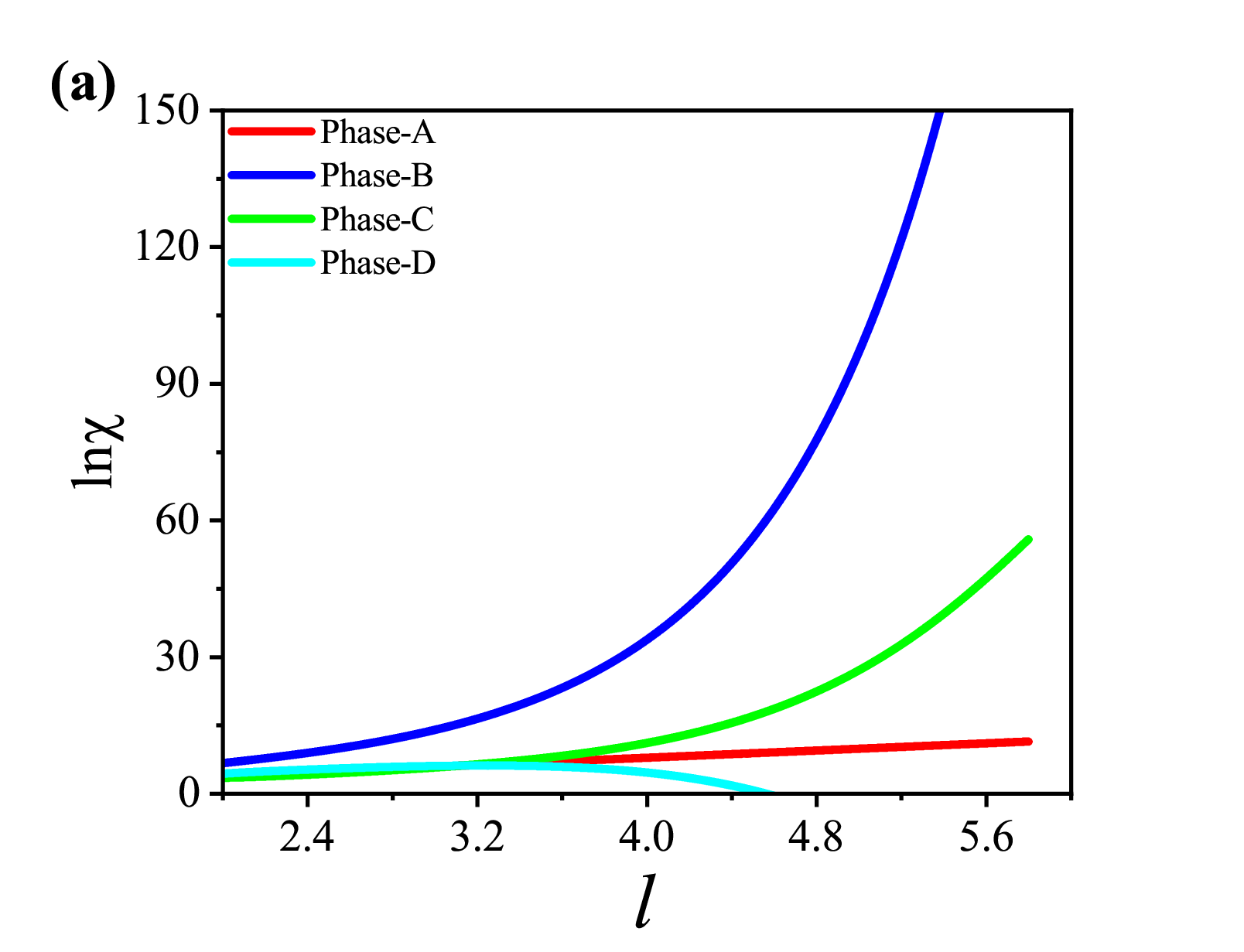}\hspace{-0.85cm}
\includegraphics[width=2.5in]{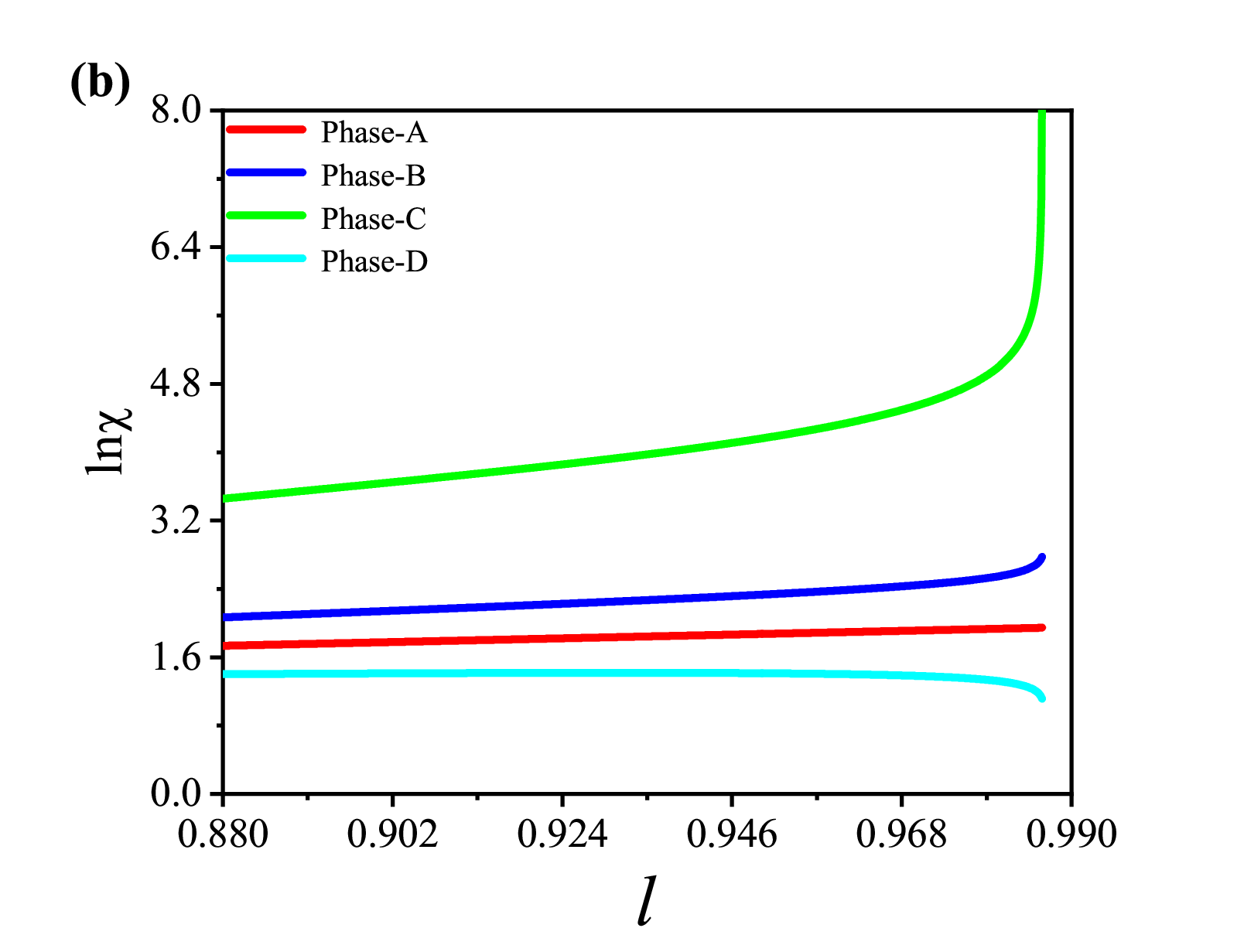}\hspace{-0.85cm}
\includegraphics[width=2.5in]{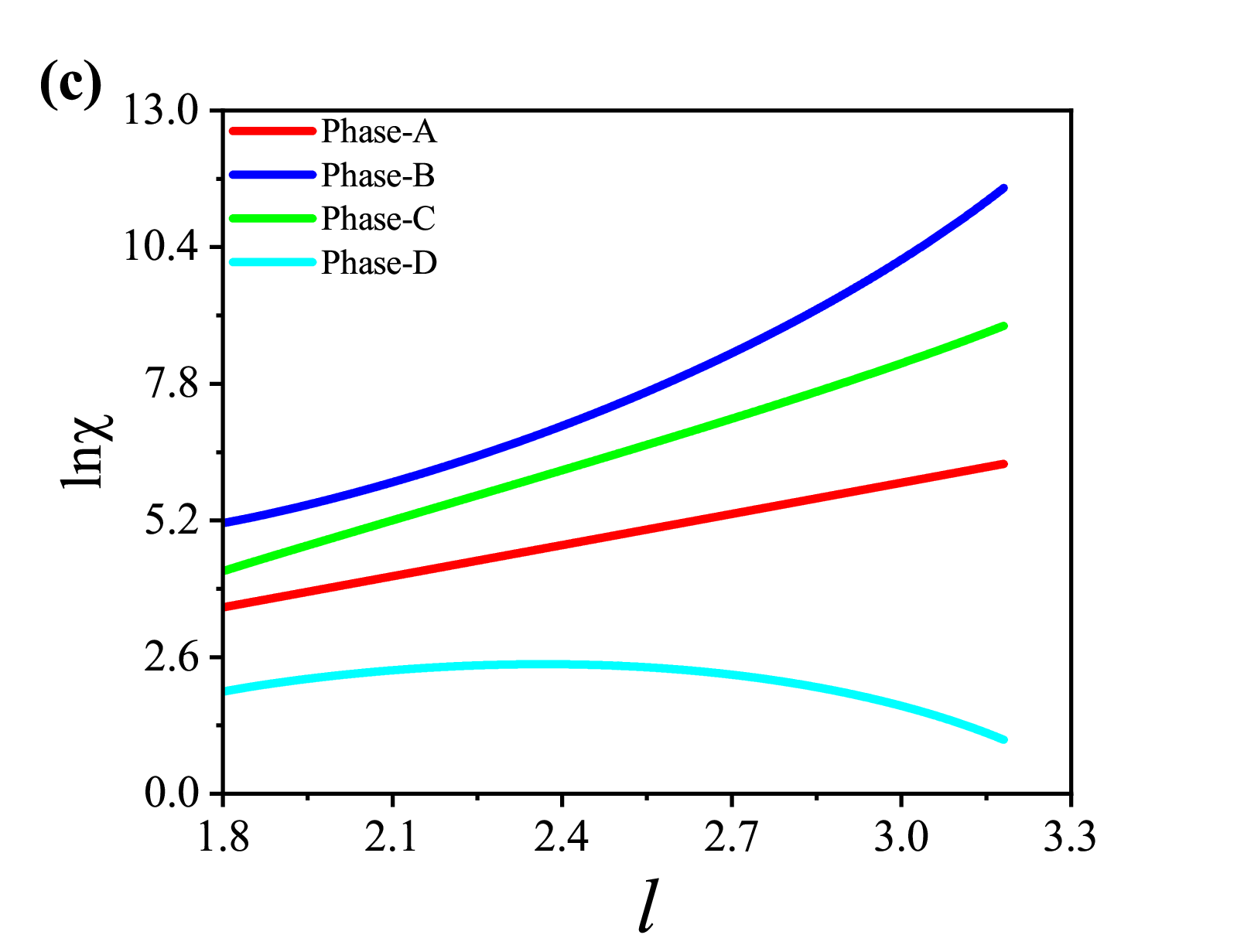}
\vspace{-0.35cm}
\caption{(Color online) Energy-dependent susceptibilities of four different phases as approaching distinct
types of FPs with $\zeta=0.5$: (a) type-I ($v_z(0)/v(0)=0.72$), (b) type-II ($v_z(0)/v(0)=0.15$),
and (c) type-III ($v(0)/v_z(0)=0.60$) FPs.}
\label{Fig_four-chi_three-types}
\end{figure*}

Generally, the instability serves as an indicator of certain symmetry breaking and is accompanied by the development of
specific fermionic bilinears~\cite{Maiti2010PRB,Halboth2000PRL,Halboth2000RPB,Nandkishore2012NP,
Cvetkovic2012PRB,Wang2020NPB}. To investigate the potential instability around the FPs,
we introduce the following source terms to denote the potential phase $X$~\cite{Vafek2010PRB,Murray2014PRB,Roy2009.05055}
\begin{eqnarray}
S_{\mathrm{sou}}
&=&\int d\tau\int d^{3}\mathbf{x}
\Big\{\sum_{i=1}\Delta_{i}^{\mathrm{PH}}\psi^{\dag}\mathcal{M}_{i}^{\mathrm{PH}}\psi\nonumber\\
&&+\sum_{i=1}\left[\Delta_{i}^{\mathrm{PP}}\psi^\dagger\mathcal{M}_{i}^{\mathrm{PP}}\psi^{\ast}+\mathrm{h.c.}\right]\Big\},\label{Eq_source-terms}
\end{eqnarray}
where $\mathcal{M}^{\mathrm{PH}}_{i}$ and $\mathcal{M}^{\mathrm{PP}}_{i}$ denote the related matrices associated with
the fermionic bilinears in the particle-hole and particle-particle channels, and
$\Delta^{\mathrm{PH}}_{i}$ and $\Delta^{\mathrm{PP}}_{i}$ specify the strengths of the corresponding source terms, respectively.
In our model, the primary candidates of instability and phase transition are outlined in
Table~\ref{Table_candidate-phases}~\cite{Roy2018PRX,Ruhman2019PRX}.
To proceed, we incorporate the source terms~(\ref{Eq_source-terms}) into the effective
action~(\ref{Eq_eff-action}) and calculate the
one-loop corrections to $\Delta^{\mathrm{PH}}_{i}$ and $\Delta^{\mathrm{PP}}_{i}$. This yields the related RG
equations for $\Delta^{\mathrm{PH}}_{i}$ and $\Delta^{\mathrm{PP}}_{i}$, which can be compactly expressed as
\begin{eqnarray}
\frac{d\Delta^{\mathrm{PH}}_{i}}{dl}&=&\mathcal{P}(\Delta^{\mathrm{PH/PP}}_{i},v,v_{z}......),\\
\frac{d\Delta^{\mathrm{PP}}_{i}}{dl}&=&\mathcal{P}(\Delta^{\mathrm{PH/PP}}_{i},v,v_{z}......),
\end{eqnarray}
where the detailed expressions for $\mathcal{P}(\Delta^{\mathrm{PH}}_{i},v,v_{z}......)$
and $\mathcal{P}(\Delta^{\mathrm{PP}}_{i},v,v_{z}......)$
are provided in Appendix~\ref{appendix-source-terms}.
Subsequently, we are now in a suitable position to evaluate the susceptibilities accompanied by the
source terms resorting to the relationship~\cite{Vafek2010PRB,Cvetkovic2012PRB,Murray2014PRB,Zhai2021NPB}
\begin{eqnarray}
\delta\chi=\frac{\partial^2 f}{\partial\Delta(0)\partial\Delta^*(0)},
\end{eqnarray}
where $f$ denotes the free-energy density. With all these in hand, we can systematically study
the susceptibilities of all candidates phases to identify the
leading phase for phase $X$, which exhibits the strongest divergence of susceptibility~\cite{Cvetkovic2012PRB}.

\begin{table}
\caption{Potential candidates for instabilities triggered by all interactions~\cite{Roy2018PRX,Ruhman2019PRX}.
SC, AP and CDW are abbreviations for superconductivity, anisotropy parameter and charge density wave, respectively.
In addition, chiral $\textrm{SC}_1$ and $\textrm{SC}_2$ specify two distinct sorts of chiral superconducting states.
The $\sigma_{0,i}$ act on the lattice space, consistent with their roles in the
Hamiltonian~(\ref{Eq H_0}) and effective theory~(\ref{Eq_eff-action}),
while $\tau_{0,i}$ with $i=x,y,z$ operate on the spin space.}
\vspace{0.3cm}
\centering{
\renewcommand\arraystretch{2}
\begin{tabular}{c|c|c}
\hline
\hline
 Order parameters & Fermionic bilinears & Potential phases\\
\hline
 $\Delta_{0}^{\mathrm{PH}}$ & $\mathcal{M}_{0}^{\mathrm{PH}}=\tau_{0}\sigma_{0}$ & density\\
\hline
 $\Delta_{1}^{\mathrm{PH}}$ & $\mathcal{M}_{1}^{\mathrm{PH}}=\tau_{0}\sigma_{x}$ & $x$-current\\
\hline
 $\Delta_{2}^{\mathrm{PH}}$ & $\mathcal{M}_{2}^{\mathrm{PH}}=\tau_{0}\sigma_{y}$ & AP\\
\hline
 $\Delta_{3}^{\mathrm{PH}}$ & $\mathcal{M}_{3}^{\mathrm{PH}}=\tau_{0}\sigma_{z}$ & CDW\\
\hline
 $\Delta_{0}^{\mathrm{PP}}$ & $\mathcal{M}_{0}^{\mathrm{PP}}=\tau_{y}\sigma_{z}$ & $s$-wave SC\\
\hline
 $\Delta_{1}^{\mathrm{PP}}$ & $\mathcal{M}_{1}^{\mathrm{PP}}=\tau_{y}\sigma_{x}$ & chiral\ $\textrm{SC}_{1}$\\
\hline
 $\Delta_{2}^{\mathrm{PP}}$ & $\mathcal{M}_{2}^{\mathrm{PP}}=\tau_{y}\sigma_{0}$ & chiral\ $\textrm{SC}_{2}$\\
\hline
 $\Delta_{0(0,1,3)}^{\mathrm{PP}}$ & $\mathcal{M}_{3i}^{\mathrm{PP}}=\tau_{(0,x,z)}\sigma_{y}$ & triplet SC\\
\hline
\hline
\end{tabular}\label{Table_candidate-phases}
}
\end{table}

To proceed, we notice from Eqs.~(\ref{Eq_Delta_1})-(\ref{Eq_Delta_2}), which characterizes the strengths of source terms,
that certain candidates in Table~\ref{Table_candidate-phases} are degenerate at the one-loop level. As a consequence,
it is convenient to categorize them into four cases: Phase-A (density, $x$-current, AP or CDW), Phase-B
($s$-wave SC or triplet $\mathrm{SC}_0$), Phase-C (chiral $\textrm{SC}_{1}$ or triplet $\mathrm{SC}_1$)
and Phase-D (chiral $\textrm{SC}_{2}$ or triplet $\mathrm{SC}_3$), respectively.
Following a numerical analysis that combines the RG equations for both interaction parameters and $\Delta^{\mathrm{PH/PP}}_{i}$,
we obtain the energy-dependent susceptibilities for all four cases, as shown in
Fig.~\ref{Fig_four-chi_three-types}, corresponding to the three types of FPs mentioned
in Table~\ref{Table_classify_FPs}. Learning from Fig.~\ref{Fig_four-chi_three-types} for several
representative initial conditions, the basic results are obtained as we approach all three distinct kinds of FPs.
At first, it is evident that Phase-C emerges as the leading instability once the system is driven to the type-I or type-II FP.
This indicates that chiral $\textrm{SC}_{1}$ or triplet $\mathrm{SC}_1$ becomes the dominant phase, and henceforth is the
preferred candidate for the phase $X$ displayed in Fig.~\ref{Fig_phase}(a)-(b).
In comparison, although Phase-C is not the predominant choice among other phases around the type-III FP,
it remains the optimal state for phase $X$ as depicted in Fig.~\ref{Fig_phase}(c). These results are
in qualitative agreement with the results in Ref.~\cite{Ruhman2019PRX}.

In this context, our analysis reveals the existence of an interaction-driven phase transition.
The onset of instability suggests the potential emergence of critical physical implications, which we elucidate
in Sec.~\ref{Sec_critical_implications}.

\begin{figure}[htpb]
%\subfigure[]{\includegraphics[width=3.7in]{fig_a-omega-05-DOS-abs.eps}}\hspace{-1.7cm}
%\subfigure[]{\includegraphics[width=3.7in]{fig_b-omega-05-DOS-abs.eps}}\\
\includegraphics[width=3.5in]{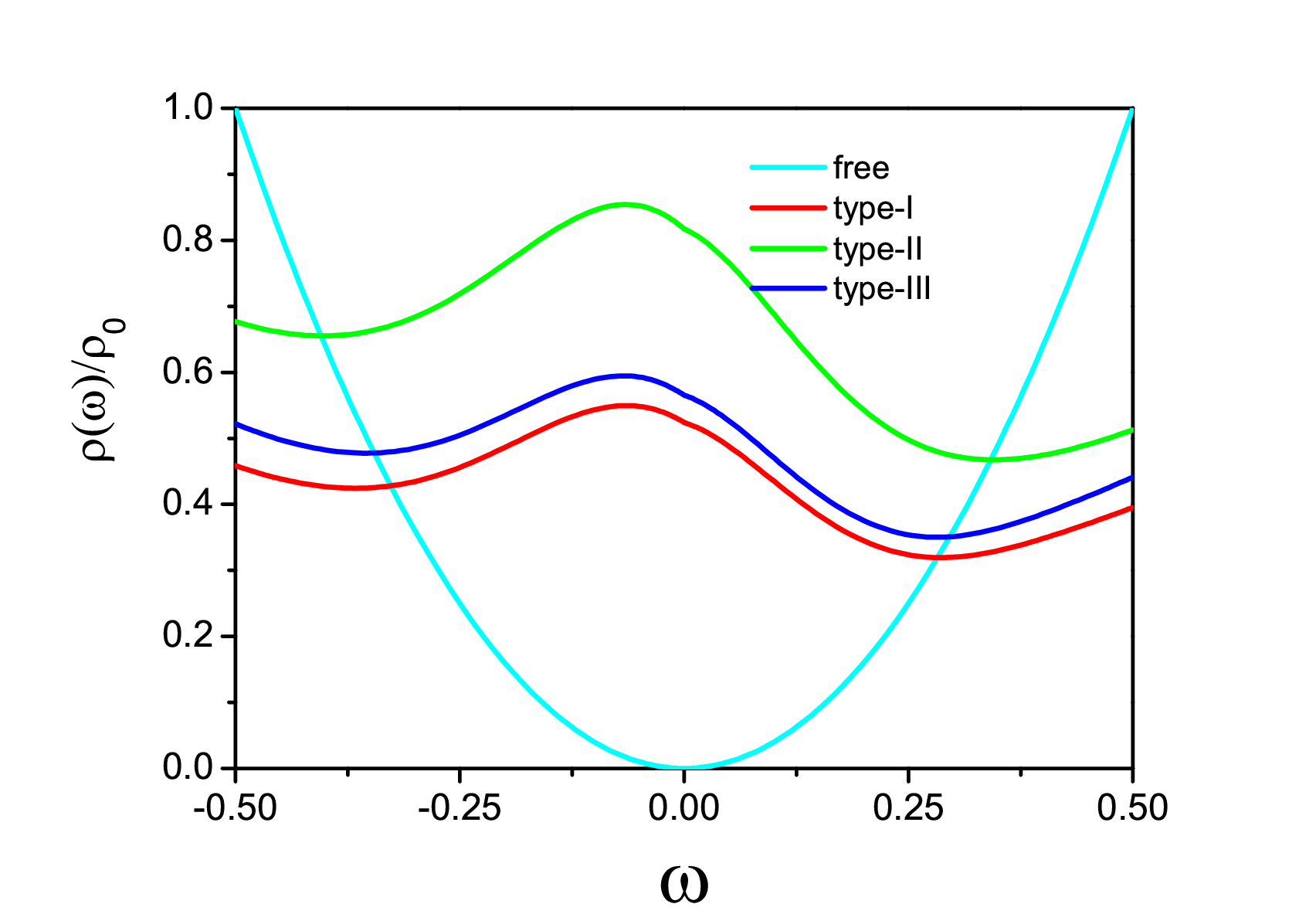}\\
\caption{(Color online) Frequency-dependent DOS $\rho(\omega)$ with $\zeta=0.5$ is depicted
around three distinct types of FPs: type-I ($v_z(0)/v(0)=0.75$), type-II ($v_z(0)/v(0)=0.15$),
and type-III ($v(0)/v_z(0)=0.40$). Hereby, $\rho_0$ is
designated by $\rho_0\equiv\rho_{\mathrm{free}}(\zeta=0,\omega=-0.5)$.}\label{Fig_DOS-all-omega-0.5}
\end{figure}

\section{Critical implications around the instabilities}\label{Sec_critical_implications}

To proceed, within this section, we examine the critical behavior of physical observables, including the
density of states (DOS), compressibility, and specific heat, as the system
approaches three types of potential FPs (instabilities) which are classified in Table~\ref{Table_classify_FPs}
of Sec.~\ref{Sec_FP-instab}. In principle, it is a very challenging task to derive the analytical expressions for
physical quantities directly from an interacting theory. Instead of delving into the exact
expressions, a suitable and operational strategy is to extract the physical implications
from the renormalized fermionic propagator~\cite{Mahan1990Book}.
Compared with their free counterparts in Eq.~(\ref{Eq_G_psi}), the fermion velocities and tilting parameter are involved in the
coupled RG equations~(\ref{Eq_RGEq-v})-(\ref{Eq_RGEq-V_T-L}) and henceforth become energy-dependent.
This energy dependence inherits the characteristics of the interactions, effectively capturing fundamental tendencies associated with potential instabilities. Following this strategic approach, we subsequently need to establish the
relationship between the physical implications and fermion velocities as well as the tilting parameter.

\subsection{Density of states and compressibility}\label{Subsection_DOS-kappa}

At first, we consider the DOS and compressibility. After performing the analytical continuation,
the renormalized retarded fermion propagator is given by~\cite{Wang2012PRD}
\begin{eqnarray}
G^{\mathrm{ret}}(\omega,\mathbf{k})
&=&\{(\omega - \zeta v_z k_z)\sigma_0
+\chi[v_z k_z\sigma_z + v(k_x\sigma_x \nonumber\\
&&+ k_y\sigma_y)]\}/\{\omega^2+i\mathrm{sgn}(\omega-\zeta v_zk_z)\delta +\zeta^2 v_z^2 k_z^2\nonumber\\
&& - 2\omega \zeta v_z k_z-[v_z^2k_z^2+v^2(k_x^2+k_y^2)]\},
\end{eqnarray}
where $v,v_z$ and $\zeta$ are treated to be energy-dependent. The DOS of fermion quasiparticles then
takes the form
\begin{eqnarray}
\frac{\rho_{\mathrm{int}}(\omega)}{\Lambda^2_0}
&=&\int\frac{d^3\mathbf{k}}{(2\pi)^3}
\mathrm{Tr}\left\{-\frac{1}{\pi}
\mathrm{Im}\left[G^{\mathrm{ret}}(\omega,\mathbf{k})\right]\right\}.
\end{eqnarray}
Carrying out some calculations, we arrive at
\begin{widetext}
\begin{small}
\begin{numcases}{\frac{\rho_{\mathrm{int}}(\omega)}{\Lambda^2_0}=}
\int_{E,\theta,\omega}\left[\left|\omega - \frac{\zeta E(-\zeta+\cos\theta)}
{1-\zeta^{2}}\right|
\delta(E-\omega)+
\left|\omega + \frac{\zeta E(\zeta+\cos\theta)}
{\zeta^{2}-1}\right|
\delta\left(E-\frac{(1-\zeta^{2})\omega}
{1+\zeta^{2}+2\zeta \cos\theta}\right)\right],\!\!\!\!\!&
$\omega>0$,\label{Eq_rho_int_1}\\
~\nonumber\\
-\int_{E,\theta,\omega}
\left[\left|\omega + \frac{\zeta (\zeta-\cos\theta)E}
{1-\zeta^{2}}\right|
\delta\left(E-\frac{(\zeta^{2}-1)\omega}
{1+\zeta^{2}-2\zeta\cos\theta}\right)+\left|\omega - \frac{\zeta(\zeta+\cos\theta)E}
{1-\zeta^{2}}\right|
\delta\left(E+\omega\right)\right],\!\!\!\!\! \!\!\!\!&
$\omega<0$, \label{Eq_rho_int_2}
\end{numcases}
\end{small}
\end{widetext}
where $N$ represents the flavor of fermions, and the notation $\int_{E,\theta,\omega}$ is designated as
\begin{eqnarray}
\int_{E,\theta,\omega}\equiv
\frac{N}{(2\pi)^2}
\int_{e^{-l_c}}^{1} dE
\int_0^\pi d\theta
\frac{E^2\sin\theta}
{v^{2} v_z (1-\zeta^{2})\omega}.
\end{eqnarray}
These expressions can be reduced to
\begin{eqnarray}
\frac{\rho_0(\omega)}{\Lambda^2_0}
=\frac{N\omega^2}{\pi^2v^2 v_z(1-\zeta^2)^2}.
\end{eqnarray}
in the noninteracting case.

As to the compressibility, it is originally defined as $\kappa=\partial V/\partial F$~\cite{Schwabl2006Book},
where $V$ and $F$ represent the volume and compression force, respectively. Hereby, it is more convenient
to introduce the chemical potential $\mu$ and then calculate
using $\kappa=\partial n/\partial \mu$, where $n$ is the number of particles per area
directly associated with the DOS~\cite{Mahan1990Book,Sarma2007PRL}. After performing the necessary calculations, we are left with
\begin{widetext}
\begin{small}
\begin{numcases}{\frac{\kappa_{\mathrm{int}}(\mu)}{\Lambda_0}= \!\!\!}
\!\!\!
\int_{E,\theta,\mu}
\left[\left|
\frac{\zeta(\zeta-\cos\theta)E}{1-\zeta^{2}}-2\mu\right|
\delta\left(E-\frac{2(1-
\zeta^{2})\mu}{1+\zeta^{2}-2\zeta \cos\theta}\right)+\left|2\mu+
\frac{\zeta (\zeta +\cos\theta)E}{1-\zeta^{2}}\right|
\delta\left(E-2\mu\right)\right],\!\!\!\!\!&
$\mu>0$,\label{Eq_kappa_int_1}\\
~\nonumber\\
\!\!\!
\int_{E,\theta,\mu}
\left[\left|
\frac{\zeta(\zeta-\cos\theta)E}{1-\zeta^{2}}-2\mu\right|
\delta\left(E+2\mu\right)+\left|2\mu+
\frac{\zeta (\zeta +\cos\theta)E}{1-\zeta^{2}}\right|
\delta\left(E+\frac{2(1-\zeta^{2})\mu}
{1+\zeta^{2}+2\zeta \cos\theta}\right)\right],\!\!\!\!\!\!\!\!\!\!\!\!\!\!\! &
$\mu<0$, \label{Eq_kappa_int_2}
\end{numcases}
\end{small}
\end{widetext}
where $\int_{E,\theta,\mu}$ is introduced as
\begin{eqnarray}
\int_{E,\theta,\mu}\equiv\frac{N}{8\pi^2}
\int_{e^{-l_c}}^{1} dE \int_0^\pi d\theta
\frac{E^2\sin\theta}{v^{2}v_z(1-\zeta^{2})\mu}.
\end{eqnarray}
Similarly, the free-limit expression can be derived by taking $l_c\rightarrow\infty$, with
$v,v_z,\zeta$ assumed as constants.

\begin{figure}[htpb]
\includegraphics[width=3.5in]{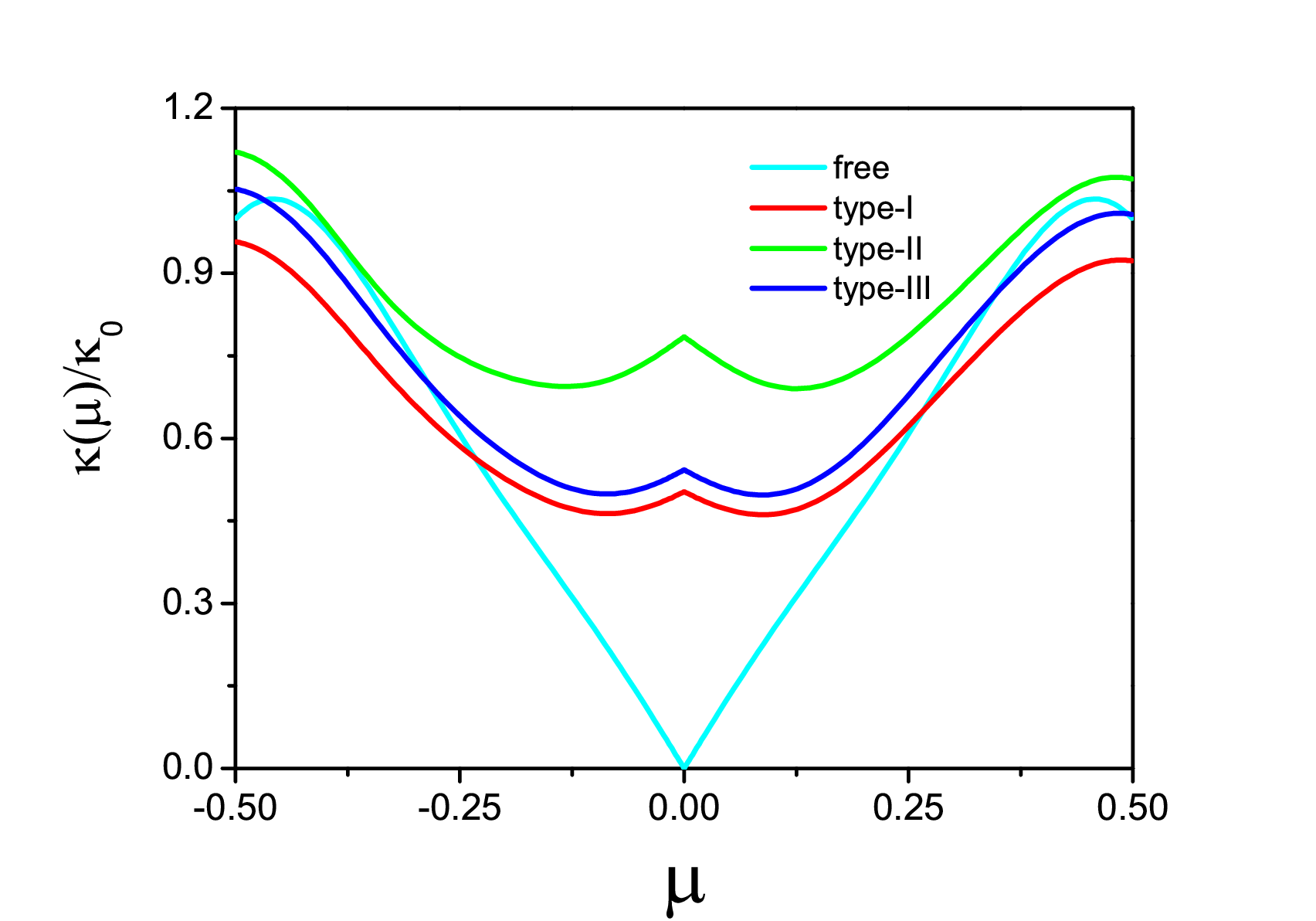}\\
\caption{(Color online) Behavior of $\kappa(\mu)$
around three distinct types of FPs with $\zeta=0.5$: type-I ($v_z(0)/v(0)=0.75$), type-II ($v_z(0)/v(0)=0.15$),
and type-III ($v_z(0)/v(0)=0.40$), respectively. Hereby, $\kappa_0$ is
designated by $\kappa_0\equiv\kappa_{\mathrm{free}}(\zeta=0,\mu=-0.5)$.
}\label{Fig_kappa-mu-0.5}
\end{figure}

With the above information in hand, we delve into the behavior of DOS and $\kappa$ both influenced by the interactions.
By combining Eqs.~(\ref{Eq_rho_int_1})-(\ref{Eq_rho_int_2}) and (\ref{Eq_kappa_int_1})-(\ref{Eq_kappa_int_2}) with the RG equations~(\ref{Eq_RGEq-v})-(\ref{Eq_RGEq-V_T-L}), their basic tendencies are obtained
as we approach three distinct kinds of instabilities. The behavior of DOS is illustrated
in Fig.~\ref{Fig_DOS-all-omega-0.5}. In the absence of interactions and
tilting parameter, the DOS exhibits a parabolic frequency dependence, $\rho_0(\omega)\propto\omega^2$,
and hence it vanishes precisely at the Dirac point. In sharp contrast, approaching the three types of
identified instabilities introduces the interaction contributions
that render the DOS a nonzero finite value at $\omega=0$.
This is qualitatively distinct from the free case and the nonzero tilting parameter further
amplifies the departure by inducing an asymmetry
for $\omega>0$ and $\omega<0$. Away from $\omega=0$, the intimate competition between the interactions and
thermal fluctuations governs the behavior of DOS. We can find that the DOS decreases at first and then ascends around
$|\omega|\gtrsim\omega_c\approx0.25$. This indicates that below the critical frequency
$\omega_c$, the thermal fluctuations are subordinate to the interactions,
while above it, the thermal fluctuations dominate. In addition, although the DOS tendencies near all three
types of instabilities are similar, the type-II instability leads to a little more corrections compared with the other two types.
Furthermore, it is worth pointing out that the interacting DOS $\rho_{\mathrm{int}}(\omega)$ for a tilted system
($\zeta\neq0$) is no longer strictly symmetric about $\omega=0$ but instead
exhibits a slight asymmetric tendency with respect to $\omega=0$ owing to the
interplay of interactions and the tilt term.

Regarding the compressibility denoted as $\kappa$,
Fig.~\ref{Fig_kappa-mu-0.5} shows the $\mu$-dependent evolution of $\kappa$ in the vicinity of
three types of instabilities. In the absence of interactions and tilting parameters, the compressibility in the free case,
$\kappa_{\mathrm{free},\zeta=0}(\mu)$, exhibits a nearly linear dependence on $\mu$ and becomes
incompressible at $\mu=0$, which is consistent with the behavior of the free DOS.
However, once the contributions from the interactions and tilting parameter are taken into account,
the behavior of $\kappa$ undergoes qualitative changes for $\mu\leq\mu_c\approx0.25$. It becomes much more compressible,
reaching its maximum value at $\mu=0$. Subsequently, the compressibility increases at $\mu>\mu_c$ and exhibits a
similar tendency to the free case for sufficiently large $\mu$.
This change would arise from the intricate interplay between interactions and thermal fluctuations
%This would be ascribed to the
%intricate interplay between interactions and thermal fluctuations.
Besides, it is observed that $\kappa$ experiences more corrections near the type-II instability.
Moreover, we notice that the tilting parameter seems to introduce only minor corrections with opposite signs to $\kappa$.

\subsection{Specific heat}

Next, we turn our attention to the specific heat of the quasiparticle.
The free-energy density $f(T)$ can be written as
\begin{eqnarray}
f(T)
&=&-\frac{T}{V}\ln Z,
\end{eqnarray}
where the partition function is associated with~\cite{Kapusta1994Book}
\begin{eqnarray}
Z&=&
\prod_{n,k,\alpha}
\int \left[d(i\psi_{\alpha,n}^\dag)\right]
\left[d\psi_{\rho,n}\right]e^{S_0},
\end{eqnarray}
with $S_0$ denoting the fermionic component of our theory. After long but straightforward calculations,
we finally obtain
\begin{widetext}
\begin{eqnarray}
\frac{f(T)}{\Lambda_0^4}
&=&-\frac{T}{4\pi^2}
\int_{e^{-l_c}}^{1}dE
\int_0^\pi d\theta
\frac{E^2\sin\theta(1-\zeta \cos\theta)}
{v^{2} v_z (1-\zeta^{2})^2}\Bigg[\ln(1+e^{-\sqrt{x^+}})+\ln(1+e^{-\sqrt{y^+}})\Bigg]\nonumber\\
&&-
\frac{T}{4\pi^2}
\int_{e^{-l_c}}^{1}dE
\int_0^\pi d\theta
\frac{E^2\sin\theta(1+\zeta \cos\theta)}
{v^{2} v_z (1-\zeta^{2})^2}\Bigg[\ln(1+e^{-\sqrt{x^-}})+\ln(1+e^{-\sqrt{y^-}})\Bigg].
\end{eqnarray}
\end{widetext}
In this expression, we utilize the transformations $E\rightarrow E/\Lambda_0$,
$T\rightarrow T/\Lambda_0$, and define $x^{\pm},y^{\pm}$ as
\begin{eqnarray}
x^+
&\equiv&
\frac{\mathcal{N}_{-}E^2}{(1-\zeta^2)^2T^2},\\
y^+
&\equiv&
\frac{(\zeta^2+1-2\zeta\cos\theta)^2E^2}{\mathcal{N}_{-}T^2},\\
x^-
&\equiv&\frac{\mathcal{N}_{+}E^2}{(1-\zeta^2)^2T^2},\\
y^-
&\equiv&
\frac{(\zeta^2+1+2\zeta\cos\theta)^2E^2}{\mathcal{N}_{+}T^2},
\end{eqnarray}
with $\mathcal{N}_{\mp}$ denominated as
\begin{eqnarray}
\mathcal{N}_{\mp}
&=&\zeta^2(1+\zeta^2) \mp2\zeta(1+\zeta^2)
\cos\theta+(1+\zeta^2\cos(2\theta))\nonumber\\
&&+\sqrt{2\zeta^2\left(\zeta \mp\cos\theta\right)^2
\left[\zeta^2\mp
4\zeta\cos\theta+\zeta^2\cos(2\theta)+2\right]}.\nonumber
\end{eqnarray}

\begin{figure}[htpb]
%\subfigure[]{\includegraphics[width=2.7in]{fig_I-T-1Em1-Cv-vz-050.eps}}\hspace{-1.7cm}
%\subfigure[]{\includegraphics[width=2.7in]{fig_II-T-1Em1-Cv-vz-015.eps}}\hspace{-1.7cm}
%\subfigure[]{\includegraphics[width=2.7in]{fig_III-T-1Em1-Cv-v-040.eps}}\\
\includegraphics[width=3.5in]{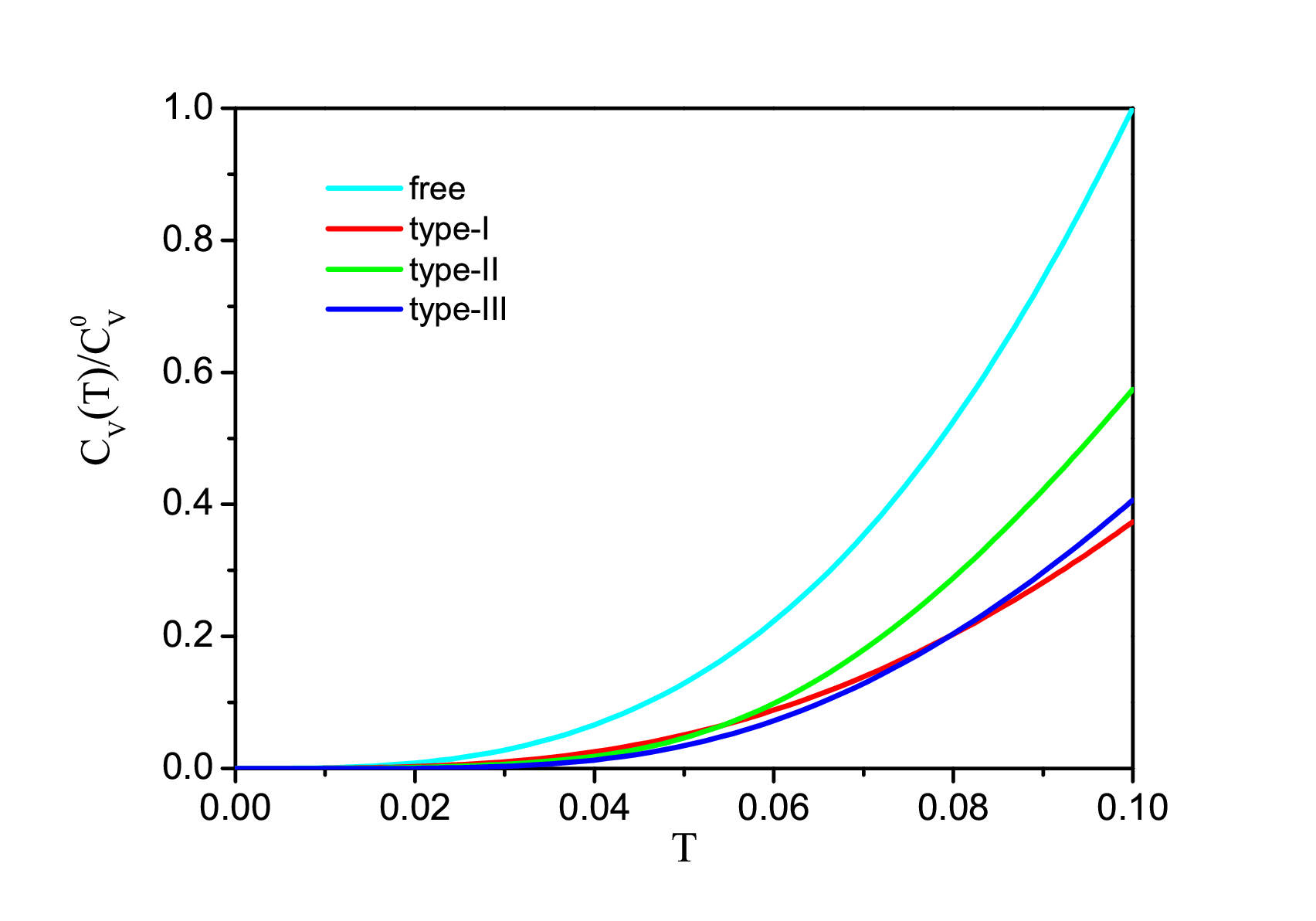}\\ \vspace{-0.5cm}
\caption{(Color online) Temperature-dependent evolutions of specific heat $C_V(T)$
around three distinct types of FPs with $\zeta=0.5$: type-I ($v_z(0)/v(0)=0.75$), type-II ($v_z(0)/v(0)=0.15$),
and type-III ($v_z(0)/v(0)=0.40$), respectively. Hereby, $C_{V0}$ is
designated by $C_{V0}\equiv C^{\mathrm{free}}_V(\zeta=0,T=0.1)$.}\label{Fig_C_V-T}
\end{figure}

In consequence, the specific heat can be derived as
\begin{eqnarray}
C_V(T)
=-T\frac{\partial^2 f(T)}{\partial T^2}.\label{Eq_Cv}
\end{eqnarray}
Once again, the free limit $C^0_V$ can be obtained by taking $l_c\rightarrow\infty$ and considering
$v,v_z,\zeta$ as specific constants.

After taking into account the RG equations of interaction parameters and the expression for
$C_V$~(\ref{Eq_Cv}), the numerical results for $C_V$ are illustrated in
Fig.~\ref{Fig_C_V-T} in the vicinity of three distinct kinds of instabilities.
At first, compared with the free case $C^0_V(T)\propto T^2$
in the absence of interactions and tilting parameter, we notice that the intimate
interplay between the tilting parameter and interactions establishes close relationships
and constraints among the low-energy quasiparticles.
As a result, this interaction-induced effect on the specific heat leads to a slight deviation of the renormalized $C_V(T)$
from the $T^2$ dependence, indicative of non-Fermi-liquid behavior~\cite{Mahan1990Book}.
In addition, although $C_V$ is suppressed as we approach all three kinds of instabilities,
it is unequivocally evident from Fig.~\ref{Fig_C_V-T} that the specific heat is less reduced in the type-II case
compared with the other two cases. This is qualitative in agreement with the basic tendencies of
the DOS and compressibility presented in Sec.~\ref{Subsection_DOS-kappa}.

To be brief, all these physical implications surrounding the potential instabilities would be of particular help to
delve deeper into the study of more quantities of the related tilted materials.

\section{Summary}\label{Sec_summary}

In summary, we study the low-energy physics of a type-I 3D tDSM by adopting the powerful RG method~\cite{Wilson1975RMP,Polchinski9210046,Shankar1994RMP}, which helps us to
unbiasedly treat all physical ingredients, including the Coulomb interactions and
electron-phonon coupling as well as phonon-phonon interactions. By considering all one-loop corrections, we derived
the coupled RG equations for all relevant parameters. After performing the numerical analysis, we systematically presented
the low-energy behavior of these interactions and their influence on potential instabilities and physical properties.

To begin with, we examine the fate of the tilting parameter and observe two distinct scenarios
depending on the initial anisotropy of fermion velocities. To ensure the self-consistency of our theory,
we only focus on the first scenario, as shown in Fig.~\ref{Fig_phase}(a).
Within this scenario, we find that the anisotropy of fermion velocities is primarily
dependent on its initial value but is insensitive to the strength of the tilting parameter.
Notably, this anisotropy may exhibit an increase, decrease, or remain nearly constant in the low-energy regime.
Regarding the ratio of dielectric constant $\epsilon_z/\epsilon$, it always flows towards either extreme anisotropy, i.e.,
$\epsilon_z/\epsilon\gg1$ or $\epsilon_z/\epsilon\ll1$, when starting from Anisotropy-I or
Anisotropy-II, respectively. This suggests the screening of Coulomb interaction in the direction-$z$ or direction-$x,y$.
Furthermore, the fate of $g/g_0$ is analogous to that of $\epsilon_z/\epsilon$.
Compared with the $\epsilon_z/\epsilon$, we notice that both
the phonon velocities and phonon-phonon interactions
can either flow towards approximate isotropy or bear similarities to
the extreme anisotropy observed in $\epsilon_z/\epsilon$ in the low-energy regime.
Analogously, the anisotropy of electron-phonon interactions $\lambda_z/\lambda$ shares the
similar tendency of phonon velocities $C_T^z/C_T$.

Subsequently, we systematically investigate the tendencies of all interaction
parameters, which give rise to three distinct types of FPs as displayed in Table~\ref{Table_classify_FPs}.
After introducing the source terms for the potential symmetry breakings and comparing their susceptibilities
accessing the FPs, we find that there exists an interaction-driven phase transition around the
FPs, with either Phase-B or Phase-C emerged as the preferred leading instability. Additionally, we delve into
the critical properties of physical quantities, including the density of states and compressibility as well as specific heat,
as the system approaches these three distinct types of FPs.
In sharp contrast to their noninteracting counterparts, they exhibit very different behavior,
particularly in the proximity of the Dirac point.  Notably, they deviate slightly from
the scope of Fermi-liquid behavior.  To recapitulate, we expect all these results will be helpful to provide relevant
clues for investigating the fascinating behavior of 3D tDSM and exploring other
related tilted materials in the future.

\section*{ACKNOWLEDGEMENTS}

We thank Y. H. Zhai, W. Liu and X. Z. Chu for useful discussions.
J.W. was partially supported by the National Natural Science Foundation of China
under Grant No. 11504360.

\appendix

\section{Collections of one-loop corrections}\label{appendix-one-loop-corrections}

After long calculations, we hereby present the one-loop corrections to
all vertex couplings in our effective theory~(\ref{Eq_eff-action}) as illustrated in Fig.~\ref{fig_appendix-1}.
These corrections arise from the interplay among the Coulomb interaction, electron-phonon interaction,
and phonon-phonon interaction, which can be formally expressed as follows,
\begin{eqnarray}
\delta g&=&\mathcal{B}gl,\label{Rq_appendix-1}\\
\delta\lambda&=&\mathcal{D}\lambda l,\\
\delta\lambda_z&=&\mathcal{D}_{z}\lambda_zl,\\
\delta V^{ij}_{T,L}&=&V^{ij}_{T,L}\mathcal{F}^{i,j}_{T,L}l,\,\,\mathrm{with}\,\,i,j=x,y,z,\label{Rq_appendix-2}
\end{eqnarray}
Here, we have neglected the unimportant constant terms for clarity.
All related coefficients that are involved in both Eqs.~(\ref{Rq_appendix-1})-(\ref{Rq_appendix-2}) and
RG equations~(\ref{Eq_RGEq-v})-(\ref{Eq_RGEq-V_T-L}) are nominated in the following.

As to the coefficients $\mathcal{A}_{1}$ and $\mathcal{A}_{2}$, we have
\begin{eqnarray}
\mathcal{A}_{1}&\equiv&\int_{0}^{\pi}d\theta(\mathcal{A}_{11}-\mathcal{A}_{12}),\label{Eq_A_1}\\
\mathcal{A}_{2}&\equiv&
\int_{0}^{\pi}d\theta\big(\mathcal{A}_{21}+\mathcal{A}_{22}\big),\label{Eq_A_2}
\end{eqnarray}
with $\mathcal{A}_{11}$, $\mathcal{A}_{12}$, $\mathcal{A}_{21}$, and $\mathcal{A}_{22}$ being
\begin{widetext}
\begin{small}
\begin{eqnarray}
\mathcal{A}_{11}&\equiv&
\frac{-\eta_{\zeta}(1-\zeta^2)g^2\epsilon}{2\pi v^4v_{z}}
\frac{(1-\zeta^2)\sin^3\theta}
{\big[\epsilon\frac{\big(1-\zeta^2\big)\sin^2\theta}{v^2}
+\epsilon_{z}\big(\frac{\cos\theta-|\zeta|}{v_{z}}
\big)^2\big]^{2}}
+\frac{\eta_{\zeta}(1-\zeta^2)}{16\pi^2v^4v_{z}}
\frac{\lambda^2\mathcal{J}\big[\frac{(1-\zeta^2)\sin^2\theta}{v^2}
+\frac{2(\cos\theta-|\zeta|)^2}{v_{z}^2}\big]\sin\theta}{(1-|\zeta|\cos\theta)^{2}
\big[\big(1-|\zeta|\cos\theta+\mathcal{I}\big)^2
-\zeta^2\big(\cos\theta-|\zeta|\big)^2\big]^2}\nonumber\\
&&\times\Big\{
2(1-|\zeta|\cos\theta)
\big[(1-2\zeta^2)\big(\cos\theta-|\zeta|\big)^2
+\big(1-\zeta^2\big)\sin^2\theta\big]
\mathcal{I}+(1-\zeta^2)^2\big[\big(\cos\theta-|\zeta|\big)^2+\sin^2\theta\big]^2
\nonumber\\
&&+\!(1\!-\!\zeta^2)\big[\big(\cos\theta\!-\!|\zeta|\big)^2\!+\!\sin^2\theta\big]
\mathcal{I}^2\!
+\!2\zeta^2\big(\cos\theta\!-\!|\zeta|\big)^2
\big[3(1\!-\!|\zeta|\cos\theta)^2
\!+\!4(1\!-\!|\zeta|\cos\theta)\mathcal{I}\!+\!\mathcal{I}^2
\!-\!\zeta^2\big(\cos\theta\!-\!|\zeta|\big)^2\big]\nonumber\\
&&+\frac{(1-\zeta^2)[(\cos\theta-|\zeta|)^2
+(1-v)\sin^2\theta]}{\mathcal{I}}
\big\{2(1-|\zeta|\cos\theta)^{3}
+\big[5(1-|\zeta|\cos\theta)^2-\zeta^2\big(\cos\theta-|\zeta|\big)^2\big]\mathcal{I}\nonumber\\
&&+4(1-|\zeta|\cos\theta)\mathcal{I}^2+\mathcal{I}^{3}\big\}\Big\}
+\frac{\eta_{\zeta}(1-\zeta^2)^2}{16\pi^2v^4v_{z}}
\frac{\lambda\mathcal{J}\bigl[\frac{\lambda_{z}
(\cos\theta-|\zeta|)^2}{v_{z}}
+\frac{\lambda(1-\zeta^2)\sin^2\theta}{4v}\bigr]
\sin^3\theta}{(1-|\zeta|\cos\theta)^{2}
\big[\big(1-|\zeta|\cos\theta+\mathcal{I}\big)^2
-\zeta^2\big(\cos\theta-|\zeta|\big)^2\big]^2\mathcal{I}}
\nonumber\\
&&\times\Big\{2(1-|\zeta|\cos\theta)^{3}
+\big[5(1-|\zeta|\cos\theta)^2-\zeta^2\big(\cos\theta-|\zeta|\big)^2\big]\mathcal{I}
+4(1-|\zeta|\cos\theta)\mathcal{I}^2+\mathcal{I}^{3}\Big\},\\
\mathcal{A}_{12}&\equiv&
\frac{\eta_{\zeta}(1-\zeta^2)}{16\pi^2v^4v_{z}}
\frac{\frac{-2\mathcal{J}\lambda_{z}^2(\cos\theta-|\zeta|)^2}
{v_{z}^2}\sin\theta}{(1-|\zeta|\cos\theta)^{2}
\big[\big(1-|\zeta|\cos\theta
+\mathcal{I'}\big)^2-\zeta^2\big(\cos\theta-|\zeta|\big)^2\big]^2}
\Big\{2(1-|\zeta|\cos\theta)
\big[(1-2\zeta^2)\big(\cos\theta-|\zeta|\big)^2\nonumber\\
&&+\big(1-\zeta^2\big)\sin^2\theta\big]\mathcal{I'}
+(1-\zeta^2)^2\big[\big(\cos\theta-|\zeta|\big)^2
+\sin^2\theta\big]^2\!+\!(1\!-\!\zeta^2)[(\cos\theta\!-\!|\zeta|)^2\!+\!\sin^2\theta]\mathcal{I'}^2\nonumber\\
&&\!+\!2\zeta^2\big(\cos\theta\!-\!|\zeta|\big)^2
\big[3(1\!-\!|\zeta|\cos\theta)^2\!+\!4(1\!-\!|\zeta|\cos\theta)\mathcal{I'}
\!+\!\mathcal{I'}^2\!-\!\zeta^2\big(\cos\theta\!-\!|\zeta|\big)^2\big]
\!+\!\frac{(1-\zeta^2)\big[(\cos\theta-|\zeta|)^2
+(1-v)\sin^2\theta\big]}{\mathcal{I'}}\nonumber\\
&&\times\big\{2(1-|\zeta|\cos\theta)^{3}
+\big[\!5(1-|\zeta|\cos\theta)^2-\zeta^2\big(\cos\theta-|\zeta|\big)^2
\big]\mathcal{I'}
+4(1-|\zeta|\cos\theta)\mathcal{I'}^2+\mathcal{I'}^{3}\big\}\Big\}\nonumber\\
&&-\frac{\eta_{\zeta}(1-\zeta^2)^2}{16\pi^2v^4v_{z}}
\frac{2\lambda\mathcal{J}\big[
\frac{\lambda_{z}(\cos\theta\!-\!|\zeta|)^2}{v_{z}}
\!+\!\frac{\lambda(1-\zeta^2)\sin^2\theta}{4v}\big]
\sin^3\theta}{(1-|\zeta|\cos\theta)^{2}
\big[\big(1-|\zeta|\cos\theta
+\mathcal{I'}\big)^2-\zeta^2\big(\cos\theta-|\zeta|\big)^2\big]^2\mathcal{I'}}\nonumber\\
&&\times\Big\{2(1-|\zeta|\cos\theta)^{3}
+\big[\!5(1-|\zeta|\cos\theta)^2-\zeta^2\big(\cos\theta-|\zeta|\big)^2
\big]\mathcal{I'}
+4(1-|\zeta|\cos\theta)\mathcal{I'}^2+\mathcal{I'}^{3}\Big\},\\
\mathcal{A}_{21}&\equiv&\frac{-\eta_{\zeta}(1-\zeta^2)\epsilon_{z}g^2}{\pi v^2v_{z}^3}\frac{\big(\cos\theta-|\zeta|\big)^2\sin\theta}
{\big[\epsilon\frac{\big(1-\zeta^2\big)\sin^2\theta}{v^2}
+\epsilon_{z}\big(\frac{\cos\theta-|\zeta|}{v_{z}}\big)^2\big]^{2}}
+\frac{\eta_{\zeta}(1-\zeta^2)}{16\pi^2v^2v_{z}^3}
\frac{2\lambda_{z}^2\mathcal{J}\frac{(1-\zeta^2)\sin^2\theta}{v^2}
\sin\theta}{(1-|\zeta|\cos\theta)^{2}
\big[\big(1-|\zeta|\cos\theta
+\mathcal{I}\big)^2-\zeta^2\big(\cos\theta-|\zeta|\big)^2
\big]^2}\nonumber\\
&&\times\Big\{(1-\zeta^2)^2\big[\big(\cos\theta-|\zeta|\big)^2+\sin^2\theta\big]^2
+2(1-|\zeta|\cos\theta)\big[(1-2\zeta^2)\big(\cos\theta-|\zeta|\big)^2
+\big(1-\zeta^2\big)\sin^2\theta\big]\mathcal{I}\nonumber\\
&&+(1-\zeta^2)\big[\big(\cos\theta-|\zeta|\big)^2
+\sin^2\theta\big]\mathcal{I}^2
+2\zeta^2(1-v_{z})\big(\cos\theta-|\zeta|\big)^2
\big[3(1-|\zeta|\cos\theta)^2+4(1-|\zeta|\cos\theta)
\mathcal{I}+\mathcal{I}^2-\zeta^2\big(\cos\theta-|\zeta|\big)^2\big]\nonumber\\
&&+\frac{\big(1-\zeta^2\big)\big[(1+2v_{z})\big(\cos\theta-|\zeta|\big)^2
+\sin^2\theta\big]}{\mathcal{I}}
\!\big\{2(1-|\zeta|\cos\theta)^{3}
+\!\big[5(1-|\zeta|\cos\theta)^2
\!\!-\!\!\zeta^2\big(\cos\theta\!-\!|\zeta|\big)^2\big]\mathcal{I}\!+\!\mathcal{I}^{3}
\!+4(1\!-\!|\zeta|\cos\theta)\mathcal{I}^2\big\}\!\Big\}\nonumber\\
&&-\frac{\eta_{\zeta}(1-\zeta^2)}{16\pi^2v^2v_{z}^3}
\frac{\mathcal{J}v_{z}(1-\zeta^2)\lambda
\lambda_{z}\sin^2\theta\big(\cos\theta-|\zeta|\big)^2\sin\theta}
{(1-|\zeta|\cos\theta)^{2}
\big[\big(1-|\zeta|\cos\theta
+\mathcal{I}\big)^2-\zeta^2\big(\cos\theta-|\zeta|\big)^2
\big]^2}\nonumber
\Big\{\zeta^2\big[3(1-|\zeta|\cos\theta)^2+4(1-|\zeta|\cos\theta)
\mathcal{I}+\mathcal{I}^2\nonumber\\
&&-\zeta^2\big(\cos\theta-|\zeta|\big)^2\big]
-\frac{1}{\mathcal{I}}\big\{2(1-|\zeta|\cos\theta)^{3}
+\big[5(1-|\zeta|\cos\theta)^2-\zeta^2\big(\cos\theta-|\zeta|\big)^2
\big]\mathcal{I}
+4(1-|\zeta|\cos\theta)\mathcal{I}^2
+\mathcal{I}^{3}\big\}\Big\},\\
\mathcal{A}_{22}&\equiv&\frac{\eta_{\zeta}(1-\zeta^2)}{16\pi^2v^2v_{z}^3}
\frac{2\mathcal{J}\big[-\frac{\lambda^2(1-\zeta^2)\sin^2\theta}{v^2}
+\frac{\lambda_{z}^2(\cos\theta-|\zeta|)^2}{v_{z}^2}\big]
\sin\theta}{(1-|\zeta|\cos\theta)^{2}
\big[\big(1-|\zeta|\cos\theta+\mathcal{I'}\big)^2
-\zeta^2\big(\cos\theta-|\zeta|\big)^2\big]^2}
\Big\{(1-\zeta^2)^2\big[\big(\cos\theta-|\zeta|\big)^2
+\sin^2\theta\big]^2\nonumber\\
&&+2(1-|\zeta|\cos\theta)
\big[(1-2\zeta^2)\big(\cos\theta-|\zeta|\big)^2
+\big(1-\zeta^2\big)\sin^2\theta\big]
\mathcal{I'}
+(1-\zeta^2)\big[\big(\cos\theta-|\zeta|\big)^2+\sin^2\theta\big]\mathcal{I'}^2+2\zeta^2(1-v_{z})\nonumber\\
&&\times\big(\cos\theta-|\zeta|\big)^2
\big[3(1-|\zeta|\cos\theta)^2+4(1-|\zeta|\cos\theta)
\mathcal{I'}
+\mathcal{I'}^2-\zeta^2\big(\cos\theta-|\zeta|\big)^2\big]\!
+\!\frac{(1-\zeta^2)\big[(1+2v_{z})\big(\cos\theta-|\zeta|\big)^2
+\sin^2\theta\big]}{\mathcal{I'}}\nonumber\\
&&\!\times\!\big\{2(1-|\zeta|\cos\theta)^{2}
+\big[5(1-|\zeta|\cos\theta)^2-\zeta^2\big(\cos\theta-|\zeta|\big)^2\big]\mathcal{I'}
+4(1-|\zeta|\cos\theta)\mathcal{I'}^2+\mathcal{I'}^{3}\big\}\Big\}+\frac{\eta_{\zeta}(1-\zeta^2)}{16\pi^2v^2v_{z}^3}\nonumber\\
&&\times
\frac{2\mathcal{J}(1-\zeta^2)\lambda\lambda_{z}v_{z}\sin^2\theta
\big(\cos\theta-|\zeta|\big)^2\sin\theta}
{(1-|\zeta|\cos\theta)^{2}
\big[\big(1-|\zeta|\cos\theta+\mathcal{I'}\big)^2
-\zeta^2\big(\cos\theta-|\zeta|\big)^2\big]^2}
\Big\{\zeta^2\big[3(1-|\zeta|\cos\theta)^2+4(1-|\zeta|\cos\theta)
\mathcal{I'}+\mathcal{I'}^2-\zeta^2\big(\cos\theta-|\zeta|\big)^2\big]\nonumber\\
&&-\frac{1}{\mathcal{I'}}
\big\{2(1-|\zeta|\cos\theta)^{3}
+\big[5(1-|\zeta|\cos\theta)^2-\zeta^2\big(\cos\theta-|\zeta|\big)^2\big]\mathcal{I'}
+4(1-|\zeta|\cos\theta)\mathcal{I'}^2+\mathcal{I'}^{3}\big\}\Big\},
\end{eqnarray}
and for the coefficients $\mathcal{A}_{0}$ and $\mathcal{A}_{3}$,
\begin{eqnarray}
\mathcal{A}_{0}
&=&-\frac{\eta_{\zeta}(1-\zeta^2)}{4\pi^2v^2v_{z}}
\int_{0}^{\pi}d\theta\sin\theta
\Big\{\frac{\big[(2\lambda^2+\lambda_{z}^2)(1-|\zeta|\cos\theta)
-\mathcal{J'}\big]}
{\big\{\big[(1-|\zeta|\cos\theta)+\mathcal{I}\big]^2
-\zeta^2\big(\cos\theta-|\zeta|\big)^2\big\}\mathcal{I}}
+\frac{\mathcal{J'}(1-|\zeta|\cos\theta)}
{\big\{\big[(1-|\zeta|\cos\theta)+\mathcal{I'}\big]^2
-\zeta^2\big(\cos\theta-|\zeta|\big)^2\big\}\mathcal{I'}}\Big\},\label{Rq_appendix-A0}\\
\mathcal{A}_{3}&\equiv&
\frac{\eta_{\zeta}(1-\zeta^2)}{4\pi^2v^2v_{z}}
\int_{0}^{\pi}d\theta\Bigl\{\frac{\sin\theta\big[(2\lambda^2+\lambda_{z}^2)
-\lambda^2\mathcal{J\frac{(1-\zeta^2)\sin^2\theta}{v^2}
+\lambda_{z}^2\frac{(\cos\theta-|\zeta|)^2}{v_{z}^2}}\big]}
{(1-|\zeta|\cos\theta)^{2}
\big[\big(1-|\zeta|\cos\theta+\mathcal{I}\big)^2
-\zeta^2(\cos\theta-|\zeta|)^2\big]^2}
\Bigl\{-\frac{1}{2}
\big\{(1-\zeta^2)^2\big[(\cos\theta-|\zeta|)^2
+\sin^2\theta\big]^2\nonumber\\
&&+2(1-|\zeta|\cos\theta)
\big[(1-2\zeta^2)\big(\cos\theta-|\zeta|\big)^2
+\big(1-\zeta^2\big)\sin^2\theta\big]\mathcal{I}
+(1-\zeta^2)\big[\big(\cos\theta-|\zeta|\big)^2
+\sin^2\theta\big]\mathcal{I}^2\big\}
+(1-\zeta^2)(\cos\theta-|\zeta|)^2\nonumber\\
&&\times\big[3(1-|\zeta|\cos\theta)^2
+4(1-|\zeta|\cos\theta)\mathcal{I}
+\mathcal{I}^2-\zeta^2(\cos\theta-|\zeta|)^2\big]+\frac{\big(1-\zeta^2\big)
\big[\sin^2\theta-(\cos\theta-|\zeta|)^2\big]}
{2\mathcal{I}}
\big\{2(1-|\zeta|\cos\theta)^{3}\nonumber\\
&&+\big[5(1-|\zeta|\cos\theta)^2-\zeta^2(\cos\theta-|\zeta|)^2\big]\mathcal{I}
+4(1-|\zeta|\cos\theta)\mathcal{I}^2+\mathcal{I}^{3}
\big\}\Bigr\}
+\frac{\sin\theta\mathcal{J'}}{(1-|\zeta|\cos\theta)^{2}
\big[\big(1-|\zeta|\cos\theta+\mathcal{I'}\big)^2
-\zeta^2(\cos\theta-|\zeta|)^2\big]^2}\nonumber\\
&&\times\Bigl\{-\frac{1}{2}
\big\{(1-\zeta^2)^2\big[(\cos\theta-|\zeta|)^2
+\sin^2\theta\big]^2+2(1-|\zeta|\cos\theta)
\big[(1-2\zeta^2)(\cos\theta-|\zeta|)^2
+\big(1-\zeta^2\big)\sin^2\theta\big]\mathcal{I'}\nonumber\\
&&+(1-\zeta^2)\big[(\cos\theta-|\zeta|)^2
+\sin^2\theta\big]\mathcal{I'}^2\big\}+(1-\zeta^2)(\cos\theta-|\zeta|)^2
\times\big[3(1-|\zeta|\cos\theta)^2+4(1-|\zeta|\cos\theta)\mathcal{I'}
+\mathcal{I'}^2-\zeta^2(\cos\theta-|\zeta|)^2\big]\nonumber\\
&&+\frac{\big(1-\zeta^2\big)
\big[\sin^2\theta-(\cos\theta-|\zeta|)^2\big]}
{2\mathcal{I'}}
\!\big\{2(1-|\zeta|\cos\theta)^{3}
+\big[5(1-\!|\zeta|\cos\theta)^2\!-\!\zeta^2(\cos\theta\!-\!|\zeta|)^2\big]\mathcal{I'}
+4(1\!-\!|\zeta|\cos\theta)\mathcal{I'}^2\!+\!\mathcal{I'}^{3}
\big\}\Bigr\}\!\Bigl\}.
\end{eqnarray}

\begin{figure*}
\centering
\includegraphics[width=6.3in]{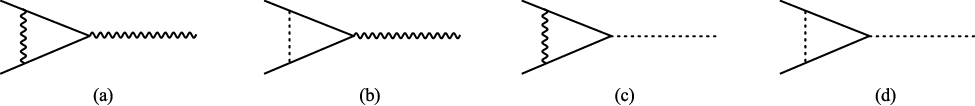}\\ \vspace{0.35cm}
\includegraphics[width=6.3in]{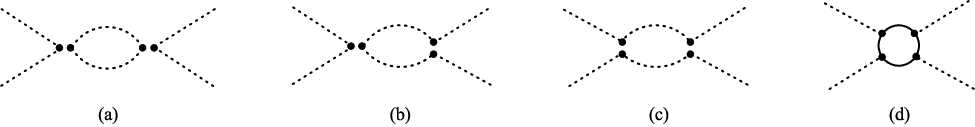}
\vspace{-0.15cm}
\caption{One-loop corrections to (a)-(b) the Coulomb interaction, (c)-(d) the electron-phonon interaction,
and (e)-(h) the phonon-phonon interactions, respectively (the solid, wavy, and dashed lines
denote the free fermionic, auxiliary bosonic, and phonon propagators).}\label{fig_appendix-1}
\end{figure*}

With respect to the other coefficients, we get
\begin{eqnarray}
\mathcal{B}&\equiv&
-\int_{0}^{\pi}d\theta
\frac{\sin\theta}{(1-|\zeta|\cos\theta)^2
\big[(1-|\zeta|\cos\theta+\mathcal{I})^2
-\zeta^2(\cos\theta-|\zeta|)^2\big]^2}
\frac{\eta_{\zeta}(1-\zeta^2)}{16\pi^2v^2v_{z}}\Big\{
2\big[(2\lambda^2+\lambda_{z}^2)
-\mathcal{J'}\big]\nonumber\\
&&\times\big\{\mathcal{I}^2(1-\zeta^2)
\big[(\cos\theta-|\zeta|)^2+\sin^2\theta\big]
+(1-\zeta^2)^2\big[(\cos\theta-|\zeta|)^2+\sin^2\theta\big]^2
+2\mathcal{I}(1-|\zeta|\cos\theta)
\big[(1-2\zeta^2)(\cos\theta-|\zeta|)^2\nonumber\\
&&+(1-\zeta^2)\sin^2\theta\big]\big\}
-\frac{1}{2\mathcal{I}}\big\{\mathcal{I}^{3}
+2\big(1-|\zeta|\cos\theta\big)^{3}
+4\mathcal{I}^2(1-|\zeta|\cos\theta)
+\mathcal{I}\big[5(1-|\zeta|\cos\theta)^2
-\zeta^2(\cos\theta-|\zeta|)^2\big]\big\}\nonumber\\
&&\times\big\{2\lambda^2(1-\zeta^2)\sin^2\theta+2\lambda_{z}^2
(\cos\theta-|\zeta|)^2-\frac{\lambda^2\mathcal{J}(1-\zeta^2)^2\sin^4\theta}{2v^2}
-\frac{2\lambda_{z}^2\mathcal{J}(\cos\theta-|\zeta|)^4}{v_{z}^2}\nonumber\\
&&-\frac{\frac{1}{2}\lambda^2(1-\zeta^2)^2\sin^4\theta
+4\lambda\lambda_{z}(1-\zeta^2)\sin^2\theta
(\cos\theta-|\zeta|)^2}{vv_{z}\big[\frac{(1-|\zeta|\cos\theta)^2}{v^2}
+\frac{(\cos\theta-|\zeta|)^2}{v_{z}^2}\big]}
+2\zeta^2(\cos\theta-|\zeta|)^2\big[2\lambda^2\!+\!\lambda_{z}^2
\!-\!\frac{\lambda^2(1-\zeta^2)\sin^2\theta}{v^2
\big[\frac{(1-\zeta^2)\sin^2\theta}{v^2}
\!+\frac{(\cos\theta-|\zeta|)^2}{v_{z}^2}\big]}\nonumber\\
&&-\frac{\lambda_{z}^2\mathcal{J}(\cos\theta-|\zeta|)^2}{v_{z}^2}\big]
\big\}
+\zeta^2\big[(2\lambda^2+\lambda_{z}^2)
-\mathcal{J'}](\cos\theta-|\zeta|)^2
\big[\mathcal{I}^2+4\mathcal{I}(1-|\zeta|\cos\theta)\nonumber\\
&&
+3(1-|\zeta|\cos\theta)^2-\zeta^2
(\cos\theta-|\zeta|)^2\big]\Big\}
-
\int_{0}^{\pi}d\theta\frac{\sin\theta}
{(1-|\zeta|\cos\theta)^{2}\big[
\big(\mathcal{I'}+1-|\zeta|\cos\theta\big)^2
-\zeta^2(\cos\theta-|\zeta|)^2\big]^2}\nonumber\\
&&\times\frac{\eta_{\zeta}(1-\zeta^2)}{16\pi^2v^2v_{z}}
\Big\{2\mathcal{J'}\big\{\mathcal{I'}^2(1-\zeta^2)
\big[(\cos\theta-|\zeta|)^2+\sin^2\theta\big]
+(1-\zeta^2)^2
\big[(\cos\theta-|\zeta|)^2+\sin^2\theta\big]^2\nonumber\\
&&+2\mathcal{I'}
(1-|\zeta|\cos\theta)
\big[(1-2\zeta^2)(\cos\theta-|\zeta|)^2
+(1-\zeta^2)\sin^2\theta\big]\big\}
-\frac{1}{2\mathcal{I'}}\big\{\mathcal{I'}^{3}
+2(1-|\zeta|\cos\theta)^{3}
+4\mathcal{I'}^2(1-|\zeta|\cos\theta)\nonumber\\
&&+\mathcal{I'}\big[5(1-|\zeta|\cos\theta)
-\zeta^2(\cos\theta-|\zeta|)^2\big]\big\}
\big\{\frac{\lambda^2\mathcal{J}(1-\zeta^2)^2\sin^4\theta}{2v^2}
+\frac{2\lambda_{z}^2\mathcal{J}(\cos\theta-|\zeta|)^4}{v_{z}^2}
\nonumber\\
&&+\frac{\frac{1}{2}\lambda^2(1-\zeta^2)^2\sin^4\theta
+4\lambda\lambda_{z}(1-\zeta^2)\sin^2\theta
(\cos\theta-|\zeta|)^2}{vv_{z}\big[\frac{(1-|\zeta|\cos\theta)^2}{v^2}
+\frac{(\cos\theta-|\zeta|)^2}{v_{z}^2}\big]}
+4\big\{
\frac{\lambda^2(1-\zeta^2)\sin^2\theta}{v^2
\big[\frac{(1-\zeta^2)\sin^2\theta}{v^2}
+\frac{(\cos\theta-|\zeta|)^2}{v_{z}^2}\big]}
+\frac{\lambda_{z}^2(\cos\theta-|\zeta|)^2}{v_{z}^2
\big[\frac{(1-\zeta^2)\sin^2\theta}{v^2}
+\frac{(\cos\theta-|\zeta|)^2}{v_{z}^2}\big]}\big\}\nonumber\\
&&\times\zeta^2(\cos\theta-|\zeta|)^2
\big\}
+\zeta^2\mathcal{J'}(\cos\theta-|\zeta|)^2
\big[\mathcal{I'}^2+4\mathcal{I'}(1-|\zeta|\cos\theta)
+3(1-|\zeta|\cos\theta)^2-\zeta^2(\cos\theta-|\zeta|)^2
\big]\Big\},\\
\mathcal{D}&\equiv&
\frac{-\lambda_{z}^2(1-\zeta^2)^2\eta_{\zeta}}{8\pi^2v^4v_{z}}
\int_{0}^{\pi}d\theta\frac{\sin^{3}\theta}{(1-|\zeta|\cos\theta)^{2}\mathcal{J}^{-1}
\big[(1-|\zeta|\cos\theta+\mathcal{I})^2
-\zeta^2(\cos\theta-|\zeta|)^2\big]^2}\nonumber\\
&&\times\Big\{\mathcal{I}^2
\big[(1-|\zeta|\cos\theta)^2-\zeta^2(\cos\theta-|\zeta|)^2\big]
+2\mathcal{I}(1-|\zeta|\cos\theta)
\big[(1-|\zeta|\cos\theta)^2-2\zeta^2(\cos\theta-|\zeta|)^2\big]
\nonumber\\
&&+\big[(1-|\zeta|\cos\theta)^2-\zeta^2(\cos\theta-|\zeta|)^2\big]^2\Big\}
-\frac{\lambda_{z}^2(1-\zeta^2)\eta_{\zeta}}{32\pi^2v^2v_{z}^3}
\int_{0}^{\pi}d\theta
\frac{\sin\theta(\cos\theta-|\zeta|)^{2}}{(1-|\zeta|\cos\theta)^{2}\mathcal{J}^{-1}
\big[(1-|\zeta|\cos\theta+\mathcal{I'})^2-\zeta^2(\cos\theta-|\zeta|)^2\big]^2}\nonumber\\
&&\times\Big\{\mathcal{I'}^2
\big[(1-|\zeta|\cos\theta)^2-\zeta^2(\cos\theta-|\zeta|)^2\big]
+2\mathcal{I'}(1-|\zeta|\cos\theta)
\big[(1-|\zeta|\cos\theta)^2-2\zeta^2(\cos\theta-|\zeta|)^2\big]\nonumber\\
&&+\big[(1-|\zeta|\cos\theta)^2-\zeta^2(\cos\theta-|\zeta|)^2\big]^2\Big\},\\
\mathcal{D}_{z}&\equiv&
\frac{(1-\zeta^2)\eta_{\zeta}}{8\pi^2v^2v_{z}}
\int_{0}^{\pi}d\theta
\frac{
\big\{\lambda_{z}^2\frac{(1-\zeta^2)\sin^{2}\theta}{v^2}
-\lambda^2\big[\frac{(1-\zeta^2)\sin^{2}\theta}{v^2}
+\frac{(\cos\theta-|\zeta|)^{2}}{v_{z}^2}\big]\big\}\sin\theta}
{(1-|\zeta|\cos\theta)^{2}\mathcal{J}^{-1}
\big[(1-|\zeta|\cos\theta+\mathcal{I})^2
-\zeta^2(\cos\theta-|\zeta|)^2\big]^2}\nonumber\\
&&\times\Big\{\mathcal{I}^2
\big[(1-|\zeta|\cos\theta)^2-\zeta^2(\cos\theta-|\zeta|)^2\big]
+2\mathcal{I}(1-|\zeta|\cos\theta)
\big[(1-|\zeta|\cos\theta)^2-2\zeta^2(\cos\theta-|\zeta|)^2\big]
+\big[(1-|\zeta|\cos\theta)^2\nonumber\\
&&-\zeta^2(\cos\theta-|\zeta|)^2\big]^2\Big\}
+\frac{(1-\zeta^2)\eta_{\zeta}}{32\pi^2v^2v_{z}}
\int_{0}^{\pi}d\theta
\frac{\sin\theta\big(\lambda_{z}^2\frac{(\cos\theta-|\zeta|)^{2}}{v_{z}^2}
-2\lambda^2\frac{(1-\zeta^2)
\sin^{2}\theta}{v^2}\big)}{(1-|\zeta|\cos\theta)^{2}\mathcal{J}^{-1}
\big[(1-|\zeta|\cos\theta+\mathcal{I'})^2
-\zeta^2(\cos\theta-|\zeta|)^2\big]^2}\nonumber\\
&&\times\Big\{\mathcal{I'}^2
\big[(1-|\zeta|\cos\theta)^2-\zeta^2(\cos\theta-|\zeta|)^2\big]
+2\mathcal{I'}(1-|\zeta|\cos\theta)
\big[(1-|\zeta|\cos\theta)^2-2\zeta^2(\cos\theta-|\zeta|)^2\big]
\nonumber\\
&&+\big[(1-|\zeta|\cos\theta)^2-\zeta^2(\cos\theta-|\zeta|)^2\big]^2
\Big\},\\
\mathcal{F}^{xx}_{T}&=&\frac{\eta_{\zeta}(1-\zeta^2)
}{8\pi^2v^2v_{z}}\int_{0}^{\pi}d\theta
\frac{\mathcal{J}^{2}\sin\theta(1-|\zeta|\cos\theta)}{\mathcal{I}^3}
\Big[\big(\frac{7}{4}V_{T}^{xx}
+\frac{(V_{T}^{xy})^2}{V_{T}^{xx}}
+8\frac{(V_{T}^{xz})^2}{V_{T}^{xx}}
+\frac{5}{2}V_{T}^{xy}
\big)\frac{(1-\zeta^2)^2\sin^{4}\theta}{4v^4}\nonumber\\
&&+\frac{\big[7(V_{T}^{xx})^2+4(V_{T}^{xy})^2\big](-|\zeta|+\cos\theta)^4}{2V_{T}^{xx}v_{z}^4}
+\big(\frac{7}{2}V_{T}^{xx}
+2\frac{(V_{T}^{xy})^2}{V_{T}^{xx}}
+\frac{5}{2}V_{T}^{xz}+2\frac{V_{T}^{xy}V_{T}^{xz}}{V_{T}^{xx}}\big)
\frac{(-|\zeta|+\cos\theta)^2
(1-\zeta^2)\sin^{2}\theta}{v^2v_{z}^2}
\Big],\\
\mathcal{F}^{zz}_{T}
&=&\frac{\eta_{\zeta}(1-\zeta^2)
}{8\pi^2v^2v_{z}}\int_{0}^{\pi}d\theta
\frac{\mathcal{J}^{2}\sin\theta(1-|\zeta|\cos\theta)}{\mathcal{I}^3}
\Big\{
+\frac{\big[14(V_{T}^{zz})^2+4(V_{T}^{xz})^2\big](1-\zeta^2)^2\sin^{4}\theta}{4V_{T}^{zz}v^4}
+\frac{4(V_{T}^{xz})^2(-|\zeta|+\cos\theta)^4}{V_{T}^{zz}v_{z}^4}\nonumber\\
&&+
\frac{\big[5(V_{T}^{xz})^{2}+4(V_{T}^{xz})^2\big](-|\zeta|+\cos\theta)^2
(1-\zeta^2)\sin^{2}\theta}{V_{T}^{zz}v^2v_{z}^2}
\Big\},\\
\mathcal{F}^{xy}_{T}
&=&\frac{\eta_{\zeta}(1-\zeta^2)
}{8\pi^2v^2v_{z}}\int_{0}^{\pi}d\theta
\frac{\mathcal{J}^{2}\sin\theta(1-|\zeta|\cos\theta)}{\mathcal{I}^3}
\Big[\big(\frac{(3V_{T}^{xx})^2}{2V_{T}^{xy}}
+V_{T}^{xy}+\frac{V_{T}^{xx}V_{T}^{xz}}{2V_{T}^{xy}}
+8\frac{(V_{T}^{xz})^2}{V_{T}^{xy}}
+3V_{T}^{xx}
\big)\frac{(1-\zeta^2)^2\sin^{4}\theta}{4v^4}\nonumber\\
&&+\frac{6V_{T}^{xx}(-|\zeta|+\cos\theta)^4}{v_{z}^4}
+\big(6V_{T}^{xx}+2V_{T}^{xz}
+\frac{3V_{T}^{xz}V_{T}^{xx}}{V_{T}^{xy}}\big)
\frac{(-|\zeta|+\cos\theta)^2
(1-\zeta^2)\sin^{2}\theta}{v^2v_{z}^2}
\Big],\\
\mathcal{F}^{xz}_{T}
&=&\frac{\eta_{\zeta}(1-\zeta^2)
}{8\pi^2v^2v_{z}}
\int_{0}^{\pi}d\theta
\frac{\mathcal{J}^{2}\sin\theta(1-|\zeta|\cos\theta)}{\mathcal{I}^3}
\Big[\big(2V_{T}^{xy}
+3V_{T}^{xx}
+16V_{T}^{zz}
\big)\frac{(1-\zeta^2)^2\sin^{4}\theta}{4v^4}\nonumber\\
&&+\frac{\big(3V_{T}^{xx}+2V_{T}^{xy}\big)(-|\zeta|+\cos\theta)^4}{v_{z}^4}
+\big(3V_{T}^{xx}
+2V_{T}^{xz}
+2V_{T}^{xy}
+\frac{2V_{T}^{xy}V_{T}^{zz}}{V_{T}^{xz}}
+\frac{2V_{T}^{xx}V_{T}^{zz}}{V_{T}^{xz}}\big)
\frac{(-|\zeta|+\cos\theta)^2
(1-\zeta^2)\sin^{2}\theta}{v^2v_{z}^2}
\Big],\\
\mathcal{F}^{xx}_{L}
&=&\frac{\eta_{\zeta}(1-\zeta^2)
}{8\pi^2v^2v_{z}}\int_{0}^{\pi}d\theta
\frac{\mathcal{J}^{2}\sin\theta(1-|\zeta|\cos\theta)}{\mathcal{I'}^3}
\big[\big(\frac{7}{4}V_{L}^{xx}
+\frac{(V_{L}^{xy})^2}{V_{L}^{xx}}
+\frac{5}{2}V_{L}^{xy}
\big)\frac{(1-\zeta^2)^2\sin^{4}\theta}{4v^4}
+\frac{2(V_{L}^{xz})^2(-|\zeta|+\cos\theta)^4}{V_{L}^{xx}v_{z}^4}\nonumber\\
&&+\big(\frac{5}{2}V_{L}^{xz}+2\frac{V_{L}^{xy}V_{L}^{xz}}{V_{L}^{xx}}\big)
\frac{(-|\zeta|+\cos\theta)^2
(1-\zeta^2)\sin^{2}\theta}{v^2v_{z}^2}
\Big],\\
\mathcal{F}^{zz}_{L}
&=&\frac{\eta_{\zeta}(1-\zeta^2)}{8\pi^2v^2v_{z}}\int_{0}^{\pi}d\theta
\frac{\mathcal{J}^{2}\sin\theta(1-|\zeta|\cos\theta)}{\mathcal{I'}^3}
\Big[\frac{4(V_{L}^{xz})^2(1-\zeta^2)^2\sin^{4}\theta}{4V_{L}^{zz}v^4}
+\frac{7V_{L}^{zz}(-|\zeta|+\cos\theta)^4}{2v_{z}^4}\nonumber\\
&&+\frac{5V_{L}^{xz}(-|\zeta|+\cos\theta)^2
(1-\zeta^2)\sin^{2}\theta}{v^2v_{z}^2}
\Big],\\
\mathcal{F}^{xy}_{L}&=&\frac{\eta_{\zeta}(1-\zeta^2)
}{8\pi^2v^2v_{z}}\int_{0}^{\pi}d\theta
\frac{\mathcal{J}^{2}\sin\theta(1-|\zeta|\cos\theta)}{\mathcal{I'}^3}
\Big[\big(\frac{(9V_{L}^{xx})^2}{4V_{T}^{xy}}
+V_{L}^{xy}+3V_{L}^{xx}
\big)\frac{(1-\zeta^2)^2\sin^{4}\theta}{4v^4}
+\frac{2(V_{T}^{xz})^2(-|\zeta|+\cos\theta)^4}{V_{L}^{xy}v_{z}^4}\nonumber\\
&&+\big(2V_{L}^{xz}+\frac{3V_{L}^{xx}V_{L}^{xz}}{V_{L}^{xy}}\big)
\frac{(-|\zeta|+\cos\theta)^2
(1-\zeta^2)\sin^{2}\theta}{v^2v_{z}^2}
\Big],\\
\mathcal{F}^{xz}_{L}&=&\frac{\eta_{\zeta}(1-\zeta^2)}{8\pi^2v^2v_{z}}
\int_{0}^{\pi}d\theta
\frac{\mathcal{J}^{2}\sin\theta(1-|\zeta|\cos\theta)}{\mathcal{I'}^3}
\Big[\frac{\big(4V_{L}^{xx}+2V_{L}^{xy}
\big)(1-\zeta^2)^2\sin^{4}\theta}{4v^4}
+3V_{L}^{zz}
\frac{(-|\zeta|+\cos\theta)^4}{v_{z}^4}\nonumber\\
&&+\big(2V_{L}^{xz}
+\frac{11V_{L}^{xx}V_{L}^{zz}}{4V_{L}^{xz}}
+\frac{3V_{L}^{xy}V_{L}^{zz}}{2V_{L}^{xz}}\big)
\frac{(-|\zeta|+\cos\theta)^2
(1-\zeta^2)\sin^{2}\theta}{v^2v_{z}^2}
\Big], %\\
%\mathcal{F}&\equiv&
%\frac{\eta_{\zeta}(1-\zeta^{2})}{4\pi^{2}v^{2}v_{z}}
%\int_{0^{\pi}}d\theta\frac{\sin\theta(1-|\zeta|\cos\theta)}{\mathcal{I}^{3}}
%\big\{8+\mathcal{J}^{2}\big[\frac{13(1-\zeta^{2}\sin^{4}\theta)}{2v^{4}}+6(\cos\theta-|\zeta|)^{4}
%+\frac{8(\cos\theta-|\zeta|)^{2}(1-\zeta^{2})\sin^{2}\theta}{v^{2}v_{z}^{2}}
%\big]\big\}\\
%\mathcal{F'}&\equiv&
%\frac{\eta_{\zeta}(1-\zeta^{2})}{4\pi^{2}v^{2}v_{z}}
%\int_{0^{\pi}}d\theta\frac{\sin\theta(1-|\zeta|\cos\theta)}{\mathcal{I}^{3}}
%\big\{2+\mathcal{J}^{2}\big[\frac{13(1-\zeta^{2}\sin^{4}\theta)}{2v^{4}}+6(\cos\theta-|\zeta|)^{4}
%+\frac{8(\cos\theta-|\zeta|)^{2}(1-\zeta^{2})\sin^{2}\theta}{v^{2}v_{z}^{2}}
%\big]\big\}
\end{eqnarray}
\end{small}
\end{widetext}
with $\mathcal{I}$, $\mathcal{I'}$, $\mathcal{J}$, and $\mathcal{J'}$ being defined as
\begin{small}
\begin{eqnarray}
\mathcal{I}&\equiv& \sqrt{C_{T}^{2}\frac{\left(1-\zeta^2\right)\sin^2\theta}
{v^2}+C_{T_{z}}^{2}\frac{\left(\cos\theta-|\zeta|\right)^2}
{v_{z}^2}},\\
\mathcal{I'}&\equiv& \sqrt{C_{L}^{2}\frac{\left(1-\zeta^2\right)\sin^2\theta}
{v^2}+C_{L_{z}}^{2}\frac{\left(\cos\theta-|\zeta|\right)^2}
{v_{z}^2}},\\
\mathcal{J}&\equiv& \big[\frac{(\cos\theta-|\zeta|)^2}{v_{z}^2}}{\frac{(1-\zeta^2)\sin^2\theta}{v^2}\big]^{-1},\\
\mathcal{J'}&\equiv& \frac{\lambda^2\frac{(1-\zeta^2)\sin^2\theta}{v^2}
+\lambda_{z}^2\frac{(\cos\theta-|\zeta|)^2}{v_{z}^2}}{\frac{(1-\zeta^2)\sin^2\theta}{v^2}
+\frac{(\cos\theta-|\zeta|)^2}{v_{z}^2}}.
\end{eqnarray}
\end{small}

\begin{figure}
\centering
\includegraphics[width=3.1in]{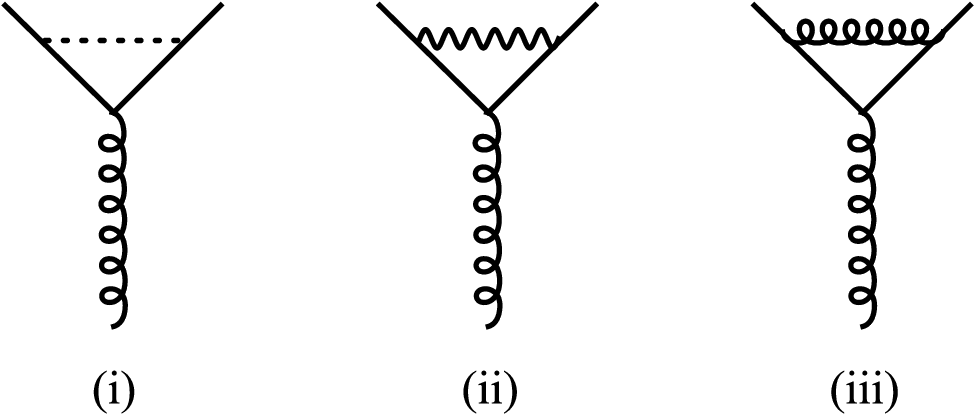}
\vspace{0cm}
\caption{One-loop corrections to the strength of source term $\Delta_{i}$
due to (a) the electron-phonon interaction, (b) the Coulomb interaction, and (c) the source-term itself, respectively.}\label{Fig_1L-source-terms}
\end{figure}

\section{Energy-dependent evolutions of source terms}\label{appendix-source-terms}

After completing the calculations for the one-loop corrections to the strengths of the source terms $\Delta^{\mathrm{PH/PP}}_{i}$
shown in Fig.~\ref{Fig_1L-source-terms}, we then can derive the RG equations with the help of
RG rescalings~(\ref{Eq_RG-rescaling-omega})-(\ref{Eq_RG-rescaling-u-L}). The concrete RG equations are listed as follows,
\begin{small}
\begin{eqnarray}
\frac{d\Delta^{\mathrm{PH}}_{0}}{dl}
\!\!&=&\!\!(1-2\eta_\psi)\Delta^{\mathrm{PH}}_{0}\label{Eq_Delta_1},\\
\frac{d\Delta^{\mathrm{PH}}_{1}}{dl}
\!\!&=&\!\!\left[1-2\eta_\psi+(\mathcal{R}_{1}^{+}-\mathcal{R}_{1}^{-})(-g^{2}-\lambda_{z}^{2})\right]\!\!\Delta^{\mathrm{PH}}_{1},\\
\frac{d\Delta^{\mathrm{PH}}_{2}}{dl}
&=&\left[1-2\eta_\psi+(\mathcal{R}_{2}^{+}-\mathcal{R}_{2}^{-})(-g^{2}-\lambda_{z}^{2})\right]\!\!\Delta^{\mathrm{PH}}_{2},\\
\frac{d\Delta^{\mathrm{PH}}_{3}}{dl}
\!\!&=&\!\!\left[1-2\eta_\psi\!+\!(\mathcal{R}_{3}^{+}\!-\!\mathcal{R}_{3}^{-})
(\lambda_{z}^{2}-g^{2}\!-2\lambda^{2})\right]\!\!\Delta^{\mathrm{PH}}_{3},
\end{eqnarray}
\end{small}
for the particle-hole channel, and
\begin{small}
\begin{eqnarray}
\frac{d\Delta^{\mathrm{PP}}_{0}}{dl}
\!\!&=&\!\!\left[1-2\eta_\psi+(\mathcal{T}_{0}^{+}-\mathcal{T}_{0}^{-})(-g^{2}-\lambda_{z}^{2})\right]\!\!\Delta^{\mathrm{PP}}_{0},\\
\frac{d\Delta^{\mathrm{PP}}_{1}}{dl}
\!\!&=&\!\!\left[1-2\eta_\psi+(\mathcal{T}_{1}^{+}-\mathcal{T}_{1}^{-})(\lambda_{z}^{2}-g^{2}-2\lambda^{2})\right]\!\!\Delta^{\mathrm{PP}}_{1},\\
\frac{d\Delta^{\mathrm{PP}}_{2}}{dl}
\!\!&=&\!\!\left[1-2\eta_\psi+(\mathcal{T}_{2}^{+}-\mathcal{T}_{2}^{-})(-g^{2}-\lambda_{z}^{2})\right]\!\!\Delta^{\mathrm{PP}}_{2},\\
\frac{d\Delta^{\mathrm{PP}}_{30}}{dl}
\!\!&=&\!\!\left[1-2\eta_\psi+(\mathcal{T}_{0}^{+}-\mathcal{T}_{0}^{-})(-g^{2}-\lambda_{z}^{2})
\right]\Delta^{\mathrm{PP}}_{30},\\
\frac{d\Delta^{\mathrm{PP}}_{31}}{dl}
\!\!&=&\!\!\left[1-2\eta_\psi
+(\mathcal{T}_{1}^{+}-\mathcal{T}_{1}^{-})(\lambda_{z}^{2}-g^{2}-2\lambda^{2})\right]\!\!\Delta^{\mathrm{PP}}_{31},\\
\frac{d\Delta^{\mathrm{PP}}_{33}}{dl}
\!\!&=&\!\!\left[1-2\eta_\psi
+(\mathcal{T}_{2}^{+}-\mathcal{T}_{2}^{-})(-g^{2}-\lambda_{z}^{2})\right]\!\!\Delta^{\mathrm{PP}}_{33},\label{Eq_Delta_2}
\end{eqnarray}
\end{small}
for the particle-particle channel, where all the related coefficients are designated as
\begin{widetext}
\begin{small}
\begin{eqnarray}
\mathcal{R}_{1}^{+}&=&\int_{0}^{\pi}\frac
{(1-\zeta\cos{\theta})|\zeta^{2}-1|\sin{\theta}(\cos^{2}{\theta}-\zeta^{2}\cos^{2}{\theta}+\zeta^{2}-4\zeta\sqrt{1-\zeta^{2}}\cos{\theta}+1)}
{16\pi^{2}(\zeta^{2}-1)^{2}(1-2\zeta\sqrt{1-\zeta^{2}}\cos{\theta})^{\frac{3}{2}}}d\theta,\\
\mathcal{R}_{1}^{-}&=&\int_{0}^{\pi}\frac
{(1+\zeta\cos{\theta})|\zeta^{2}-1|\sin{\theta}(\cos^{2}{\theta}-\zeta^{2}\cos^{2}{\theta}+\zeta^{2}+4\zeta\sqrt{1-\zeta^{2}}\cos{\theta}+1)}
{16\pi^{2}(\zeta^{2}-1)^{2}(1+2\zeta\sqrt{1-\zeta^{2}}\cos{\theta})^{\frac{3}{2}}}d\theta,\\
\mathcal{R}_{2}^{+}&=&\int_{0}^{\pi}\frac
{(1-\zeta\cos{\theta})|\zeta^{2}-1|\sin{\theta}(\cos^{2}{\theta}-\zeta^{2}\cos^{2}{\theta}+\zeta^{2}-4\zeta\sqrt{1-\zeta^{2}}\cos{\theta}+1)}
{16\pi^{2}(\zeta^{2}-1)^{2}(1-2\zeta\sqrt{1-\zeta^{2}}\cos{\theta})^{\frac{3}{2}}}d\theta,\\
\mathcal{R}_{2}^{-}&=&\int_{0}^{\pi}\frac
{(1+\zeta\cos{\theta})|\zeta^{2}-1|\sin{\theta}(\cos^{2}{\theta}-\zeta^{2}\cos^{2}{\theta}+\zeta^{2}+4\zeta\sqrt{1-\zeta^{2}}\cos{\theta}+1)}
{16\pi^{2}(\zeta^{2}-1)^{2}(1+2\zeta\sqrt{1-\zeta^{2}}\cos{\theta})^{\frac{3}{2}}}d\theta,\\
\mathcal{R}_{3}^{+}&=&\int_{0}^{\pi}\frac
{(1-\zeta\cos{\theta})|\zeta^{2}-1|\sin{\theta}(\cos^{2}{\theta}-1)}
{16\pi^{2}(\zeta^{2}-1)(1-2\zeta\sqrt{1-\zeta^{2}}\cos{\theta})^{\frac{3}{2}}}d\theta,\\
\mathcal{R}_{3}^{-}&=&\int_{0}^{\pi}\frac
{(1+\zeta\cos{\theta})|\zeta^{2}-1|\sin{\theta}(\cos^{2}{\theta}-1)}
{16\pi^{2}(\zeta^{2}-1)(1+2\zeta\sqrt{1-\zeta^{2}}\cos{\theta})^{\frac{3}{2}}}d\theta,
\end{eqnarray}
and
\begin{eqnarray}
\mathcal{T}_{0}^{+}&=&\int_{0}^{\pi}\frac{\sin^{3}{\theta}(1-\zeta\cos{\theta})}
{16\pi^{2}\zeta(1-\zeta)^{2}(\frac{\cos{\theta}}{\sqrt{1-\zeta^{2}}}-\frac{\zeta}{1-\zeta^{2}})
\left[\sin^{2}{\theta}-(\zeta^{2}-1)(\cos^{2}{\theta}-\frac{2\zeta}{\sqrt{1-\zeta^{2}}}+\frac{\zeta^{2}}{1-\zeta^{2}})\right]}d\theta,\\
\mathcal{T}_{0}^{-}&=&\int_{0}^{\pi}\frac{\sin^{3}{\theta}(1+\zeta\cos{\theta})}
{16\pi^{2}\zeta(1-\zeta)^{2}(\frac{\cos{\theta}}{\sqrt{1-\zeta^{2}}}-\frac{\zeta}{1-\zeta^{2}})
\left[\sin^{2}{\theta}-(\zeta^{2}-1)(\cos^{2}{\theta}+\frac{2\zeta}{\sqrt{1-\zeta^{2}}}+\frac{\zeta^{2}}{1-\zeta^{2}})\right]}d\theta,\\
\mathcal{T}_{1}^{+}&=&\int_{0}^{\pi}\frac{\sin{\theta}(1-\zeta\cos{\theta})(\cos^{2}
{\theta}-\frac{2\zeta}{\sqrt{1-\zeta^{2}}}\cos{\theta}+\frac{\zeta^{2}}{1-\zeta^{2}})}
{16\pi^{2}\zeta(1-\zeta^{2})^{2}(\frac{\cos{\theta}}{\sqrt{1-\zeta^{2}}}-\frac{\zeta}{1-\zeta^{2}})
\left[\sin^{2}{\theta}-(\zeta^{2}-1)(\cos^{2}{\theta}-\frac{2\zeta}{\sqrt{1-\zeta^{2}}}+\frac{\zeta^{2}}{1-\zeta^{2}})\right]}d\theta,\\
\mathcal{T}_{1}^{-}&=&\int_{0}^{\pi}\frac{\sin{\theta}(1+\zeta\cos{\theta})(\cos^{2}{\theta}+\frac{2\zeta}
{\sqrt{1-\zeta^{2}}}\cos{\theta}+\frac{\zeta^{2}}{1-\zeta^{2}})}
{16\pi^{2}\zeta(1-\zeta^{2})^{2}(\frac{\cos{\theta}}{\sqrt{1-\zeta^{2}}}-\frac{\zeta}{1-\zeta^{2}})
\left[\sin^{2}{\theta}-(\zeta^{2}-1)(\cos^{2}{\theta}+\frac{2\zeta}{\sqrt{1-\zeta^{2}}}+\frac{\zeta^{2}}{1-\zeta^{2}})\right]}d\theta,\\
\mathcal{T}_{2}^{+}&=&\int_{0}^{\pi}\frac{-\sin^{3}{\theta}(1-\zeta\cos{\theta})}
{16\pi^{2}\zeta(1-\zeta^{2})^{2}(\frac{\cos{\theta}}{\sqrt{1-\zeta^{2}}}-\frac{\zeta}{1-\zeta^{2}})
\left[\sin^{2}{\theta}-(\zeta^{2}-1)(\cos^{2}{\theta}-\frac{2\zeta}{\sqrt{1-\zeta^{2}}}+\frac{\zeta^{2}}{1-\zeta^{2}})\right]}d\theta,\\
\mathcal{T}_{2}^{-}&=&\int_{0}^{\pi}\frac{-\sin^{3}{\theta}(1+\zeta\cos{\theta})}
{16\pi^{2}\zeta(1-\zeta^{2})^{2}(\frac{\cos{\theta}}{\sqrt{1-\zeta^{2}}}-\frac{\zeta}{1-\zeta^{2}})
\left[\sin^{2}{\theta}-(\zeta^{2}-1)(\cos^{2}{\theta}+\frac{2\zeta}{\sqrt{1-\zeta^{2}}}+\frac{\zeta^{2}}{1-\zeta^{2}})\right]}d\theta.
\end{eqnarray}
\end{small}

\end{widetext}

%%%%%%%%%%%%%%%%%%%%%%%%%%%%%%%%%%%%%%%%%%%%%%%%%%%%%%%%%%%%%%%%%%%%%%%%%%%%%%%%%%%%%
%%%%%%%%%%%%%%%%%%%%%%%%%%%%%%%%%%%%%%%%%%%%%%%%%%%%%%%%%%%%%%%%%%%%%%%%%%%%%%%%%%%%%

%\end{CJK*}

\end{document}